%% file: manuscript.tex
\documentclass{article}       

\usepackage{graphicx}
\usepackage{booktabs}
\usepackage{authblk}
\usepackage{amsmath}
\usepackage{longtable}
\usepackage{epstopdf,subfigure}
\usepackage{textcomp}
\usepackage{caption}
\usepackage{multirow}
\usepackage{threeparttable}
\usepackage{newpxtext,newpxmath}
\usepackage{array, makecell,diagbox}
\usepackage{tablefootnote}
\usepackage{url}
\usepackage[table,xcdraw]{xcolor}

\title{A Structured Table of Graphs with Symmetries and Other Special Properties}
\author[1]{Yidan Zhang}
\author[2]{Xiaolong Huang}
\author[2,3]{Zhipeng Xu}
\author[2]{\\Yuefan Deng\thanks{Corresponding author: yuefan.deng@stonybrook.edu (Y. Deng)\\E-mails: yidan.zhang@stonybrook.edu (Y. Zhang), xiaolong.huang@stonybrook.edu (X. Huang), xuzhp9@mail2.sysu.edu.cn (Z. Xu)}}
\affil[1]{Mathematics Department, Stony Brook University, Stony Brook, NY 11794, USA}
\affil[2]{Department of Applied Mathematics and Statistics, Stony Brook University, Stony Brook, NY 11794, USA}
\affil[3]{School of Data and Computer Science, Sun Yat-sen University, Guangzhou, Guangdong 510006, P.R. China}

\begin{document}
\maketitle
\abstract{We organize a table of regular graphs with minimal diameters and minimal mean path lengths, large bisection widths and high degrees of symmetries, obtained by enumerations on supercomputers. These optimal graphs, many of which are newly discovered, may find wide applications, for example, in design of network topologies.}
\section{Introduction}
The next-generation supercomputers reaching exascales require sophisticated interconnection networks to couple hundreds of millions of processing units and these interconnection networks must be scalable at high bandwidth and low latency. In addition to the requirement of many-host clusters, from the aspect of micro-chips, more and more cores are integrated on one chip that require a high-performance interconnection network to couple them. The processing speed of network-on-chip (NoC), an aggregated on-chip infrastructure to enhance the energy performance among others, sensitively relies on its framework design to maximize the latency reduction and throughput. Also, data-centers in cloud computing also demand optimal architectures for dealing with expeditious growth of nodes, memory, and interconnection networks in a supercomputing system. 
\par
For this principle challenge, analyzing and optimizing network structures, of interconnection networks, graph theory is an acknowledged powerful method by transforming network components into vertices, and communication links into edges. The great focus of structure design hence is on the topologies of interconnection networks and the associated graph properties of them. The current prevailing basic modules include tree, near-neighbor mesh \cite{Barnes1968}, torus \cite{Dally1986}, star, Heawood graph, Peterson graph, hypercube \cite{Saad1988,Tripathy1997}, etc. Major operations on these typical base graphs, in searching for optimal network topology, are hierarchical interconnection graphs and Cartesian products. Hierarchical interconnection network enables large network structures to maintain desired properties, such as low diameter and low mean path length (MPL), of the basic graphs \cite{Lai2010,Guo2011}. Classic examples of hierarchical products graphs include deterministic tree \cite{Jung2002}, Dragonfly \cite{Wu1980}, and hierarchical hypercube \cite{Malluhi1994}. The Cartesian product, a straightforward graph operation that generates many eminent combined graphs, permits designers to reach a specified performance of larger-scale interconnection networks at minimal cost with parameters that can be directly required from the corresponding parameters of initial basic graphs \cite{Hammack2011}.
\par
It is obvious that most hierarchical networks are all based on Peterson graph and hypercube. Lacking diversity of orders, they limit the scales of clusters. Deng \textit{et al.} \cite{Deng2019} proposed  more optimal graphs and their benchmarks on a Beowulf cluster with these graphs prove to enhance network performance better than the mainstream networks. Xu \textit{et al.} \cite{sc19} applied these graphs to creating larger networks by using the Cartesian product, resolving scale limitations. To fill the family of optimal graphs, we use the exhaustive search as mentioned in \cite{Xu2019b} to find the graphs with the minimal MPL and other properties aiming for optimizing interconnection networks. These graphs can be used in NoC directly or, combining with hierarchical method or Cartesian products, used as the interconnects for clusters. Symmetry, as one of the important properties for graph filtering, is also an essential factor in characterizing and measuring complex networks \cite{Garlaschelli2010,Garrido2011} as demonstrated in Conder \cite{marston} who organized a series of highly symmetric graphs. Automorphism group directly represents graph symmetry, the computation of which has been extensively studied \cite{Balasubramanian1994} and applied in structural modeling and design \cite{Razinger1993,Balasubramanian2018}. The rest of this paper introduces, in Sec. 2, our criteria of optimal graphs and our structured table of the optimal base graphs and, in Sec. 3, our conclusions.
\section{Criteria and Optimal Graphs }
\subsection{Criteria }
\begin{itemize}
\item Diameter and MPL of a Regular Graph
\end{itemize}
\par
In the study of the components of interconnection networks, a graph with minimal diameter and minimal MPL markedly reduces the communication latency. Among all regular graphs, the Moore graphs and the generalized Moore graphs, with given numbers of vertices and of degrees, have the minimal diameters and minimal MPLs.
\par
Let $G$ be a connected regular graph of $N$ vertices and degree $k$. The distance $d(u,v)$ between two distinct vertices $u$ and $v$ is the length of the shortest path between these two vertices, while its diameter $D$ is the maximum $d(u,v)$ value \cite{Bapat2014}. For the generalized Moore graphs with given diameter $D$ and degree $k$, the upper bound on the number of vertices is achieved as 
\begin{equation*}
M_{k, D}=1+k \sum_{i=1}^{D}(k-1)^{i-1}=\left\{
\begin{aligned}
&\frac{k(k-1)^{D}-2}{k-2}\quad &(k \geq 3) \\
&1+2D\quad &(k=2)
\end{aligned}\right..
\end{equation*}
known as Moore bound, named by Hoffman and Singleton after E. F. Moore who first studied the problem \cite{Hoffman2003}.
By inverting the Moore bound with regard to $D$ and $k$, the lower bound of diameter $D_{k, N}$  can be attained (degree of 2 is a trivial case):
\begin{equation*}
D_{k, N}=\left\lceil\log _{k-1} \frac{N(k-2)+2}{k}\right\rceil\quad(k \geq 3).
\end{equation*}
The MPL in a graph is the average, taken over all pairs of vertices, of the shortest path lengths between two vertices. If $G$ is a connected graph with $N$ vertices, the MPL is calculated by
\begin{equation*}
\mathrm{MPL}=\frac{1}{N(N-1)} \sum_{u \neq v} d(u,v).
\end{equation*}
Cerf \textit{et al.} \cite{Cerf1974} proved a lower bound on MPL for any regular graph G of $N$ vertices and degree $k$. Moreover, Cerf \textit{et al.} \cite{Cerf1974} defines a regular graph with MPL equal to the lower bound as the generalized Moore graph.
\begin{equation*}
\mathrm{MPL}_{min}=\frac{1}{N-1}\left(k \sum_{i=1}^{D_{k, N}}(k-1)^{i-1} i-\left(M_{k, D_{k, N}}-N\right) D_{k, N}\right)\quad(k \geq 3).
\end{equation*}
\begin{itemize}
\item Bisection Bandwidth
\end{itemize}
\par
A bisection of a graph is a bipartition of its vertex set in which the number of vertices in the two parts differ by at most 1, and its size is the number of edges which go across the two parts. The bisection width of a graph is its minimum bisection size. In network topologies, bisection bandwidth is the minimum communication volume allowed into these removed links while recognizing bisection. Bisection bandwidth is commonly used to estimate the achievable throughput and latency of a network, and an increase in bisection bandwidth of a network improves the overall system performance. Let $C(N_1,N_2)$ denote the set of edges removed to divide the total vertices $N$ into two disjoint sets $N_1$ and $N_2$. The number of removed edges is $|C(N_1,N_2)|$, and the bisection width is calculated as follows \cite{William2003}:
\begin{equation*}
\mathrm{B}_{\mathrm{C}}=\min _{\text {bisections}} \left|\mathrm{C}\left(\mathrm{N}_{1}, \mathrm{N}_{2}\right)\right|.
\end{equation*}
\begin{itemize}
\item Automorphism Group Size
\end{itemize}
\par
An automorphism of a graph $G$ is a permutation map from the vertex set to another vertex set which preserves adjacency of vertices. The set of all automorphisms of a graph $G$ is defined as the automorphism group of $G$, which is denoted by $\mathrm{Aut}(G)$. Since edge symmetry in interconnection networks ameliorates load balance across the links of the network, we further analyze on our graphs by utilizing automorphism to identify the symmetries. In general, more symmetrical network leads to better performance as in routing and robustness \cite{William2003,MacArthur2008}.
\subsection{Optimal Graphs}
\par
\begin{figure}[htbp!]
	\centering
\includegraphics[width=15cm]{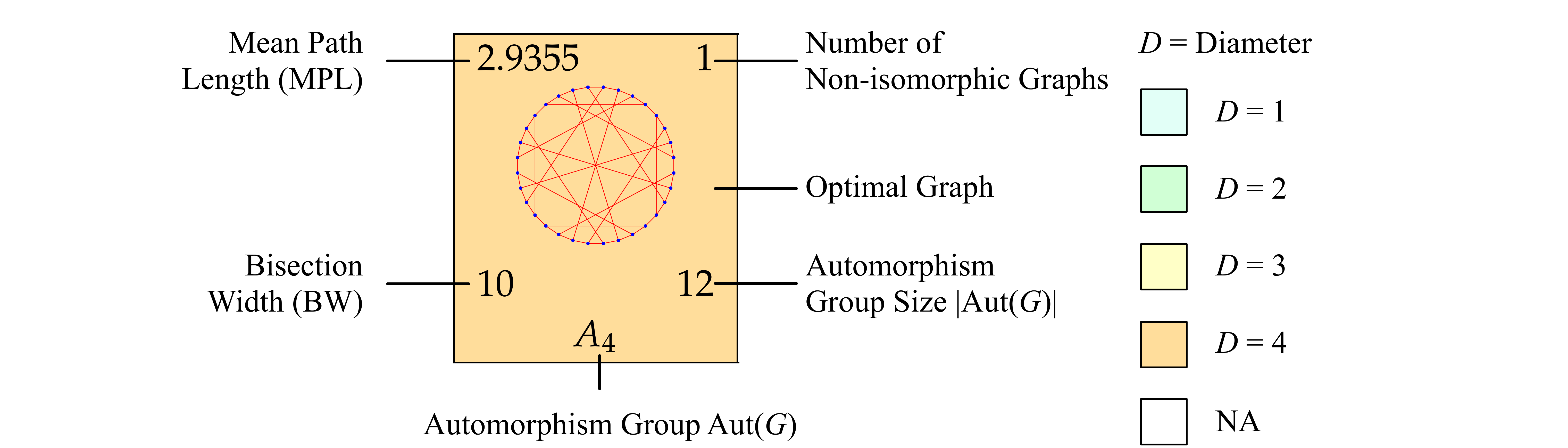} 
	\caption{Graph legends.}\label{sample}
\end{figure}

We denote regular graphs of $N$ vertices and degree $k$ as $(N,k)$ graphs. For fixed $N$ and $k$, we filter all $(N,k)$ graphs by minimizing diameter and MPL and maximizing bisection width and automorphism group size in such specific order, and define the filtered graphs as optimal $(N,k)$ graphs. The above filtering order is determined by the importance of each parameter in interconnect design as illustrated in \cite{Dally1986} and demonstrated by benchmarking performance in \cite{Deng2019}. Hence, an optimal graph has the properties of minimal diameter, minimal MPL, high throughput and high symmetry. We searched for the optimal graphs by enumeration using the same method as in \cite{Xu2019b} on supercomputers, and adopted data partially from \cite{Xu2019b}. In particular, we exhaustively calculated the diameters and MPLs of $\sim 10^{12}$ and $10^{13}$ graphs for $(21,4)$ and $(32,3)$ respectively while, for the special case of $(32,4)$, the number of non-isomorphic graphs is too massive to exhaustively search. We calculated the bisection widths using KaHIP \cite{sanders2013think} and automorphism groups using GAP \cite{GAP4} and SageMath \cite{sagemath}, both parallelized using GNU Parallel \cite{9781387509881} on supercomputers. The optimal graphs are organized in Tables \ref{r1}-\ref{rest}, with table cell background colors indicating the diameters and other parameters placed around each graph as shown in the legend (Figure \ref{sample}). The adjacency matrices of these optimal graphs can be provided upon request. These tables provide a structured representation of the optimal graphs with attributes for concise, convenient and rich references. The symbols for automorphism groups are defined in Table  \ref{symbol} with information in \cite{Dummit2003} for additional details. In the Tables \ref{r1}-\ref{tab:table2}, three $(N, k)$ pairs possess 2 or 3 optimal graphs and one such optimal graph is recorded in Tables \ref{r1}-\ref{rest} while the remainder is tabulated in the auxiliary Table \ref{rest}. The asterisk marked on the parameter indicates not being equal to the theoretical lower bound, but still minimum. Table \ref{semiproduct} provides the finite presentations of semidirect product groups in Tables  \ref{r1}-\ref{rest}. A subset of the optimal graphs are named graphs that are confirmed by comparing with existing graph database in Mathematica \cite{wolfgraph} and their properties are presented in Table \ref{tab:summary}.
\par
We also noted that many optimal graphs belong to the same category or have common properties. From Table \ref{tab:summary}, we see that many optimal graphs are Cayley or circulant or both. Some of the optimal cubic graphs are the smallest crossing number graphs \cite{pegg2009crossing}. Within the range of $(N,k)$ covered by us, cage graphs \cite{exoo2008dynamic}, and complete multipartite graphs of which the automorphism groups are wreath products of symmetric groups, are all optimal. The comparison with named graphs may lead to further study of more potential properties of the optimal graphs, which in return can expand the discovery and application. 

\begin{table}[h]
	\centering
	\caption{Illustrations of symbols used for automorphism groups.}
	\label{symbol}
	\begin{tabular}{@{}ll@{}}
		\toprule
		Symbol              & \multicolumn{1}{c}{Description}                              \\ \midrule
		$A_{n}$             & Alternating group on a set of length $n$                     \\
		$C_{n}$             & Cyclic group of order $n$                                    \\
		$D_{n}$             & Dihedral group of order $2n$                                 \\
		$\mathrm{GL}(n,p)$  & General linear group of degree $n$ over finite field $F_{p}$ \\
		$\mathrm{PGL}(n,p)$ & Projective general linear group obtained from $GL(n,p)$      \\
		$S_{n}$             & Symmetric group on a set of length $n$                       \\
		$\mathrm{Aut}(H)$ & Automorphism group of group $H$\\
		$\mathrm{Hol}(H)$ & Holomorph of group $H$\\
		$\times$            & Direct product                                               \\
		${H}^{m}$       & Direct product of $m$ copies of group $H$           \\
		$\rtimes$   & Semidirect product (For $K \rtimes H$ used here, $H$ acts faithfully on $K$)  \\
		$\wr$               & Wreath product                                               \\ \bottomrule
	\end{tabular}
\end{table}

\def\updist{0.9cm}

\input{table-color.tex}
\input{table2-color.tex}
\input{table3-color.tex}
\begin{table}[h]
	\centering
	\caption{The finite presentations of semidirect product groups in Tables \ref{r1}-\ref{rest}.}
	\label{semiproduct}
	\begin{tabular}{@{}lll@{}}
		\toprule
		$(N,k)$                 & Semidirect Product Group Structure                                       & Finite Presentation (reduced)                                    \\ \midrule
		(26,3)                  & \multirow{2}{*}{$C_{13} \rtimes C_{4}$}                           & \multirow{2}{*}{$\langle a, b ~|~ a^{13}, b^{4}, bab^{-1}a^{-5} \rangle$}    \\
		(13,4)                  &                                                                   &                                                                  \\ \midrule
		\multirow{2}{*}{(10,4)} & \multirow{2}{*}{$C_{2} \times ({C_{2}}^{4} \rtimes D_{5})$}       & $\langle a, b, c ~|~ a^{2}, b^{2}, c^{2}, (ab)^{4}, $                  \\
		&                                                                   & $(acbc)^{2}, (ca)^{4}, (cb)^{5}, (babc)^{4} \rangle$                   \\ \midrule
		\multirow{2}{*}{(20,4)} & \multirow{2}{*}{$GL(2,3) \rtimes C_{2}$}                          & $\langle a, b ~|~ a^{2}, (b^{-2}aba)^{2},$                             \\
		&                                                                   & $b^{-1}ab^{4}ab^{-3}, (ab^{-1}ab)^{3} \rangle$                         \\ \midrule
		\multirow{3}{*}{(12,7)} & \multirow{3}{*}{$(C_{2} \times (S_{4} \wr S_{2})) \rtimes C_{2}$} & $\langle a, b, c, d ~|~ a^{2}, b^{2}, c^{2}, d^{2}, (ab)^{2}, $                   \\
		&                                                                   & $(ac)^{2}, (bc)^{3}, (bd)^{4}, (da)^{4},(dadc)^{2}, $              \\
		&                                                                   & $(dbdc)^{2}, (dc)^{4}, (bdad)^{4}, (dcbdadbc)^{4} \rangle$ \\ \bottomrule
	\end{tabular}
\end{table}
\begin{longtable}[htbp!]{p{0.1\textwidth}p{0.1\textwidth}p{0.7\textwidth}cccl}
\caption{A summary of the properties of the optimal graphs in Tables  \ref{r1}-\ref{rest}.}
\label{tab:summary}\\
\toprule
\multicolumn{1}{l}{Degree $k$} & \multicolumn{1}{l}{Vertices $N$} &\multicolumn{1}{c} {Named Graphs}                                                                                                                                     \\* \midrule
\endfirsthead
\multicolumn{3}{c}%
{{\small{\bfseries Table \thetable}.\ A summary of the properties of the optimal graphs in Tables \ref{r1}-\ref{rest} (continued).}} \\
\toprule
\multicolumn{1}{l}{Degree $k$} & \multicolumn{1}{l}{Vertices $N$} & \multicolumn{1}{c}{Named Graphs}                                                                                                                                     \\* \midrule
\endhead
\bottomrule
\endfoot
\endlastfoot
         & 4                             & Tetrahedral graph; Complete; Distance-transitive; Strongly regular; Cayley; (3,3)-Cage; Circulant; Planar; Smallest cubic crossing number-0     \\*
                             & 6                             & Thomsen graph; Complete bipartite; Distance-transitive; Strongly regular; Cayley; (3,4)-Cage; Circulant; Smallest cubic crossing number-1       \\*
                             & 8                             & Wagner graph; Vertex-transitive; Cayley; Circulant; tripartite                                                                                  \\*
                             & 10                            & Petersen graph; Nonhamiltonian; Distance-transitive; Strongly regular; (3,5)-Cage; Smallest cubic crossing number-2; tripartite                 \\*
          \multirow{3}{*}{3}                    & 14                            & Heawood graph; Bipartite; Distance-transitive; Cayley; (3,6)-Cage; Smallest cubic crossing number-3                                             \\*
                             & 22                            & Smallest cubic crossing number-7; tripartite                                                                                                    \\*
                             & 24                            & McGee graph; (3,7)-Cage; Smallest cubic crossing number-8; tripartite                                                                           \\*
                             & 26                            & Generalized Petersen-(13,5); Vertex-transitive; Smallest cubic crossing number-9; tripartite                                                    \\*
                             & 28                            & Coxeter graph; Nonhamiltonian; Distance-transitive; Smallest cubic crossing number-11; tripartite                                               \\*
                             & 30                            & Levi graph; Bipartite; Distance-transitive; (3,8)-Cage; Smallest cubic crossing number-13 (to be proved)                                         \\ \midrule
         & 5                             & Pentatope graph; Complete; Distance-transitive; Strongly regular; Cayley; (4,3)-Cage; Circulant                                                 \\*
                             & 6                             & Octahedral graph; Complete tripartite; Distance-transitive; Strongly regular; Cayley; Planar; Circulant                                         \\*
                             & 8                             & Complete bipartite; Distance-transitive; Strongly regular; Cayley; (4,4)-Cage; Circulant                                                        \\*
          \multirow{2}{*}{4}                     & 9                             & Generalized quadrangle-(2,1); (2, 3)-Hamming graph; (3, 3)-rook graph; 9-Paley graph; Distance-transitive; Strongly regular; Cayley; tripartite \\*
                             & 10                            & Arc-transitive; Cayley; Circulant; tripartite                                                                                                   \\*
                  & 11                        & 4-Andrásfai graph; Vertex-transitive; Cayley; Circulant; tripartite                                                                             \\*
                             & 12                            & Vertex-transitive; Cayley; Circulant; tripartite                                                                                            \\*
                             & 13                            & 13-Cyclotomic graph; Arc-transitive; Cayley; Circulant; 4-partite                                                                               \\*
                             & 19                            & Robertson graph; (4,5)-cage; tripartite                                                                                                         \\*
                             & 21 [Tab. \ref{rest}]                           & Brinkmann graph; 4-partite                                                                                                                      \\* \midrule
           & 6                             & Complete; Distance-transitive; Strongly regular; Cayley; (5,3)-Cage; Circulant                                                                  \\*
                 5            & 8                        & (5,3)-Cone graph; 4-partite                                                                                                                     \\*
                             & 10                            & Complete bipartite; Distance-transitive; Strongly regular; Cayley; (5,4)-Cage; Circulant                                                        \\ \midrule
        & 7                             & Complete; Distance-transitive; Strongly regular; Cayley; (6,3)-Cage; Circulant                                                                  \\*
                             & 8                             & 16-Cell; Complete 4-partite; Distance-transitive; Strongly regular; Cayley; Circulant                                                           \\*
           6                  & 9                             & Complete tripartite; Distance-transitive; Strongly regular; Cayley; Circulant                                                                   \\*
                             & 10                           & (6,4)-Cone graph; tripartite                                                                                                                    \\*
                             & 12                            & Complete bipartite; Distance-transitive; Strongly regular; Cayley; (6,4)-Cage; Circulant                                                        \\* \midrule
\multirow{3}{*}{7}           & 8                             & Complete; Distance-transitive; Strongly regular; Cayley; (7,3)-Cage; Circulant                                                                  \\*
                             & 12                      & Vertex-transitive; Cayley; Circulant; 4-partite                                                                                                 \\ \bottomrule
\end{longtable}
\section{Conclusions}
Using the exhaustive search with supercomputers, we enumerate regular graphs with a wide range of orders, filter them with the diameter, MPL, bisection bandwidth, and symmetry. The benchmarks on a Beowulf also prove that the feature of symmetry affects the performance of networks, and these optimal graphs can be applied to many areas including the designs of microchips and data centers. We supply new base graphs and expansion methods to construct larger and diverse interconnection networks.
\section*{Acknowledgments}
The research of Z.X. is partially supported by the Special Project on High-Performance Computing of the National Key R\&D Program under No.2016YFB0200604.The authors thank Stony Brook Research Computing and Cyberinfrastructure, and the Institute for Advanced Computational Science at Stony Brook University for access to the high-performance SeaWulf computing system, which was made possible by a \textdollar 1.4M National Science Foundation grant (\#1531492).

\bibliographystyle{unsrt}
\bibliography{refs.bib}
\end{document}

%% file: table-color.tex
\begin{longtable}[c]{|c|lr|cr|lr|lr|lr}
\caption{The optimal graphs for $N\leq 12$ and $k\leq 7$.}
\label{r1}\\*
\hline\diagbox{$N$}{$k$}
                     & \multicolumn{2}{c|}{3}                                                                & \multicolumn{2}{c|}{4}                                                                 & \multicolumn{2}{c|}{5}                                                                 & \multicolumn{2}{c|}{6}                                                                           & \multicolumn{2}{c|}{7}                                                                           \\ \hline
\endfirsthead
\multicolumn{11}{c}%
{\small{\bfseries Table \thetable.} The optimal graphs for $N\leq 12$ and $k\leq 7$ (continued).}\\*
\hline\diagbox{$N$}{$k$}
                     & \multicolumn{2}{c|}{3}                                                                & \multicolumn{2}{c|}{4}                                                                 & \multicolumn{2}{c|}{5}                                                                 & \multicolumn{2}{c|}{6}                                                                           & \multicolumn{2}{c|}{7}                                                                           \\ \hline
\endhead
\hline
\endfoot
\endlastfoot
                     & \cellcolor[HTML]{e2fff7}1.0000              & \cellcolor[HTML]{e2fff7}1               & \multicolumn{1}{l}{}                                   &                               &                                            &                                           &                                                &                                                 &                                        & \multicolumn{1}{r|}{}                                   \\*
                     & \multicolumn{2}{c|}{\cellcolor[HTML]{e2fff7}\def\svgwidth{1.5cm}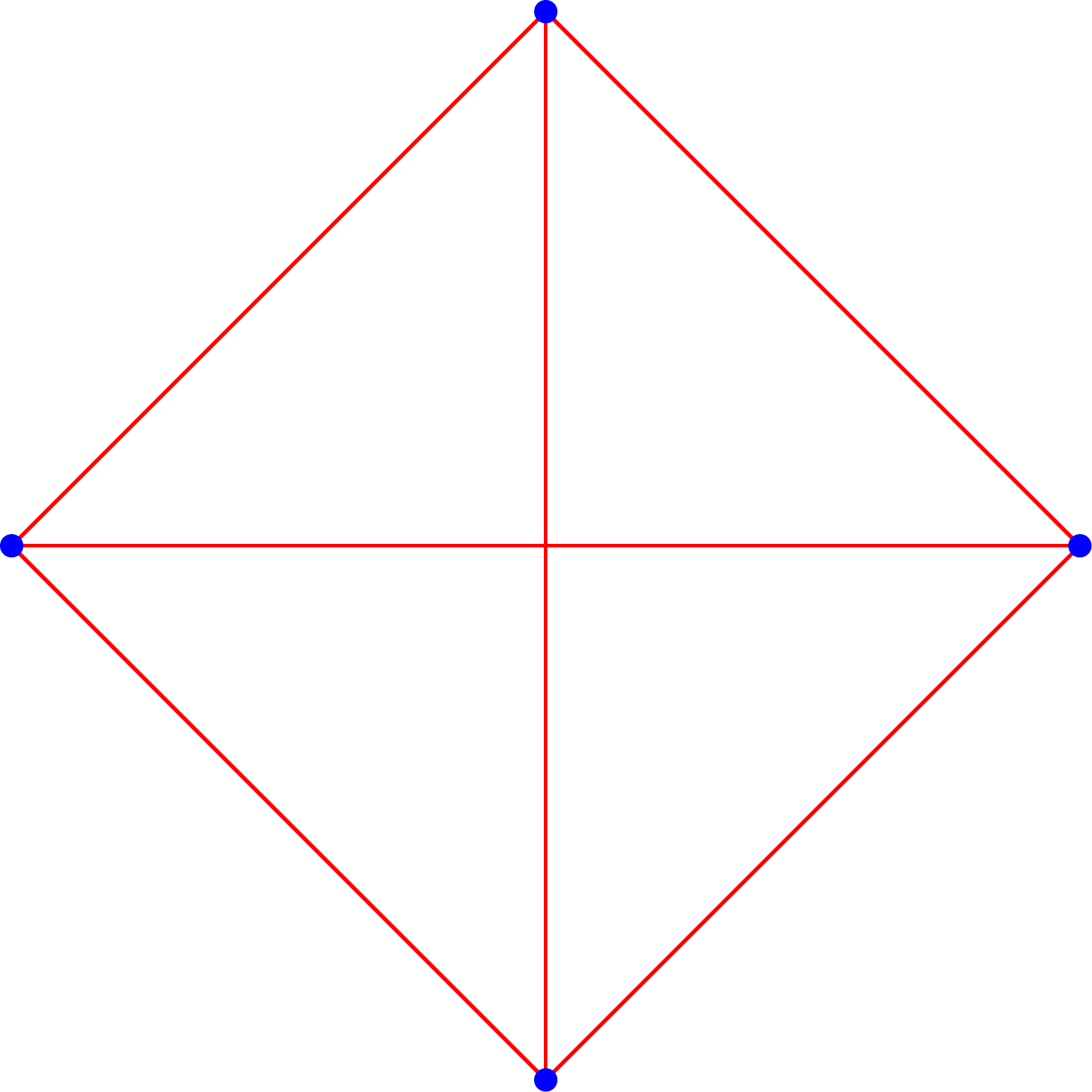}  & \multicolumn{2}{c|}{}                                                                  & \multicolumn{2}{c|}{}                                                                  & \multicolumn{2}{c|}{}                                                                            & \multicolumn{2}{c|}{}                                                                            \\*
                     & \cellcolor[HTML]{e2fff7}4                   & \cellcolor[HTML]{e2fff7}24              & \multicolumn{1}{l}{}                                   &                               &                                            &                                           &                                                &                                                 &                                        & \multicolumn{1}{r|}{}                                   \\*
\multirow{-4}{*}[0.5cm]{4}  & \multicolumn{2}{c|}{\cellcolor[HTML]{e2fff7}$S_4$}                                    & \multicolumn{1}{l}{}                                   & \multicolumn{1}{l|}{}         &                                            & \multicolumn{1}{l|}{}                     &                                                & \multicolumn{1}{l|}{}                           &                                        & \multicolumn{1}{l|}{}                                   \\ \hline
                     & \cellcolor[HTML]{FFFFFF}                    & \cellcolor[HTML]{FFFFFF}                & \multicolumn{1}{l}{\cellcolor[HTML]{e2fff7}1.0000}     & \cellcolor[HTML]{e2fff7}1     &                                            &                                           &                                                &                                                 &                                        & \multicolumn{1}{r|}{}                                   \\*
                     & \multicolumn{2}{c|}{\cellcolor[HTML]{FFFFFF}\includegraphics[width=1.5cm]{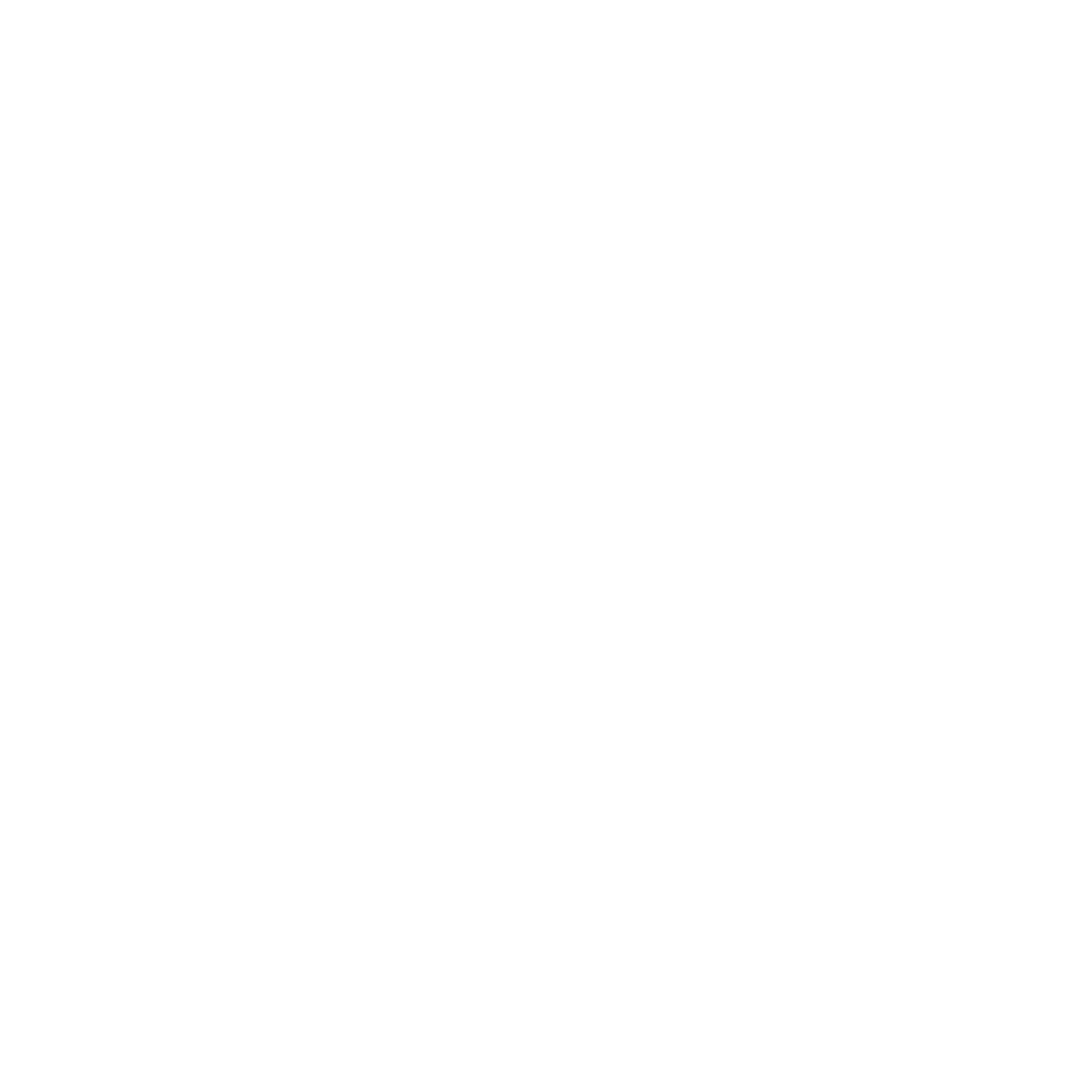}}   & \multicolumn{2}{c|}{\cellcolor[HTML]{e2fff7}\def\svgwidth{1.5cm}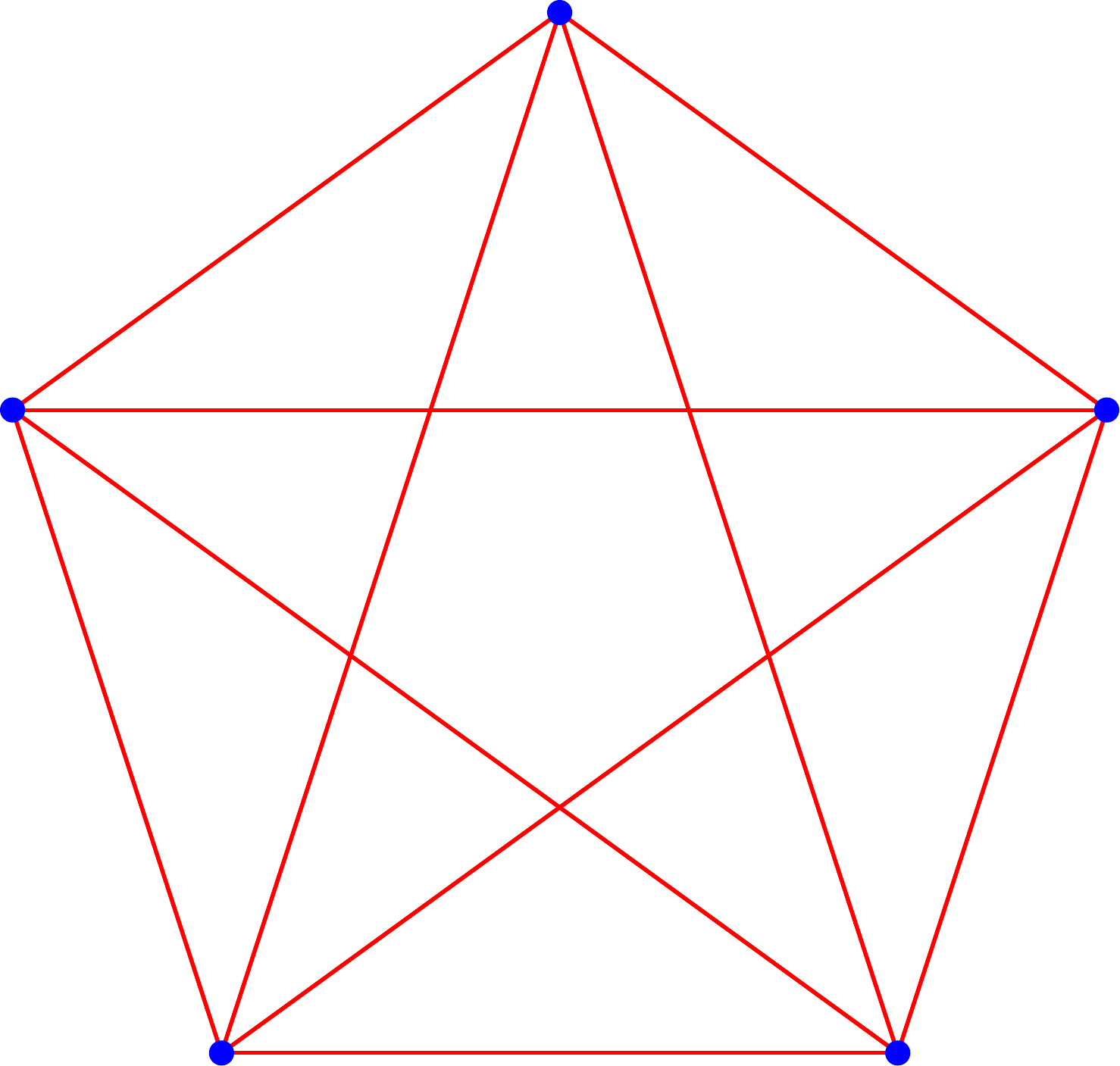}   & \multicolumn{2}{c|}{}                                                                  & \multicolumn{2}{c|}{}                                                                            & \multicolumn{2}{c|}{}                                                                            \\*
                     & \cellcolor[HTML]{FFFFFF}                    & \cellcolor[HTML]{FFFFFF}                & \multicolumn{1}{l}{\cellcolor[HTML]{e2fff7}6}          & \cellcolor[HTML]{e2fff7}120   &                                            &                                           &                                                &                                                 &                                        & \multicolumn{1}{r|}{}                                   \\*
\multirow{-4}{*}[0.5cm]{5}  &                                             & \multicolumn{1}{l|}{}                   & \multicolumn{2}{c|}{\cellcolor[HTML]{e2fff7}$S_5$}                                     &                                            & \multicolumn{1}{l|}{}                     &                                                & \multicolumn{1}{l|}{}                           &                                        & \multicolumn{1}{l|}{}                                   \\ \hline
                     & \cellcolor[HTML]{d0ffd5}1.4000              & \cellcolor[HTML]{d0ffd5}1               & \multicolumn{1}{l}{\cellcolor[HTML]{d0ffd5}1.2000}     & \cellcolor[HTML]{d0ffd5}1     & \cellcolor[HTML]{e2fff7}1.0000             & \cellcolor[HTML]{e2fff7}1                 &                                                &                                                 &                                        & \multicolumn{1}{r|}{}                                   \\*
                     & \multicolumn{2}{c|}{\cellcolor[HTML]{d0ffd5}\def\svgwidth{1.5cm}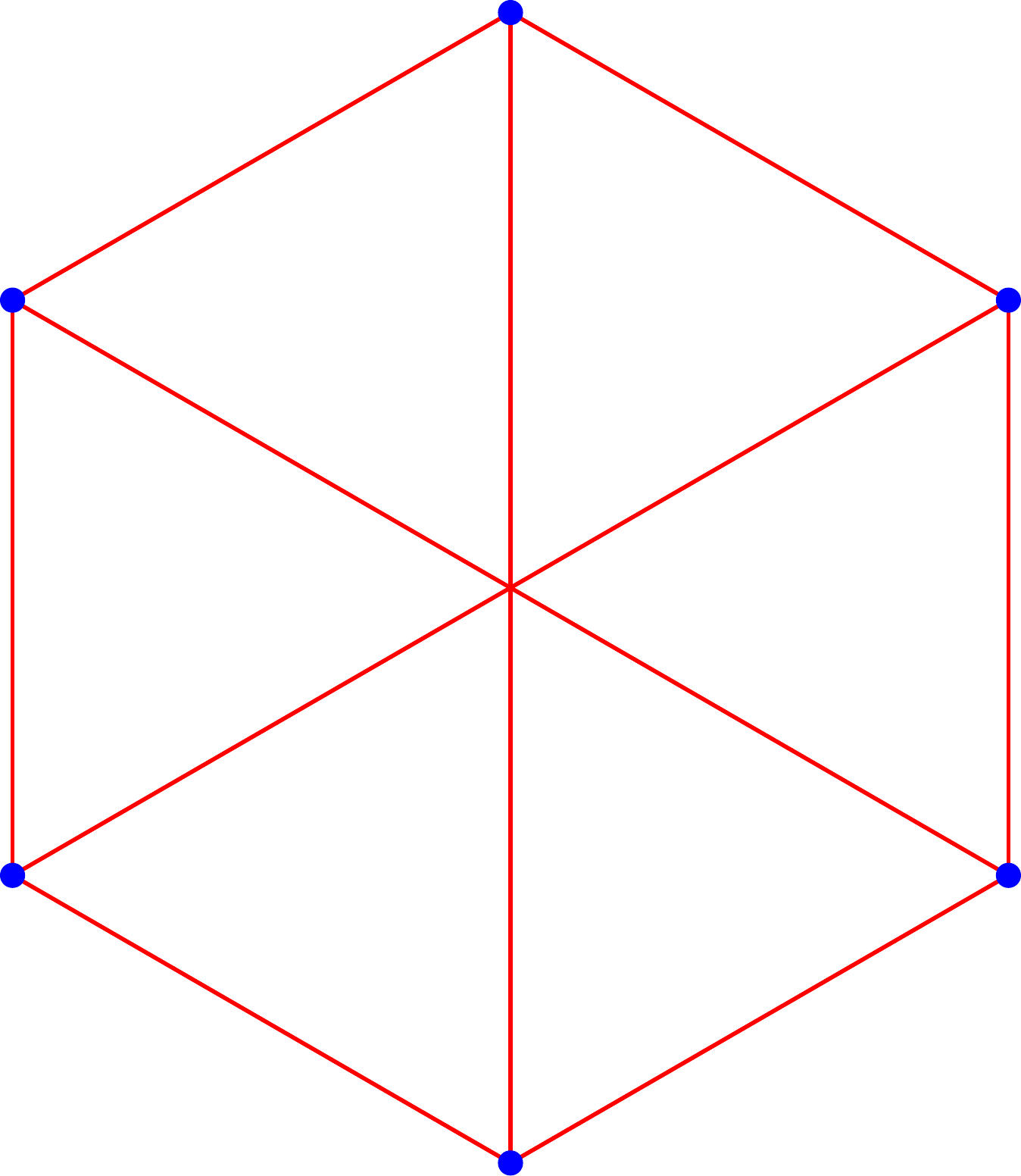}  & \multicolumn{2}{c|}{\cellcolor[HTML]{d0ffd5}\def\svgwidth{1.5cm}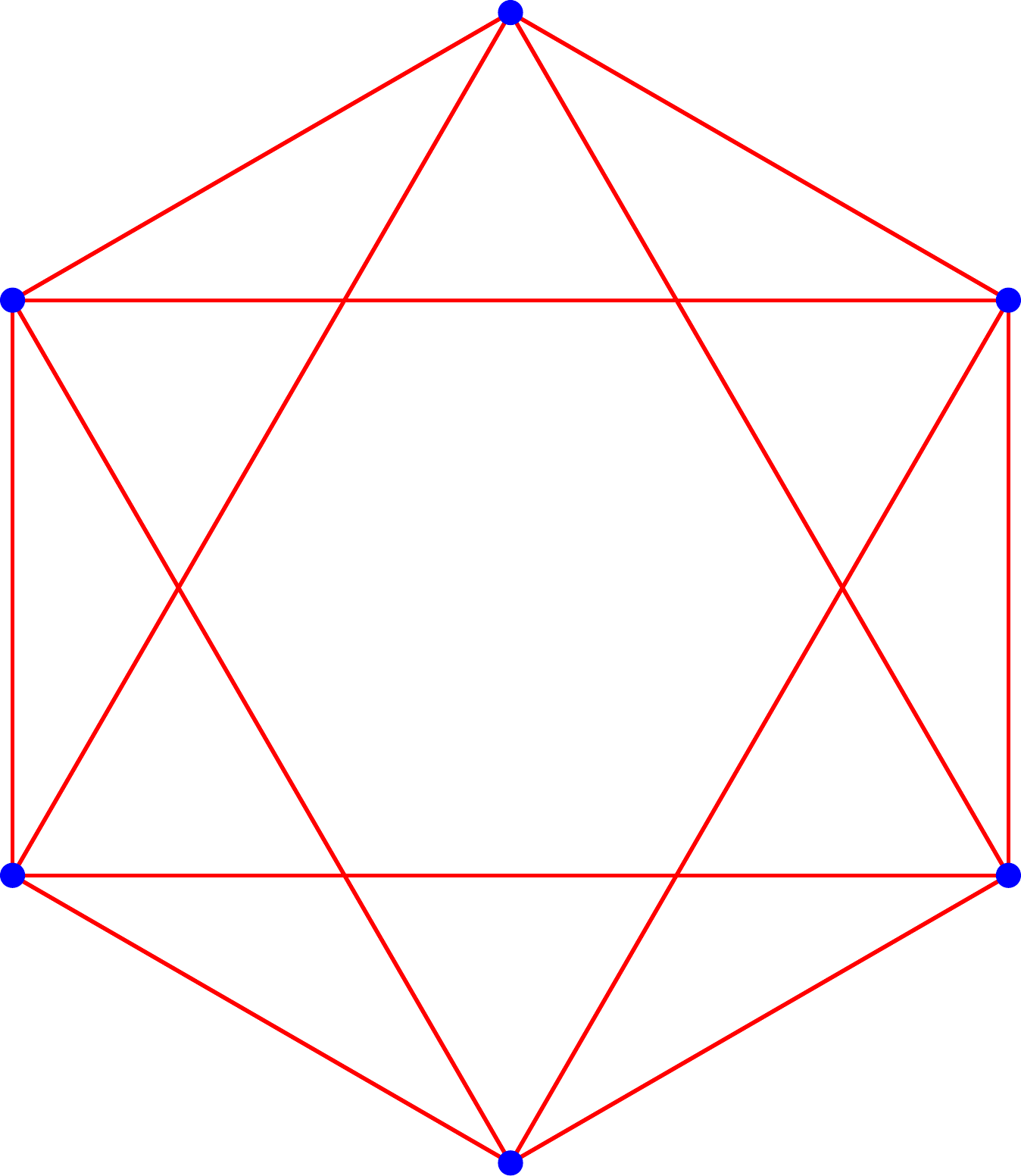}   & \multicolumn{2}{c|}{\cellcolor[HTML]{e2fff7}\def\svgwidth{1.5cm}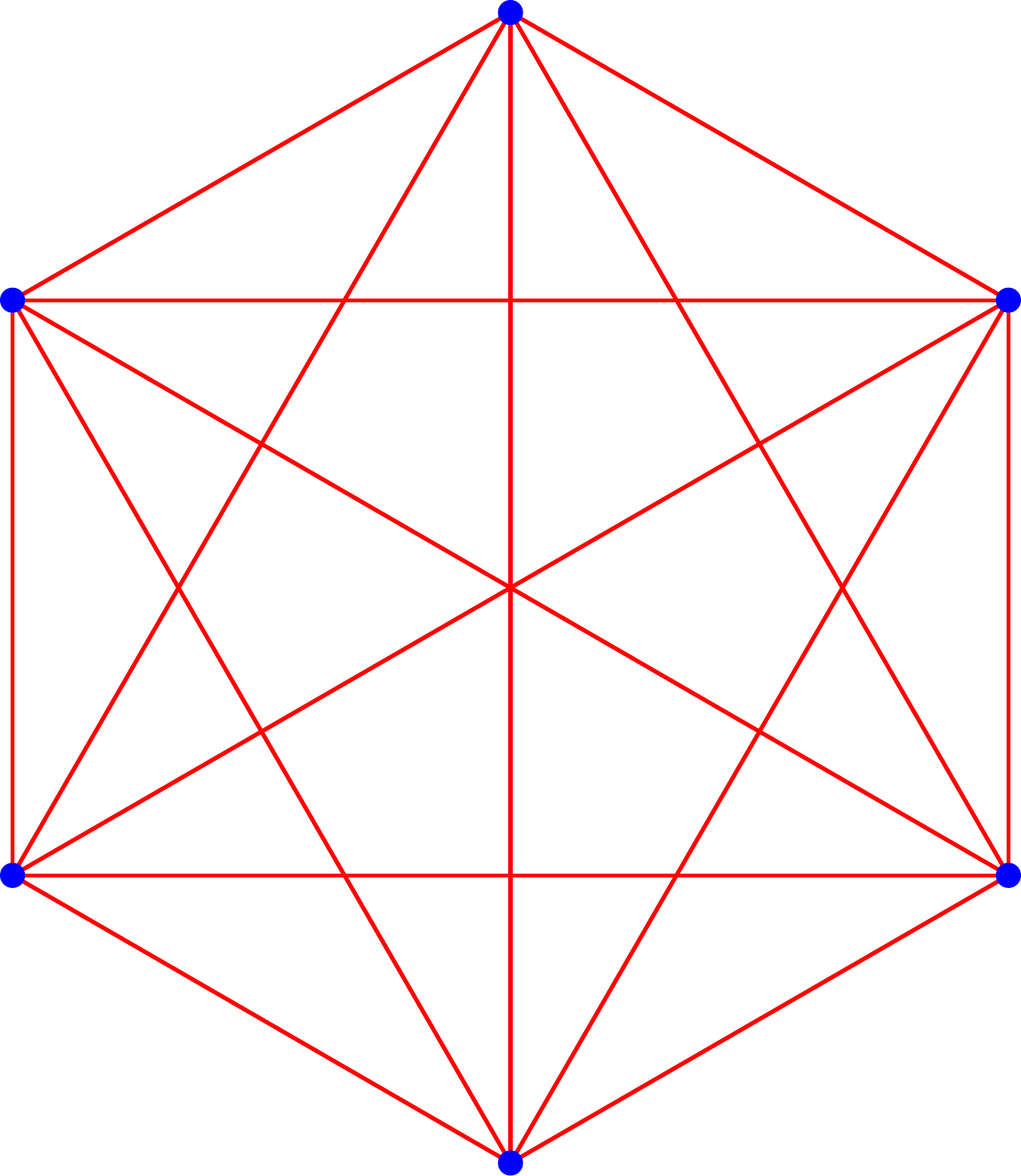}   & \multicolumn{2}{c|}{}                                                                            & \multicolumn{2}{c|}{}                                                                            \\*
                     & \cellcolor[HTML]{d0ffd5}5                   & \cellcolor[HTML]{d0ffd5}72              & \multicolumn{1}{l}{\cellcolor[HTML]{d0ffd5}6}          & \cellcolor[HTML]{d0ffd5}48    & \cellcolor[HTML]{e2fff7}9                  & \cellcolor[HTML]{e2fff7}720               &                                                &                                                 &                                        & \multicolumn{1}{r|}{}                                   \\*
\multirow{-4}{*}[0.5cm]{6}  & \multicolumn{2}{c|}{\cellcolor[HTML]{d0ffd5}$S_3\wr S_2$}                             & \multicolumn{2}{c|}{\cellcolor[HTML]{d0ffd5}$S_2 \wr S_3$}                             & \multicolumn{2}{c|}{\cellcolor[HTML]{e2fff7}$S_6$}                                     &                                                & \multicolumn{1}{l|}{}                           &                                        & \multicolumn{1}{l|}{}                                   \\ \hline
                     & \cellcolor[HTML]{FFFFFF}                    & \cellcolor[HTML]{FFFFFF}                & \multicolumn{1}{l}{\cellcolor[HTML]{d0ffd5}1.3333}     & \cellcolor[HTML]{d0ffd5}1     &                                            &                                           & \cellcolor[HTML]{e2fff7}1.0000                 & \cellcolor[HTML]{e2fff7}1                       &                                        & \multicolumn{1}{r|}{}                                   \\*
                     & \multicolumn{2}{c|}{\cellcolor[HTML]{FFFFFF}\includegraphics[width=1.5cm]{empty.pdf}}   & \multicolumn{2}{c|}{\cellcolor[HTML]{d0ffd5}\def\svgwidth{1.5cm}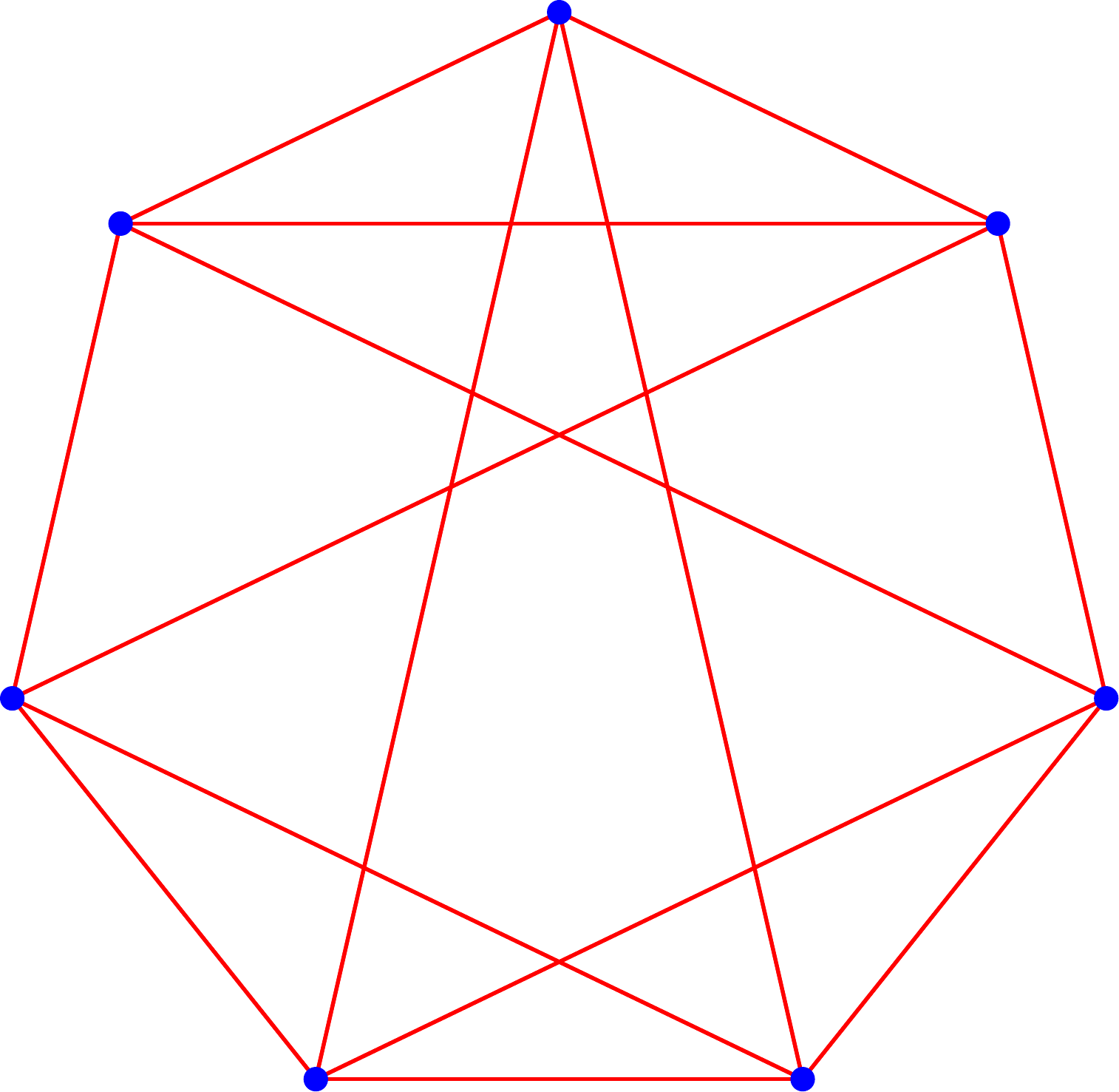}   & \multicolumn{2}{c|}{}                                                                  & \multicolumn{2}{c|}{\cellcolor[HTML]{e2fff7}\def\svgwidth{1.5cm}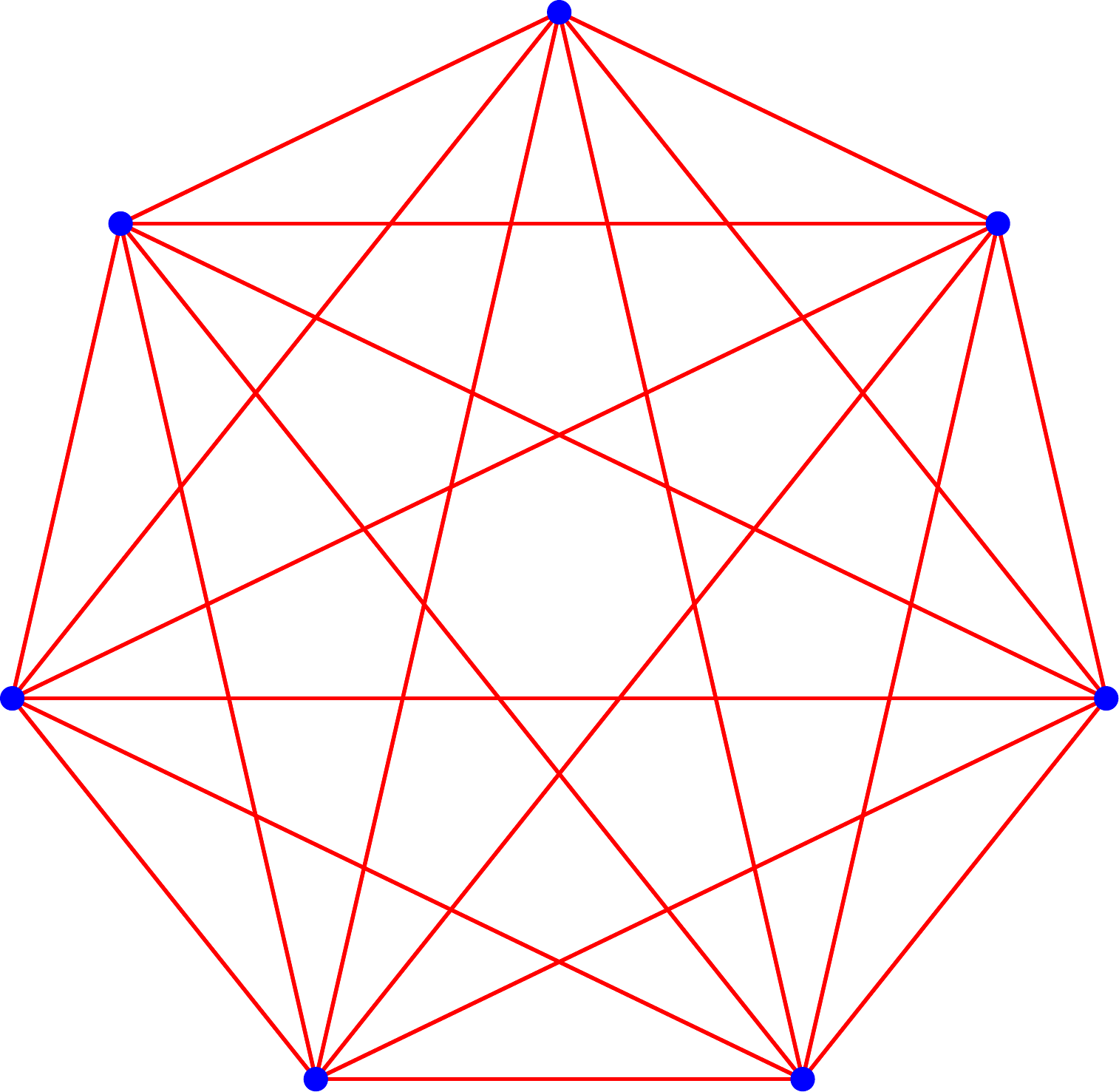}             & \multicolumn{2}{c|}{}                                                                            \\*
                     & \cellcolor[HTML]{FFFFFF}                    & \cellcolor[HTML]{FFFFFF}                & \multicolumn{1}{l}{\cellcolor[HTML]{d0ffd5}6}          & \cellcolor[HTML]{d0ffd5}48    &                                            &                                           & \cellcolor[HTML]{e2fff7}12                     & \cellcolor[HTML]{e2fff7}5040                    &                                        & \multicolumn{1}{r|}{}                                   \\*
\multirow{-4}{*}[0.5cm]{7}  & \multicolumn{2}{l|}{}                                                                 & \multicolumn{2}{c|}{\cellcolor[HTML]{d0ffd5}$D_4 \times S_3$}                          &                                            & \multicolumn{1}{l|}{}                     & \multicolumn{2}{c|}{\cellcolor[HTML]{e2fff7} $S_7$}                                          &                                        & \multicolumn{1}{l|}{}                                   \\ \hline
                     & \cellcolor[HTML]{d0ffd5}1.5714              & \cellcolor[HTML]{d0ffd5}1               & \multicolumn{1}{l}{\cellcolor[HTML]{d0ffd5}1.4286}     & \cellcolor[HTML]{d0ffd5}1     & \cellcolor[HTML]{d0ffd5}1.2857             & \cellcolor[HTML]{d0ffd5}1                 & \cellcolor[HTML]{d0ffd5}1.1429                 & \cellcolor[HTML]{d0ffd5}1                       & \cellcolor[HTML]{e2fff7}1.0000         & \multicolumn{1}{r|}{\cellcolor[HTML]{e2fff7}1}          \\*
                     & \multicolumn{2}{c|}{\cellcolor[HTML]{d0ffd5}\def\svgwidth{1.5cm}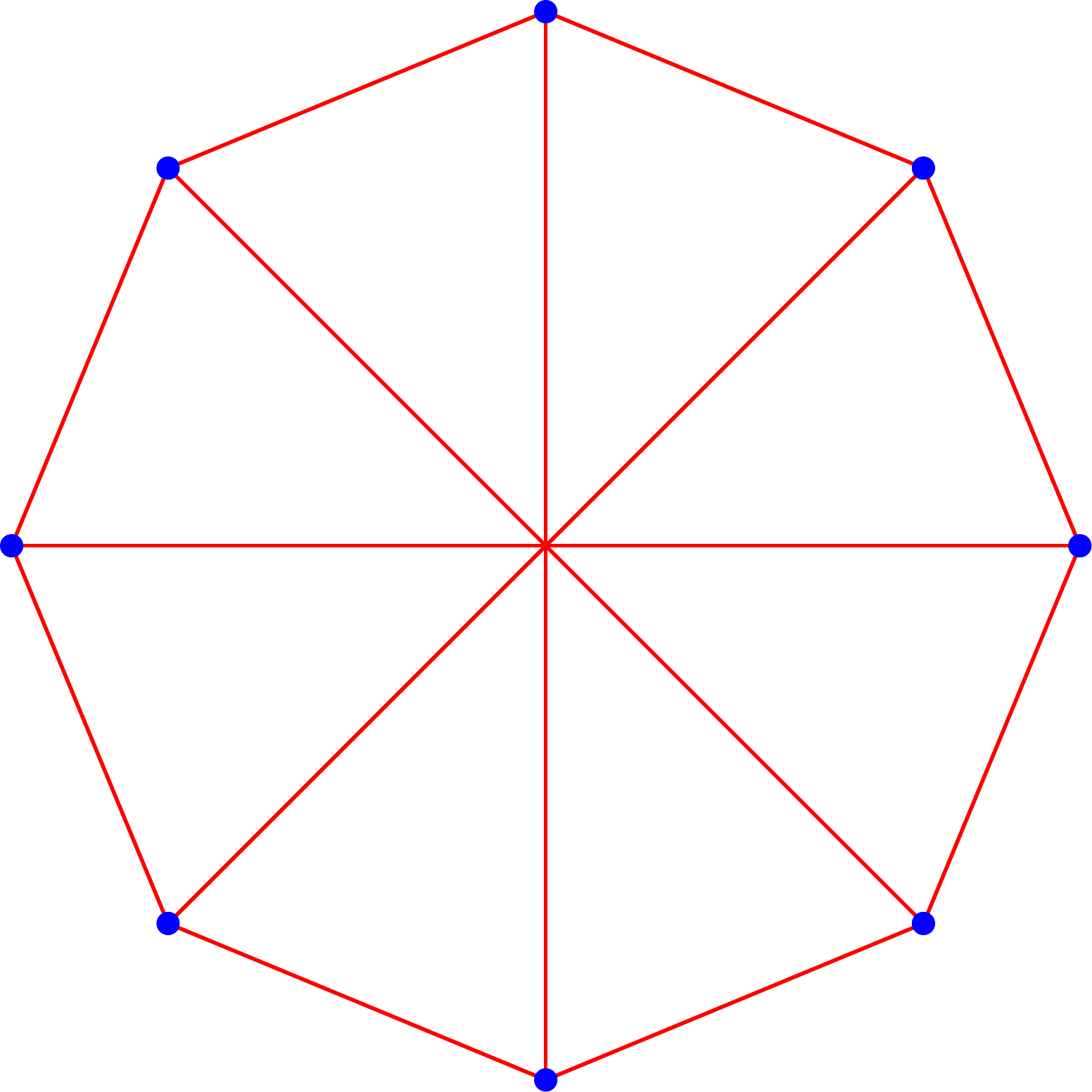}  & \multicolumn{2}{c|}{\cellcolor[HTML]{d0ffd5}\def\svgwidth{1.5cm}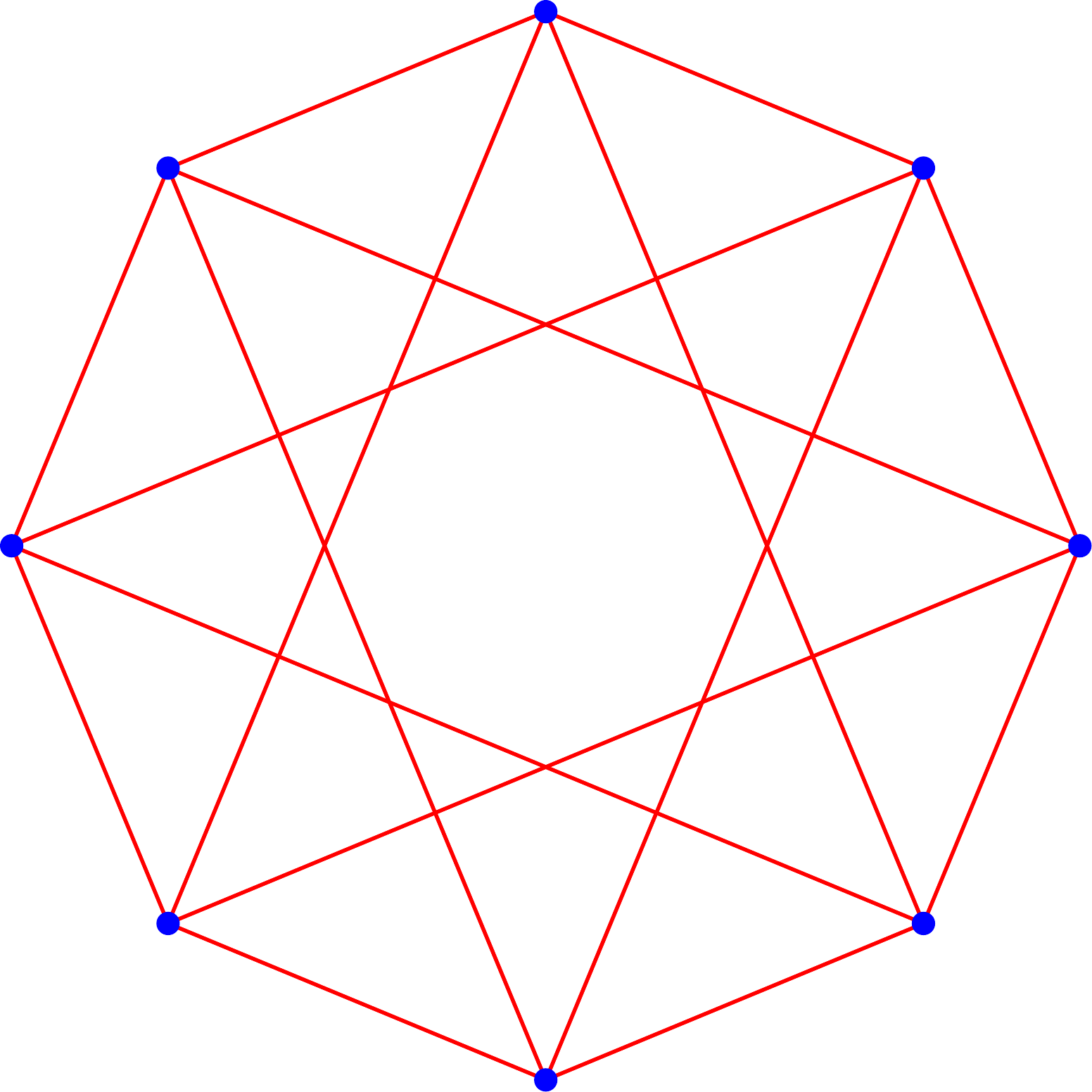}   & \multicolumn{2}{c|}{\cellcolor[HTML]{d0ffd5}\def\svgwidth{1.5cm}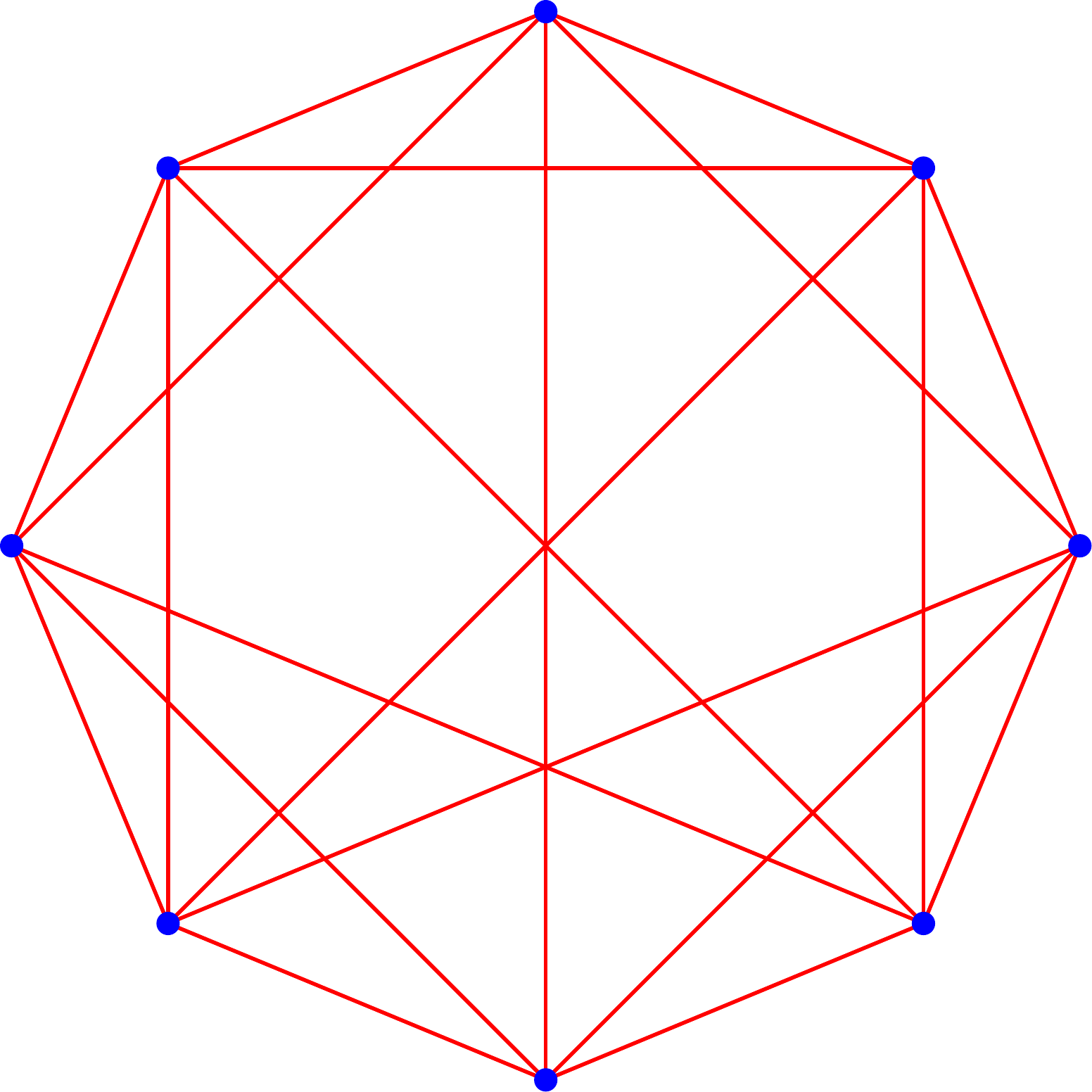}   & \multicolumn{2}{c|}{\cellcolor[HTML]{d0ffd5}\def\svgwidth{1.5cm}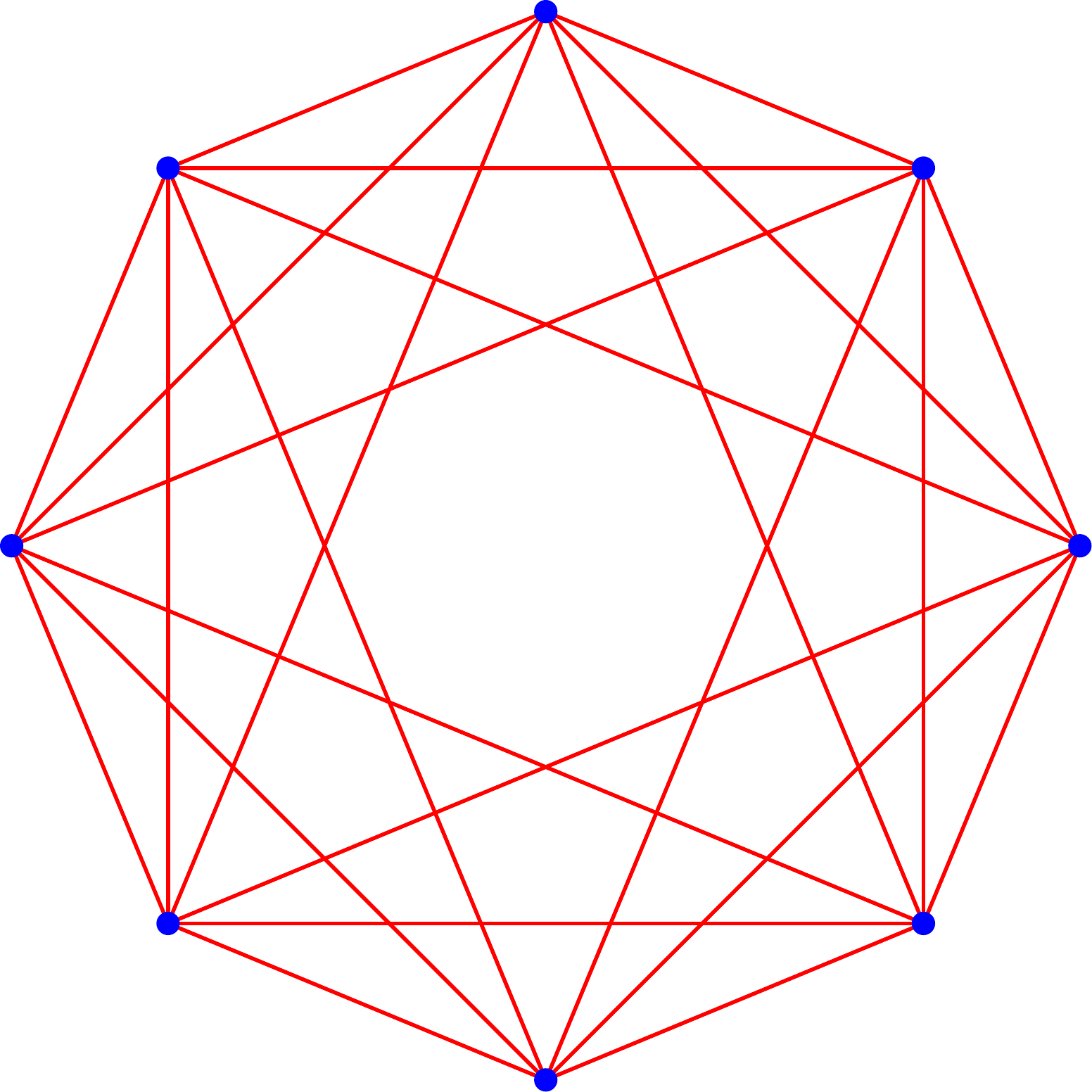}             & \multicolumn{2}{c|}{\cellcolor[HTML]{e2fff7}\def\svgwidth{1.5cm}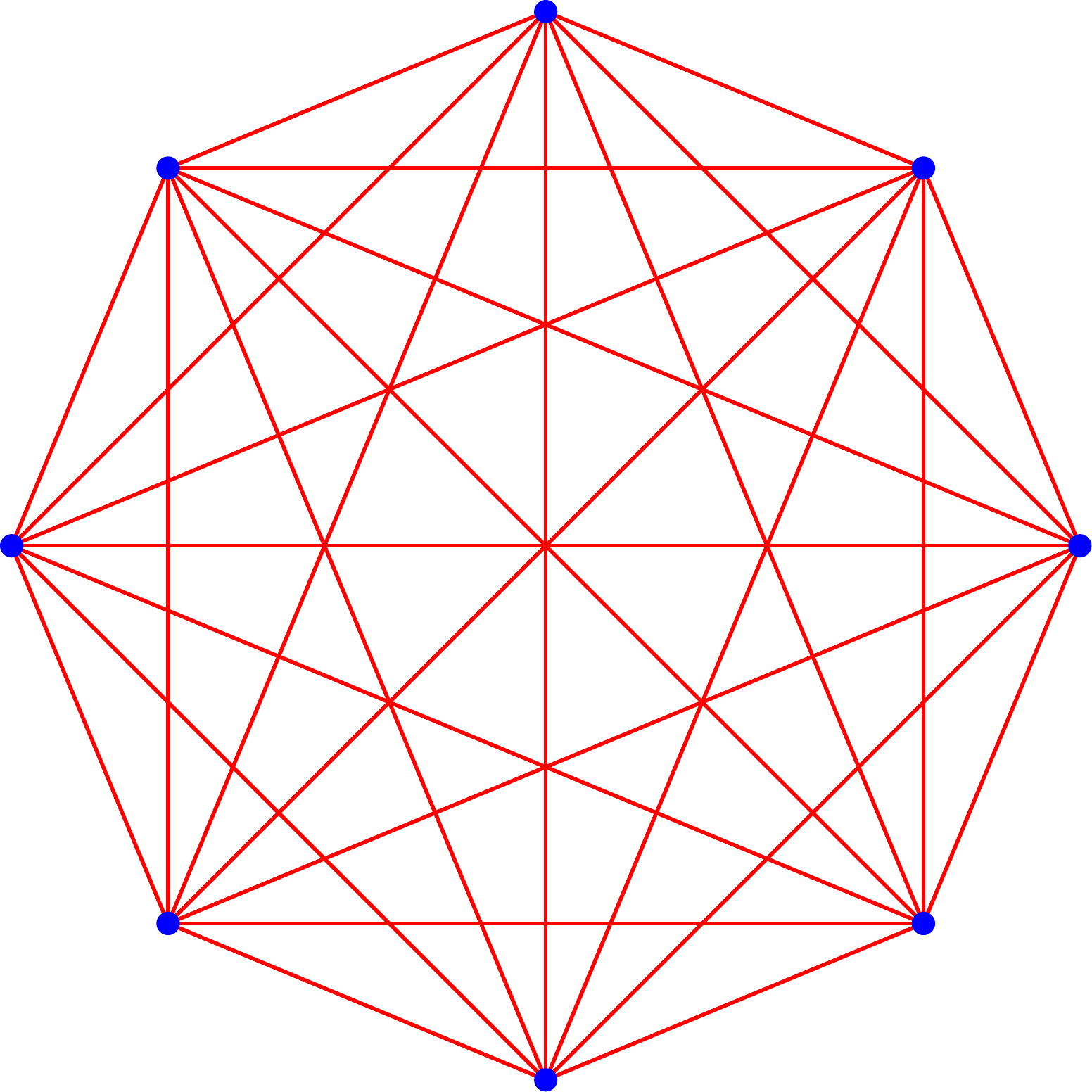}             \\*
                     & \cellcolor[HTML]{d0ffd5}4                   & \cellcolor[HTML]{d0ffd5}16              & \multicolumn{1}{l}{\cellcolor[HTML]{d0ffd5}8}          & \cellcolor[HTML]{d0ffd5}1152  & \cellcolor[HTML]{d0ffd5}10                 & \cellcolor[HTML]{d0ffd5}60                & \cellcolor[HTML]{d0ffd5}12                     & \cellcolor[HTML]{d0ffd5}384                     & \cellcolor[HTML]{e2fff7}16             & \multicolumn{1}{r|}{\cellcolor[HTML]{e2fff7}40320}      \\*
\multirow{-4}{*}[0.5cm]{8}  & \multicolumn{2}{c|}{\cellcolor[HTML]{d0ffd5}$D_8$}                                    & \multicolumn{2}{c|}{\cellcolor[HTML]{d0ffd5}$S_4 \wr S_2$}                             & \multicolumn{2}{c|}{\cellcolor[HTML]{d0ffd5}$S_3 \times D_5$}                          & \multicolumn{2}{c|}{\cellcolor[HTML]{d0ffd5}$S_2 \wr S_4$}                                       & \multicolumn{2}{c|}{\cellcolor[HTML]{e2fff7}$S_8$}                                               \\ \hline
                     & \cellcolor[HTML]{FFFFFF}                    & \cellcolor[HTML]{FFFFFF}                & \multicolumn{1}{l}{\cellcolor[HTML]{d0ffd5}1.5000}     & \cellcolor[HTML]{d0ffd5}1     &                                            &                                           & \cellcolor[HTML]{d0ffd5}1.2500                 & \cellcolor[HTML]{d0ffd5}1                       &                                        & \multicolumn{1}{r|}{}                                   \\*
                     & \multicolumn{2}{c|}{\cellcolor[HTML]{FFFFFF}\includegraphics[width=1.5cm]{empty.pdf}}   & \multicolumn{2}{c|}{\cellcolor[HTML]{d0ffd5}\def\svgwidth{1.5cm}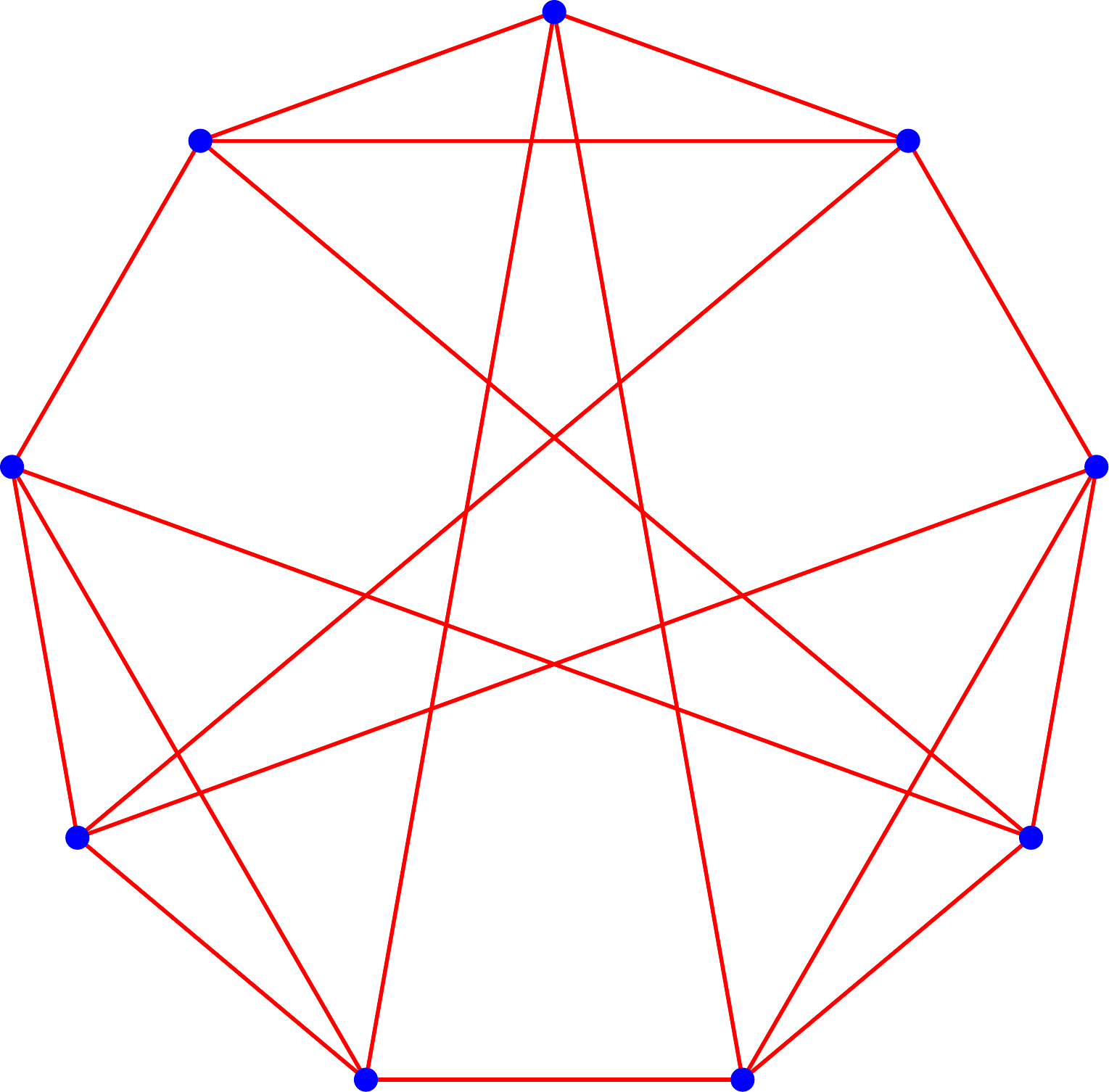}  & \multicolumn{2}{c|}{}                                                                  & \multicolumn{2}{c|}{\cellcolor[HTML]{d0ffd5}\def\svgwidth{1.5cm}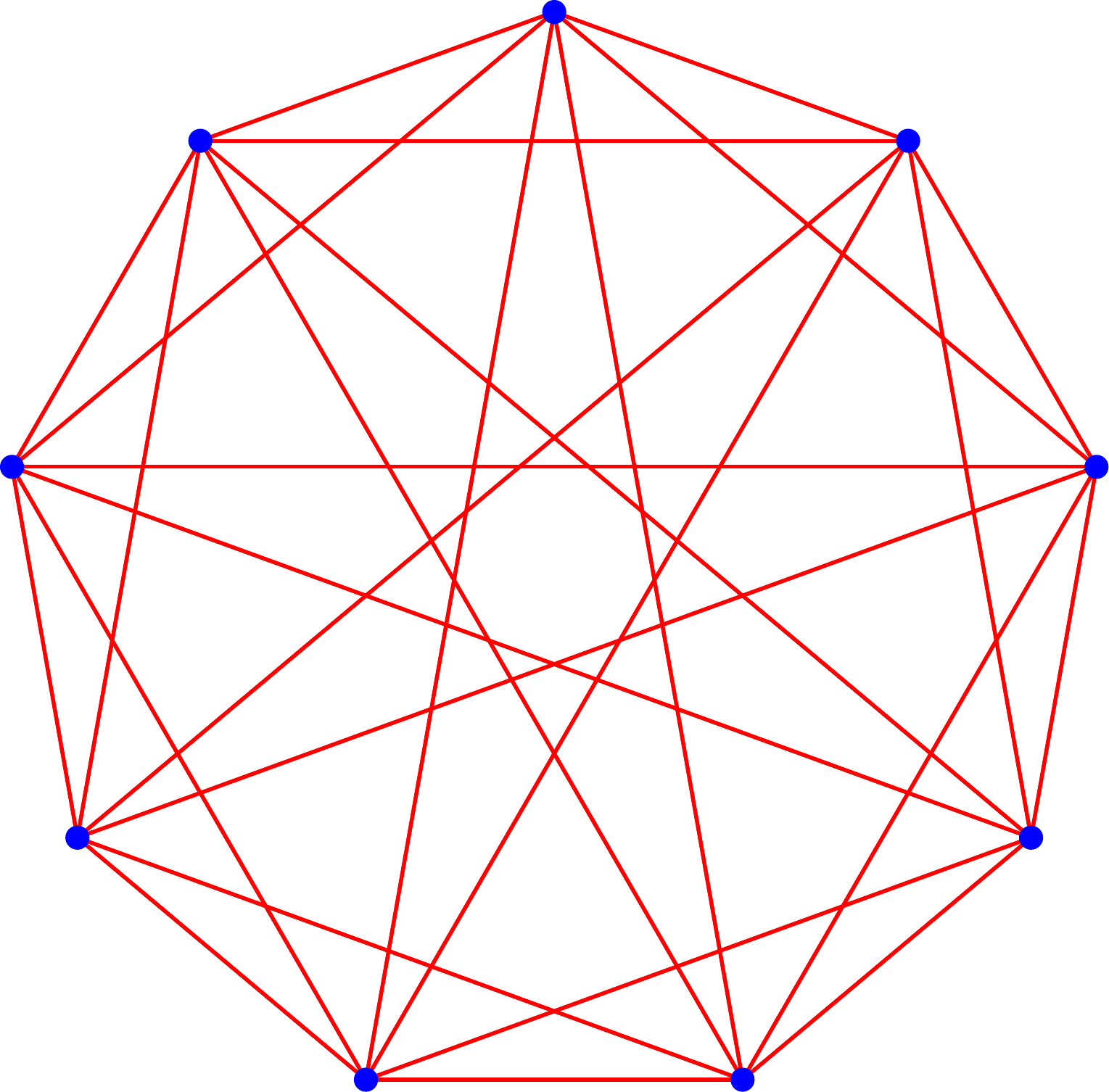}             & \multicolumn{2}{c|}{}                                                                            \\*
                     & \cellcolor[HTML]{FFFFFF}                    & \cellcolor[HTML]{FFFFFF}                & \multicolumn{1}{l}{\cellcolor[HTML]{d0ffd5}8}          & \cellcolor[HTML]{d0ffd5}72    &                                            &                                           & \cellcolor[HTML]{d0ffd5}14                     & \cellcolor[HTML]{d0ffd5}1296                    &                                        & \multicolumn{1}{r|}{}                                   \\*
\multirow{-4}{*}[0.5cm]{9}  & \multicolumn{2}{l|}{}                                                                 & \multicolumn{2}{c|}{\cellcolor[HTML]{d0ffd5}$S_3 \wr S_2$}                             &                                            & \multicolumn{1}{l|}{}                     & \multicolumn{2}{c|}{\cellcolor[HTML]{d0ffd5}$S_3 \wr S_3$}                                       &                                        & \multicolumn{1}{l|}{}                                   \\ \hline
                     & \cellcolor[HTML]{d0ffd5}1.6667              & \cellcolor[HTML]{d0ffd5}1               & \multicolumn{1}{l}{\cellcolor[HTML]{d0ffd5}1.5556}     & \cellcolor[HTML]{d0ffd5}1     & \cellcolor[HTML]{d0ffd5}1.4444             & \cellcolor[HTML]{d0ffd5}1                 & \cellcolor[HTML]{d0ffd5}1.3333                 & \cellcolor[HTML]{d0ffd5}1                       & \cellcolor[HTML]{d0ffd5}1.2222         & \multicolumn{1}{r|}{\cellcolor[HTML]{d0ffd5}1}          \\*
                     & \multicolumn{2}{c|}{\cellcolor[HTML]{d0ffd5}\def\svgwidth{1.5cm}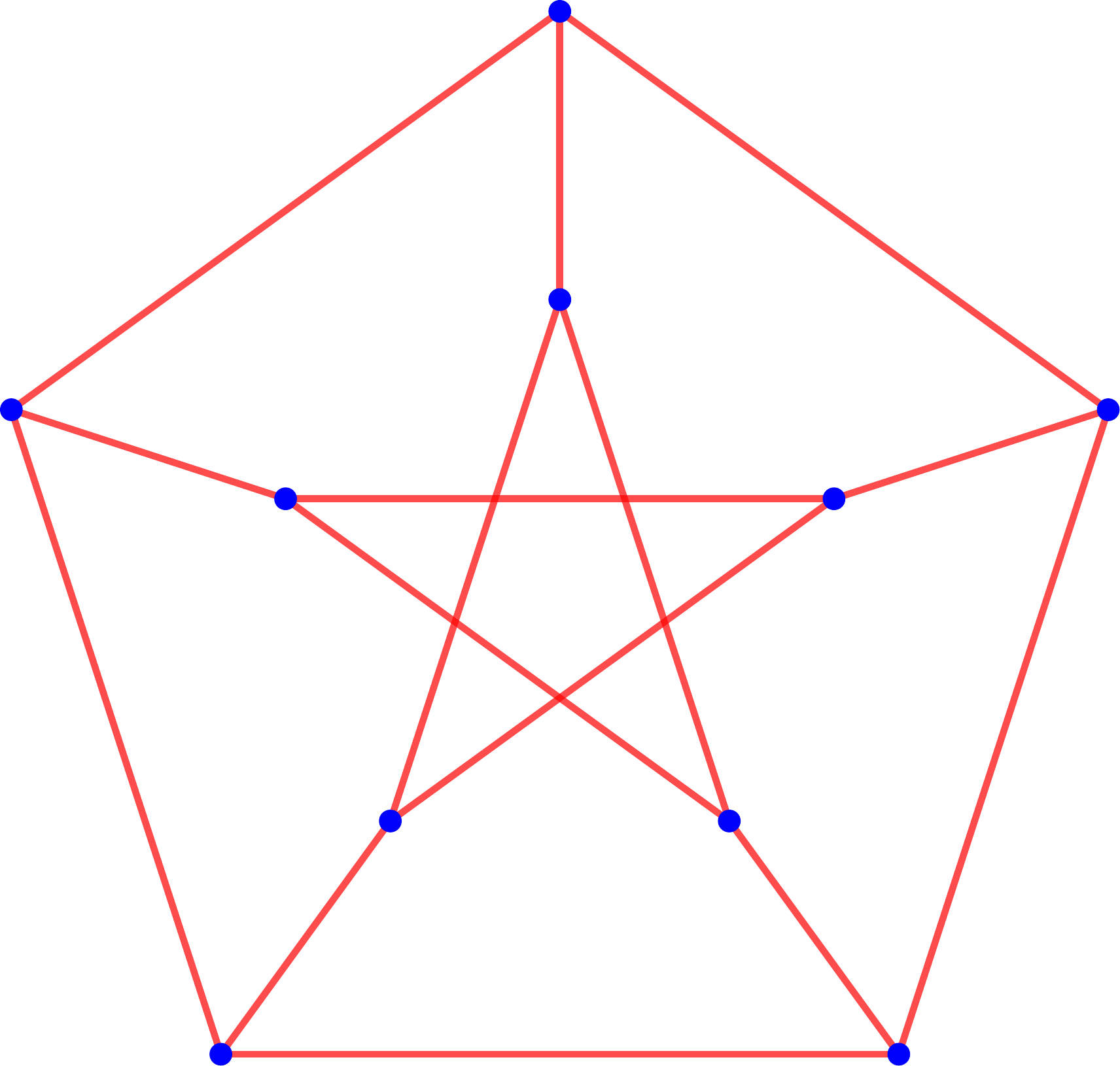} & \multicolumn{2}{c|}{\cellcolor[HTML]{d0ffd5}\def\svgwidth{1.5cm}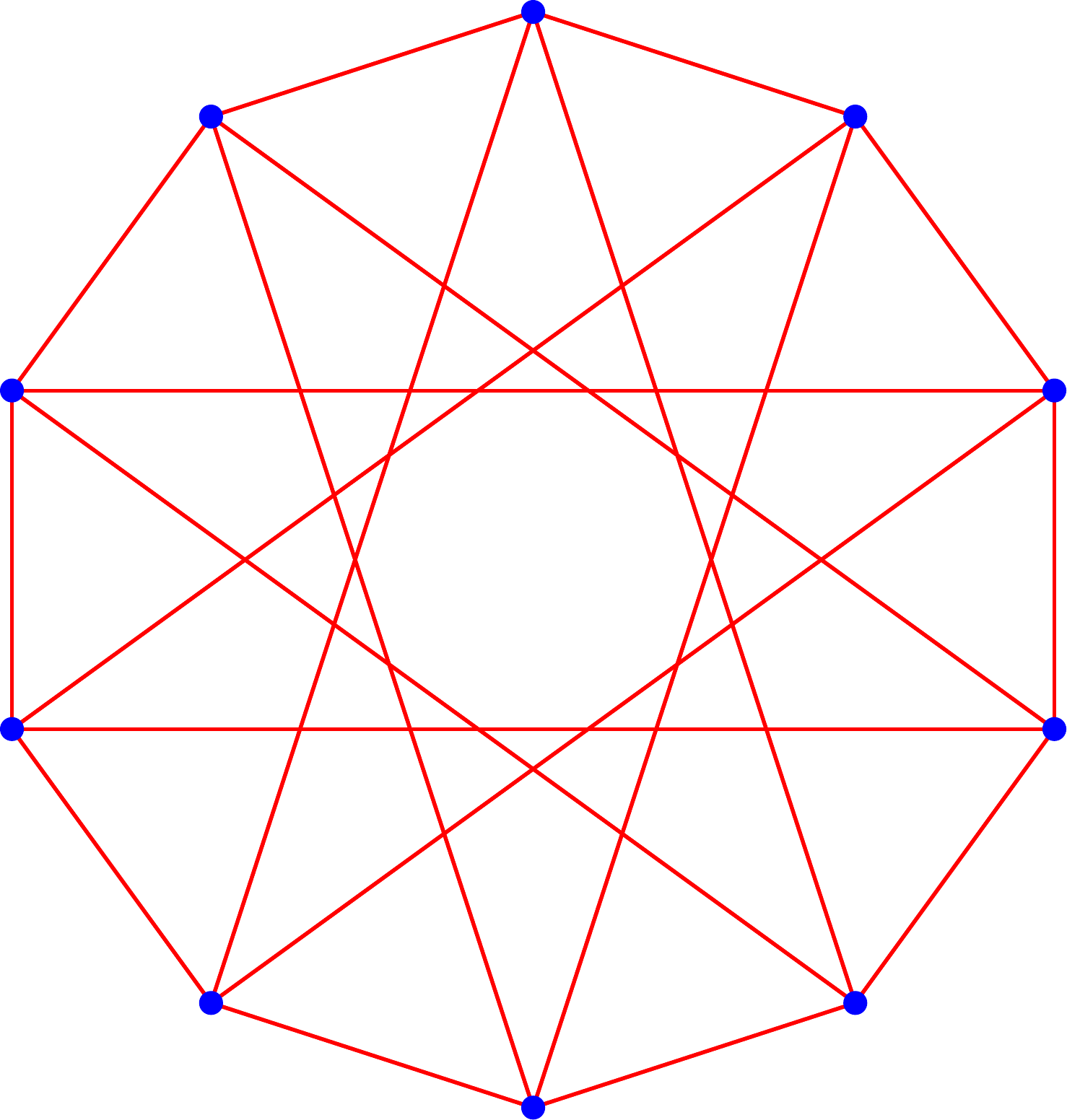} & \multicolumn{2}{c|}{\cellcolor[HTML]{d0ffd5}\def\svgwidth{1.5cm}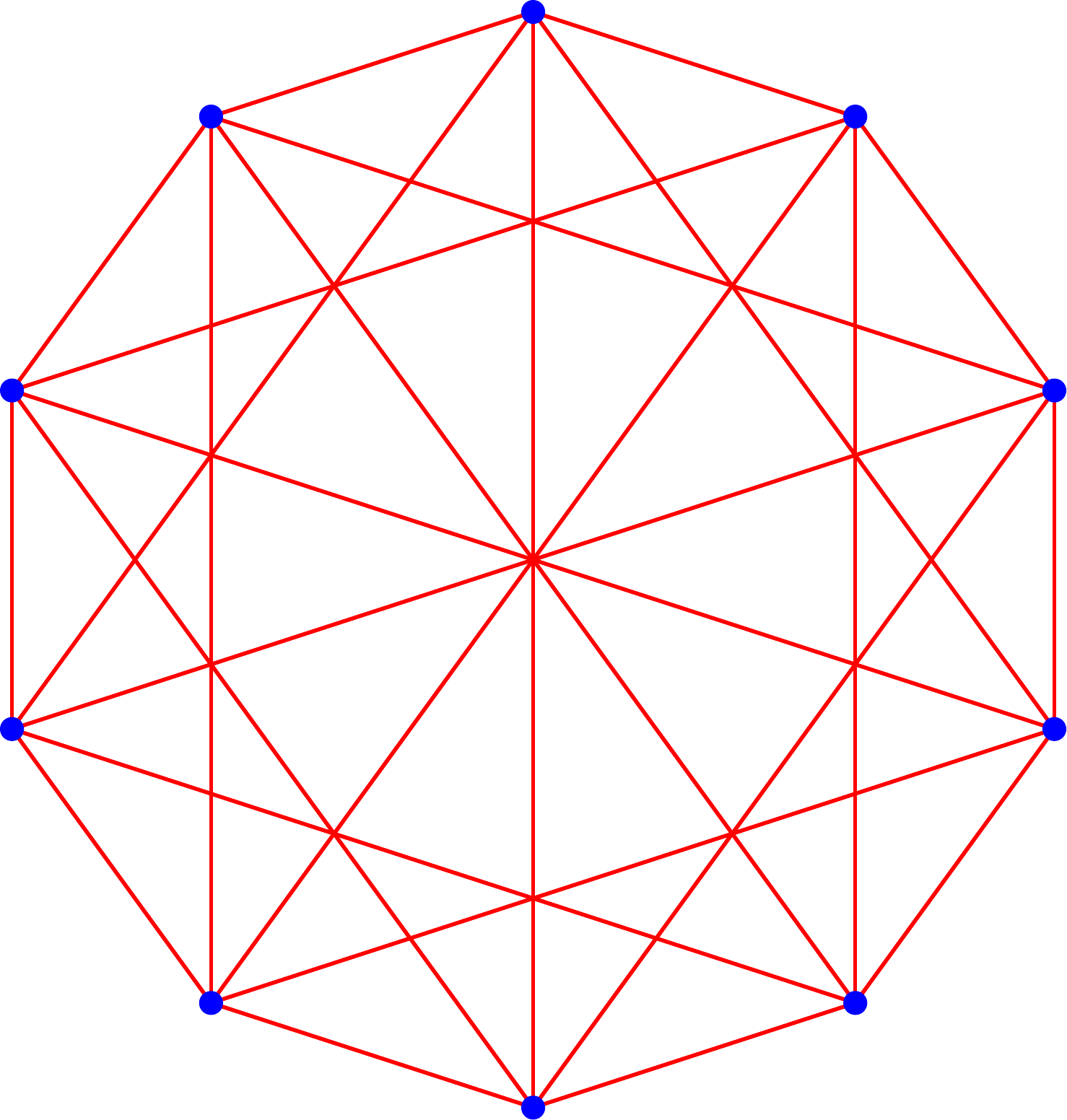} & \multicolumn{2}{c|}{\cellcolor[HTML]{d0ffd5}\def\svgwidth{1.5cm}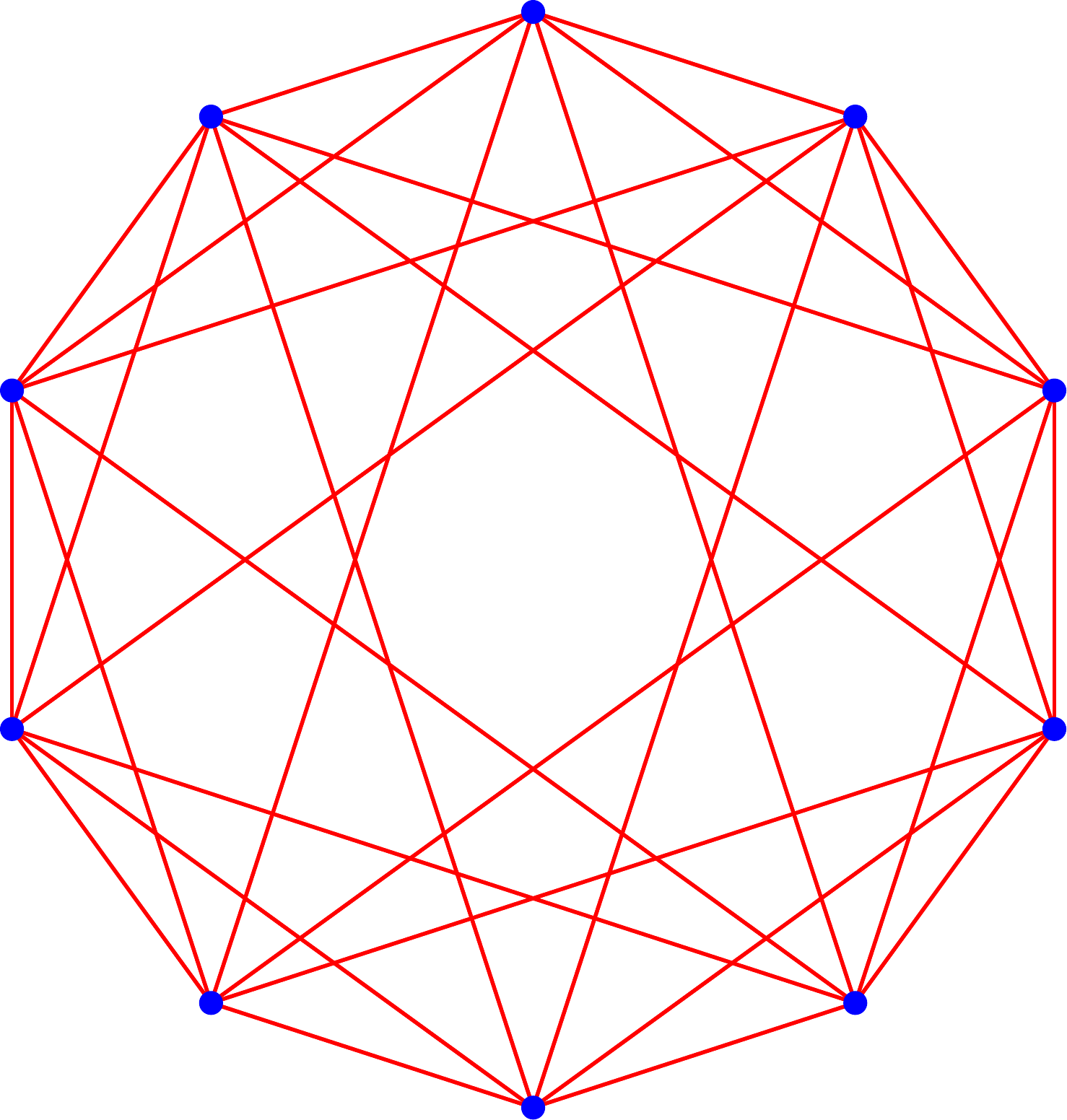}           & \multicolumn{2}{c|}{\cellcolor[HTML]{d0ffd5}\def\svgwidth{1.5cm}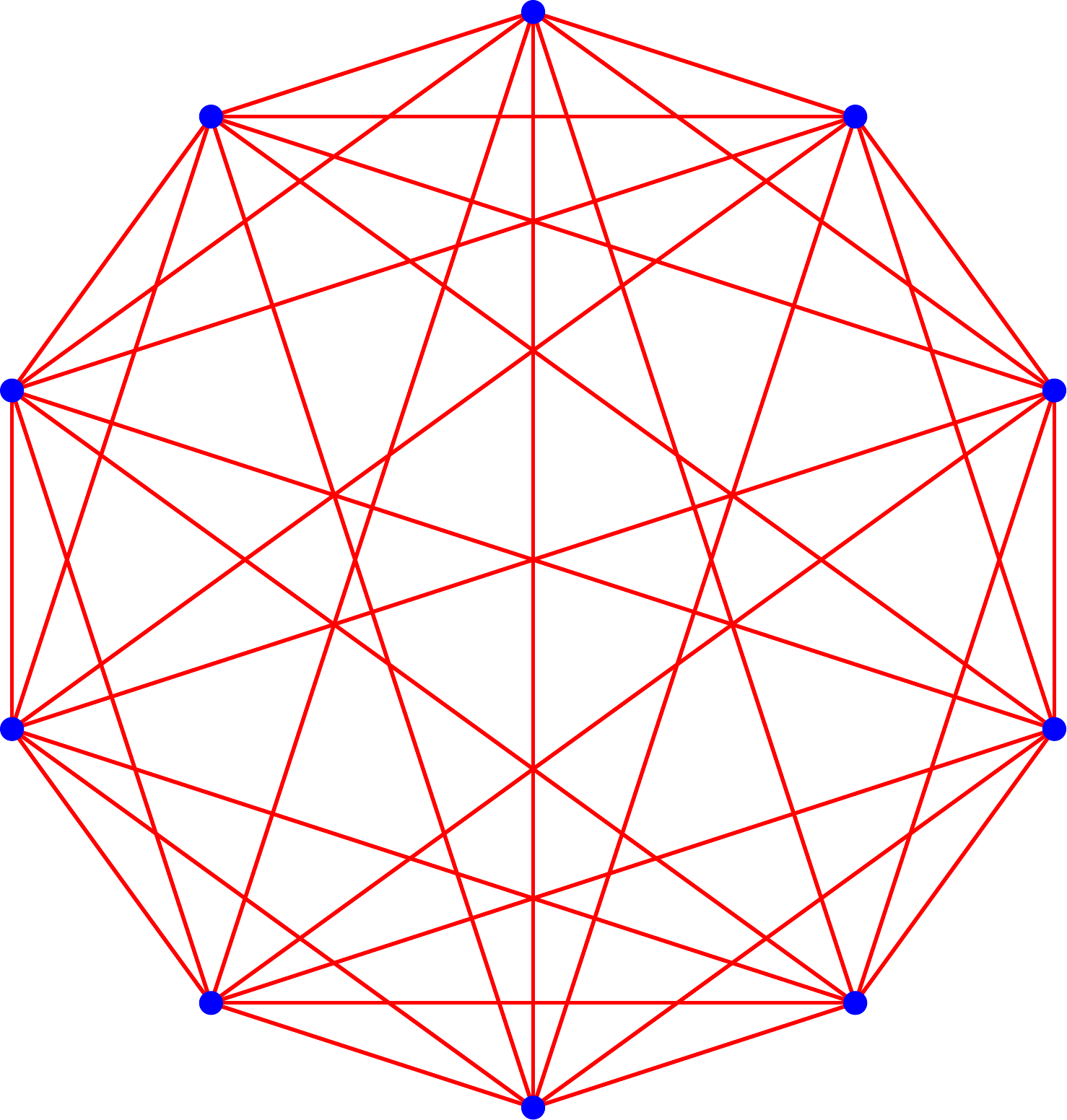}            \\*
                     & \cellcolor[HTML]{d0ffd5}5                   & \cellcolor[HTML]{d0ffd5}120             & \multicolumn{1}{l}{\cellcolor[HTML]{d0ffd5}8}          & \cellcolor[HTML]{d0ffd5}320   & \cellcolor[HTML]{d0ffd5}13                 & \cellcolor[HTML]{d0ffd5}28800             & \cellcolor[HTML]{d0ffd5}14                     & \cellcolor[HTML]{d0ffd5}288                     & \cellcolor[HTML]{d0ffd5}17             & \multicolumn{1}{r|}{\cellcolor[HTML]{d0ffd5}576}        \\*
\multirow{-4}{*}[0.5cm]{10} & \multicolumn{2}{c|}{\cellcolor[HTML]{d0ffd5}$S_{5}$}                & \multicolumn{2}{c|}{\cellcolor[HTML]{d0ffd5}$C_2\times ({C_2}^4\rtimes D_5)$}           & \multicolumn{2}{c|}{\cellcolor[HTML]{d0ffd5}$S_{5} \wr S_{2}$}       & \multicolumn{2}{c|}{\cellcolor[HTML]{d0ffd5}$C_{2} \times S_{3} \times S_{4}$} & \multicolumn{2}{c|}{\cellcolor[HTML]{d0ffd5}$ D_{4} \times (S_{3} \wr S_{2})$} \\ \hline
                     &                                             &                                         & \multicolumn{1}{l}{\cellcolor[HTML]{d0ffd5}1.6000}     & \cellcolor[HTML]{d0ffd5}1     &                                            &                                           & \cellcolor[HTML]{d0ffd5}1.4000                 & \cellcolor[HTML]{d0ffd5}1                       &                                        & \multicolumn{1}{r|}{}                                   \\*
                     & \multicolumn{2}{c|}{\includegraphics[width=1.5cm]{empty.pdf}}                           & \multicolumn{2}{c|}{\cellcolor[HTML]{d0ffd5}\def\svgwidth{1.5cm}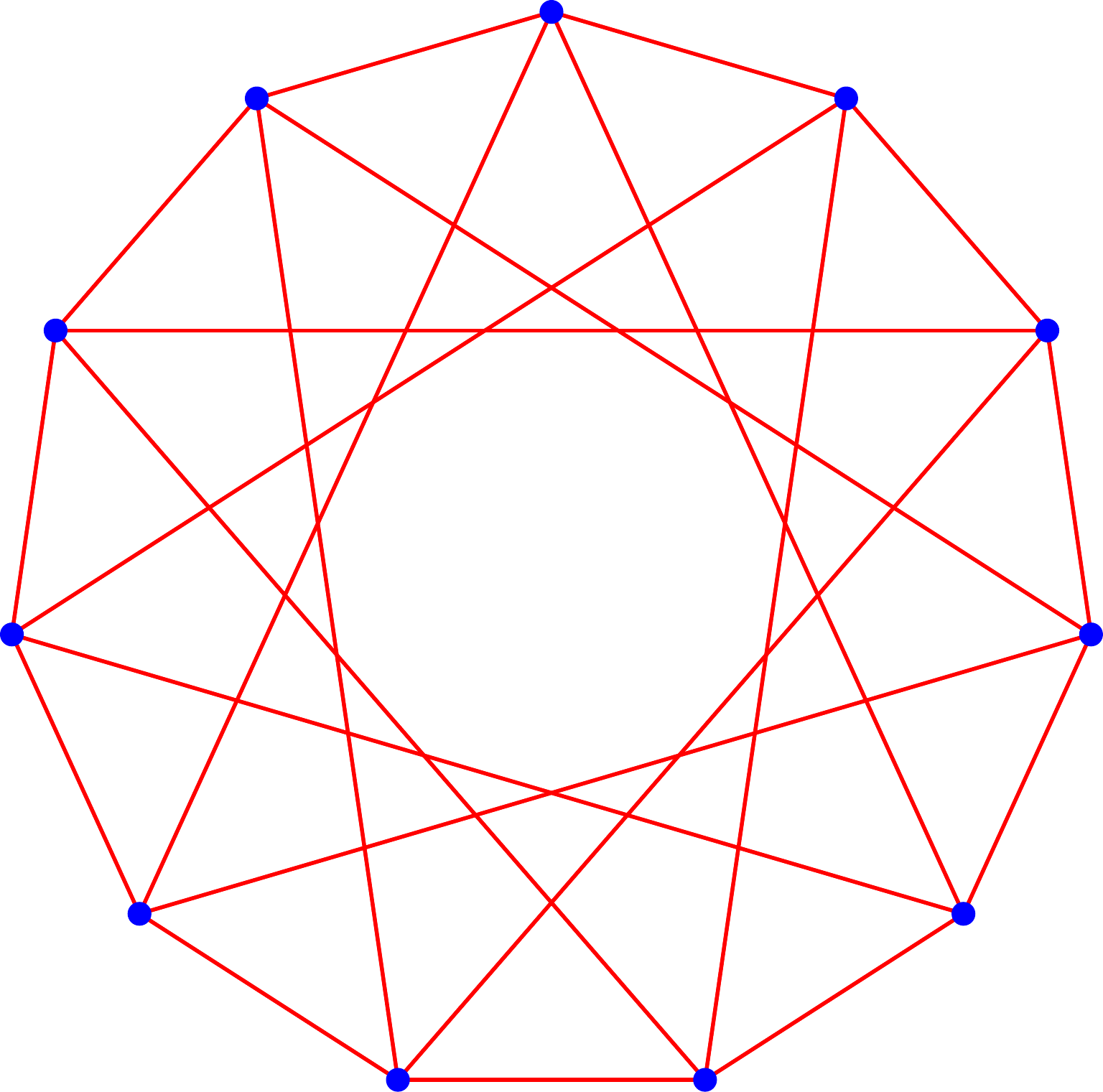} & \multicolumn{2}{c|}{}                                                                  & \multicolumn{2}{c|}{\cellcolor[HTML]{d0ffd5}\def\svgwidth{1.5cm}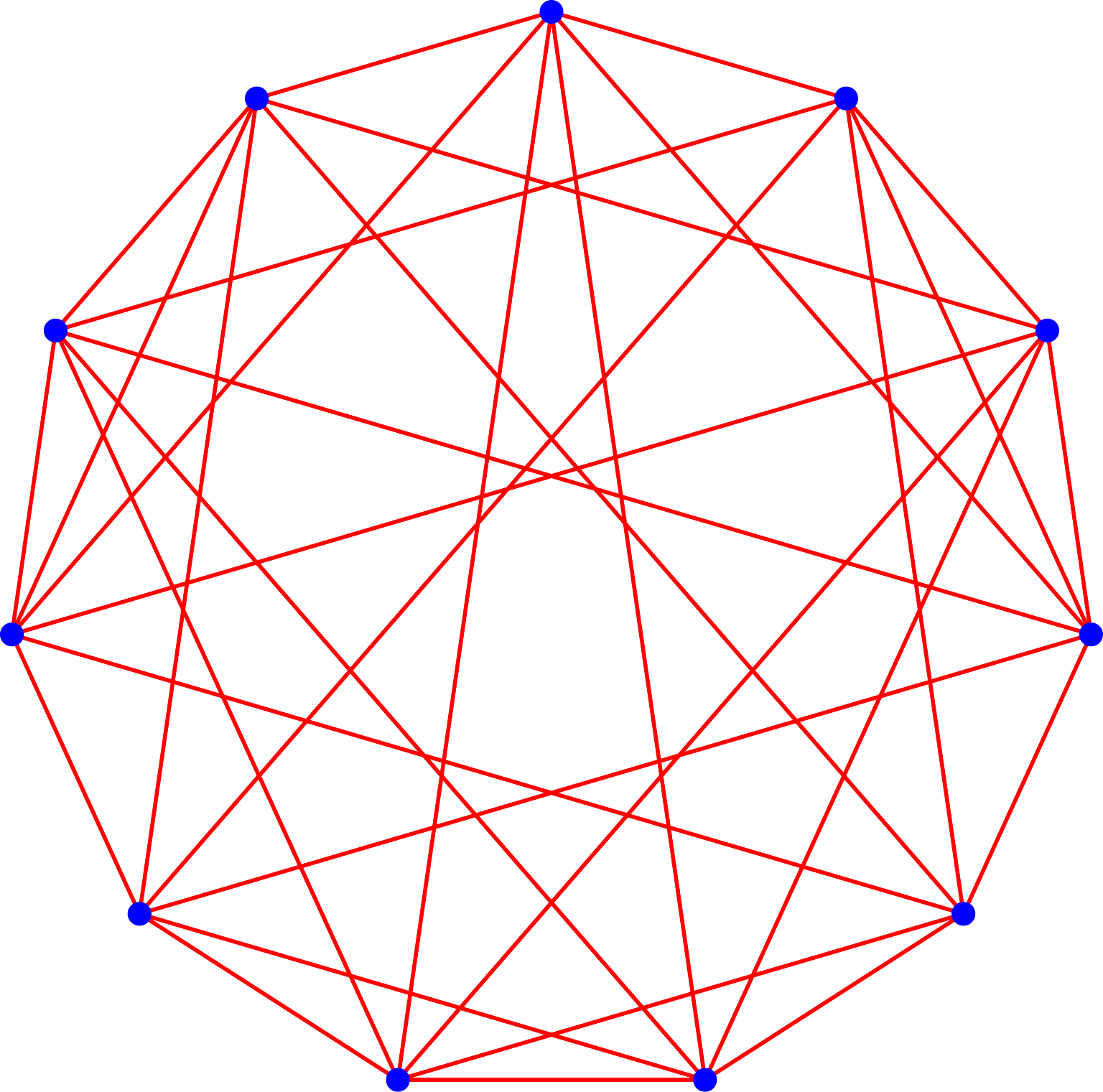}           & \multicolumn{2}{c|}{}                                                                            \\*
                     &                                             &                                         & \multicolumn{1}{l}{\cellcolor[HTML]{d0ffd5}8}          & \cellcolor[HTML]{d0ffd5}22    &                                            &                                           & \cellcolor[HTML]{d0ffd5}16                     & \cellcolor[HTML]{d0ffd5}5760                    &                                        & \multicolumn{1}{r|}{}                                   \\*
\multirow{-4}{*}[0.5cm]{11} &                                             & \multicolumn{1}{l|}{}                   & \multicolumn{2}{c|}{\cellcolor[HTML]{d0ffd5}$D_{11}$}                                  &                                            & \multicolumn{1}{l|}{}                     & \multicolumn{2}{c|}{\cellcolor[HTML]{d0ffd5}$C_{2} \times S_{4} \times S_{5}$}                   &                                        & \multicolumn{1}{l|}{}                                   \\ \hline
                     & \cellcolor[HTML]{ffffc7}1.9091              & \cellcolor[HTML]{ffffc7}1               & \multicolumn{1}{l}{\cellcolor[HTML]{d0ffd5}1.6364}     & \cellcolor[HTML]{d0ffd5}1     & \cellcolor[HTML]{d0ffd5}1.5455             & \cellcolor[HTML]{d0ffd5}1                 & \cellcolor[HTML]{d0ffd5}1.4545                 & \cellcolor[HTML]{d0ffd5}1                       & \cellcolor[HTML]{d0ffd5}1.3636         & \multicolumn{1}{r|}{\cellcolor[HTML]{d0ffd5}1}          \\*
                     & \multicolumn{2}{c|}{\cellcolor[HTML]{ffffc7}\def\svgwidth{1.5cm}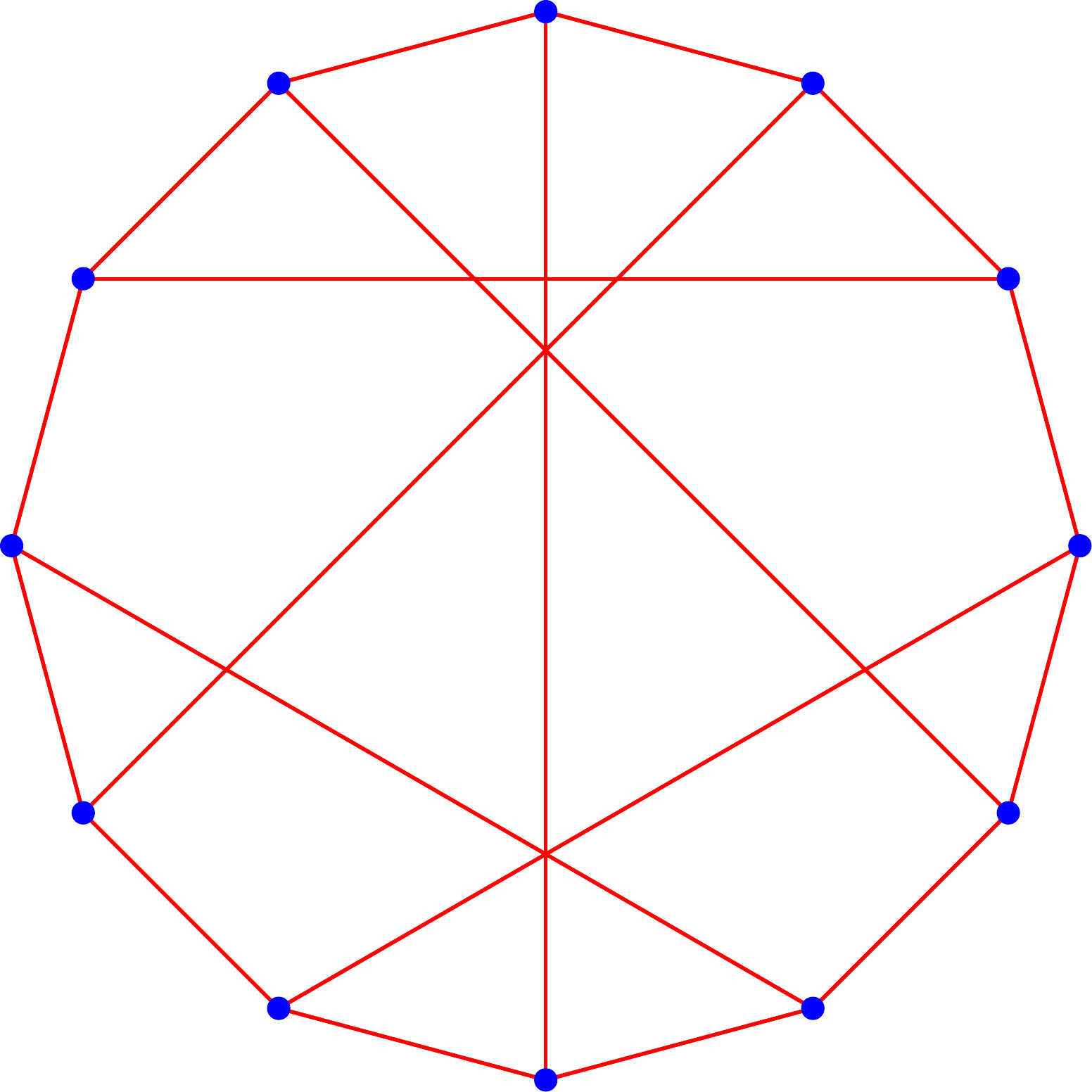} & \multicolumn{2}{c|}{\cellcolor[HTML]{d0ffd5}\def\svgwidth{1.5cm}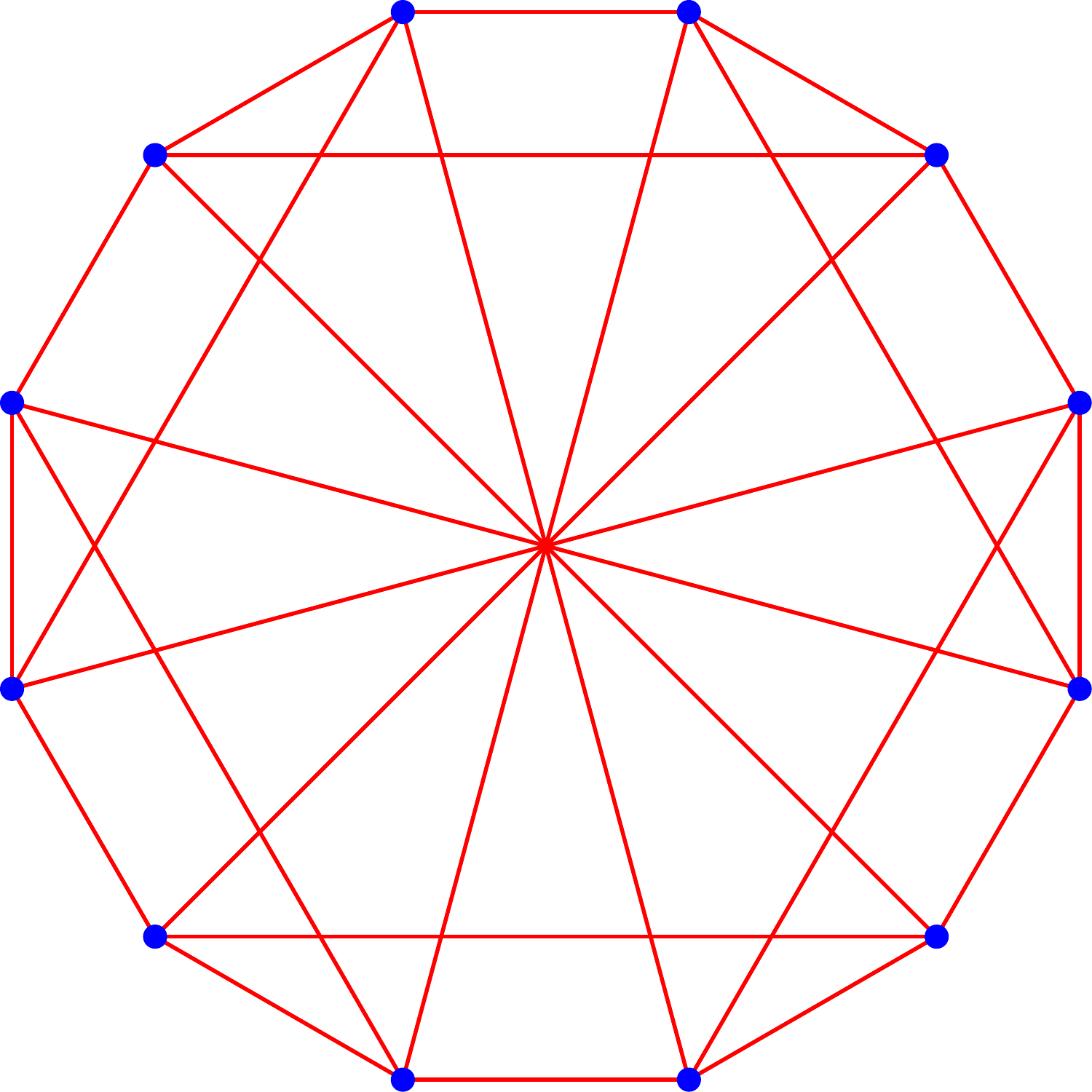} & \multicolumn{2}{c|}{\cellcolor[HTML]{d0ffd5}\def\svgwidth{1.5cm}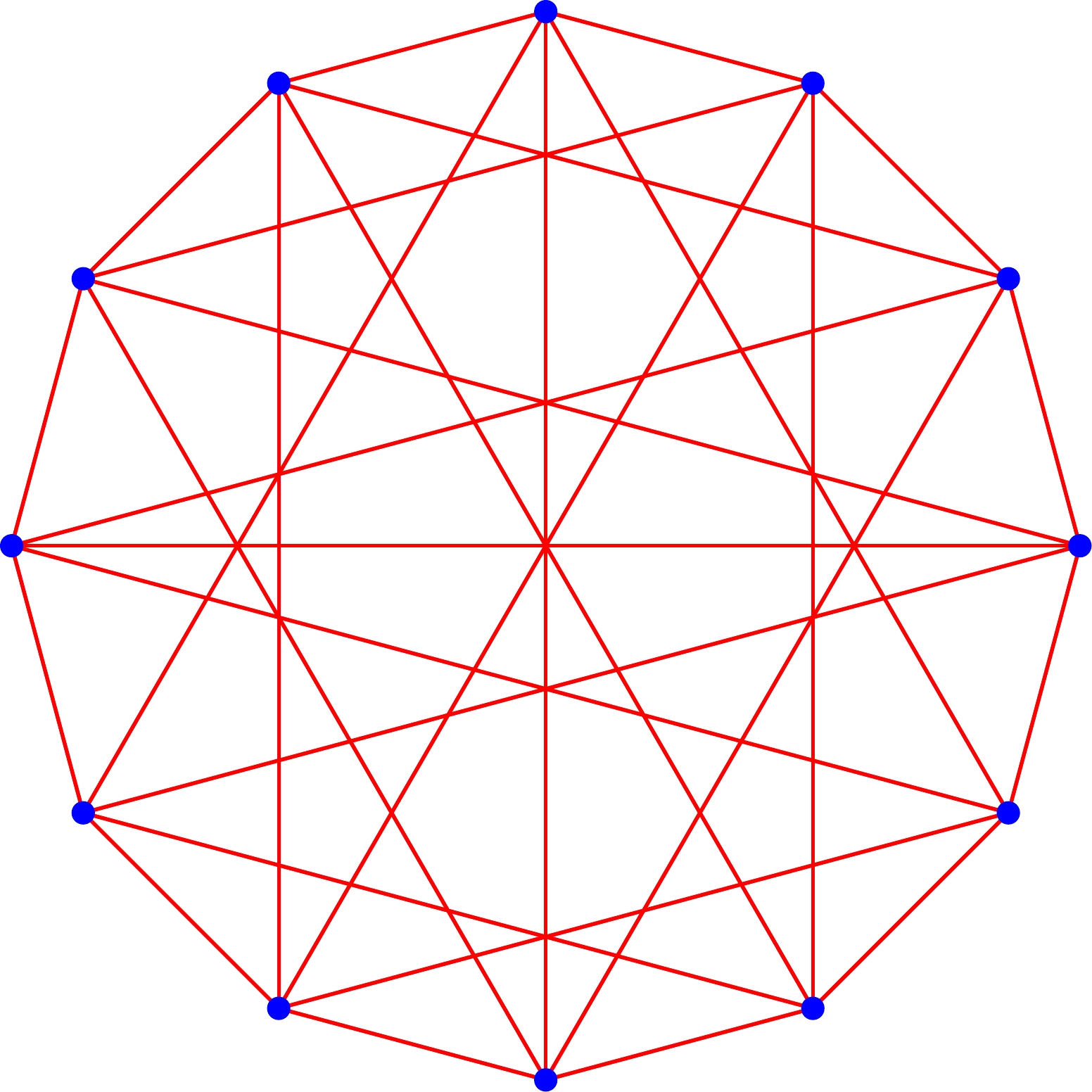} & \multicolumn{2}{c|}{\cellcolor[HTML]{d0ffd5}\def\svgwidth{1.5cm}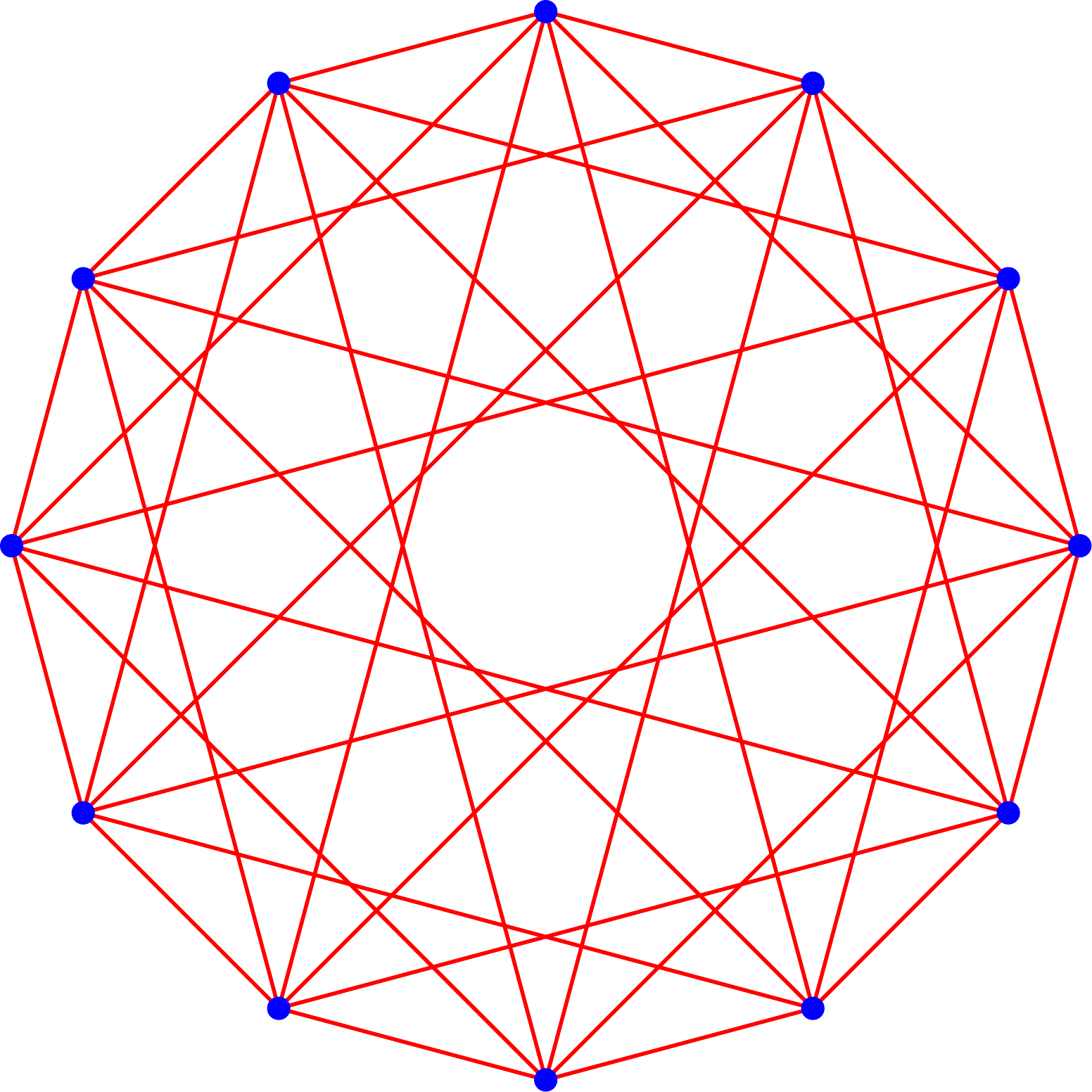}         & \multicolumn{2}{c|}{\cellcolor[HTML]{d0ffd5}\def\svgwidth{1.5cm}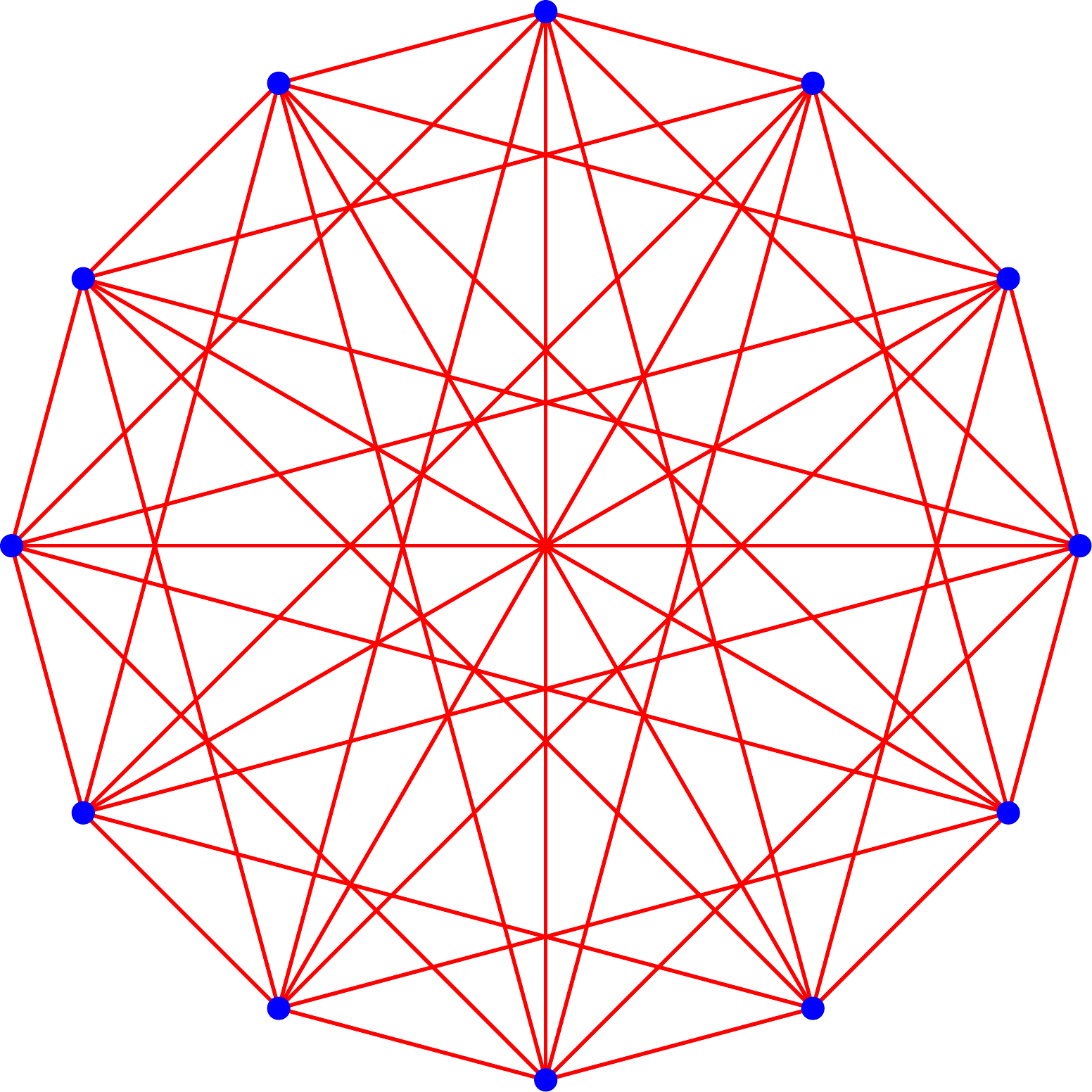}          \\*
                     & \cellcolor[HTML]{ffffc7}6                   & \cellcolor[HTML]{ffffc7}18              & \multicolumn{1}{l}{\cellcolor[HTML]{d0ffd5}10}         & \cellcolor[HTML]{d0ffd5}48    & \cellcolor[HTML]{d0ffd5}12                 & \cellcolor[HTML]{d0ffd5}576               & \cellcolor[HTML]{d0ffd5}18                     & \cellcolor[HTML]{d0ffd5}1036800                 & \cellcolor[HTML]{d0ffd5}20             & \multicolumn{1}{r|}{\cellcolor[HTML]{d0ffd5}4608}       \\*
                     & \multicolumn{2}{c|}{\cellcolor[HTML]{ffffc7}}                                         & \multicolumn{2}{c|}{\cellcolor[HTML]{d0ffd5}}                                          &  \multicolumn{2}{c|}{\cellcolor[HTML]{d0ffd5}$((C_2\times S_3)\wr S_2)$}                         & \multicolumn{2}{c|}{\cellcolor[HTML]{d0ffd5}}                                                    & \multicolumn{2}{c|}{\cellcolor[HTML]{d0ffd5}$(C_2\times (S_4\wr S_2))$}     \\*
\multirow{-5}{*}[0.5cm]{12} & \multicolumn{2}{c|}{\multirow{-2}{*}{\cellcolor[HTML]{ffffc7}$D_{9}$}}                & \multicolumn{2}{c|}{\multirow{-2}{*}{\cellcolor[HTML]{d0ffd5}$D_{4} \times S_{3}$}}    & \multicolumn{2}{c|}{\cellcolor[HTML]{d0ffd5}$\times C_2$}                         & \multicolumn{2}{c|}{\multirow{-2}{*}{\cellcolor[HTML]{d0ffd5}$S_{6} \wr S_{2}$}}                 & \multicolumn{2}{c|}{\cellcolor[HTML]{d0ffd5}$\rtimes C_2$}                      \\ \hline
\end{longtable}

%% file: 4k3_1.pdf_tex
\begingroup%
  \makeatletter%
  \providecommand\color[2][]{%
    \errmessage{(Inkscape) Color is used for the text in Inkscape, but the package 'color.sty' is not loaded}%
    \renewcommand\color[2][]{}%
  }%
  \providecommand\transparent[1]{%
    \errmessage{(Inkscape) Transparency is used (non-zero) for the text in Inkscape, but the package 'transparent.sty' is not loaded}%
    \renewcommand\transparent[1]{}%
  }%
  \providecommand\rotatebox[2]{#2}%
  \newcommand*\fsize{\dimexpr\f@size pt\relax}%
  \newcommand*\lineheight[1]{\fontsize{\fsize}{#1\fsize}\selectfont}%
  \ifx\svgwidth\undefined%
    \setlength{\unitlength}{446.80005026bp}%
    \ifx\svgscale\undefined%
      \relax%
    \else%
      \setlength{\unitlength}{\unitlength * \real{\svgscale}}%
    \fi%
  \else%
    \setlength{\unitlength}{\svgwidth}%
  \fi%
  \global\let\svgwidth\undefined%
  \global\let\svgscale\undefined%
  \makeatother%
  \begin{picture}(1,1)%
    \lineheight{1}%
    \setlength\tabcolsep{0pt}%
    \put(0,0){\includegraphics[width=\unitlength,page=1]{4k3_1.pdf}}%
  \end{picture}%
\endgroup%

%% file: 5k4_1.pdf_tex
\begingroup%
  \makeatletter%
  \providecommand\color[2][]{%
    \errmessage{(Inkscape) Color is used for the text in Inkscape, but the package 'color.sty' is not loaded}%
    \renewcommand\color[2][]{}%
  }%
  \providecommand\transparent[1]{%
    \errmessage{(Inkscape) Transparency is used (non-zero) for the text in Inkscape, but the package 'transparent.sty' is not loaded}%
    \renewcommand\transparent[1]{}%
  }%
  \providecommand\rotatebox[2]{#2}%
  \newcommand*\fsize{\dimexpr\f@size pt\relax}%
  \newcommand*\lineheight[1]{\fontsize{\fsize}{#1\fsize}\selectfont}%
  \ifx\svgwidth\undefined%
    \setlength{\unitlength}{425.39955026bp}%
    \ifx\svgscale\undefined%
      \relax%
    \else%
      \setlength{\unitlength}{\unitlength * \real{\svgscale}}%
    \fi%
  \else%
    \setlength{\unitlength}{\svgwidth}%
  \fi%
  \global\let\svgwidth\undefined%
  \global\let\svgscale\undefined%
  \makeatother%
  \begin{picture}(1,0.9521551)%
    \lineheight{1}%
    \setlength\tabcolsep{0pt}%
    \put(0,0){\includegraphics[width=\unitlength,page=1]{5k4_1.pdf}}%
  \end{picture}%
\endgroup%

%% file: 6k3_1.pdf_tex
\begingroup%
  \makeatletter%
  \providecommand\color[2][]{%
    \errmessage{(Inkscape) Color is used for the text in Inkscape, but the package 'color.sty' is not loaded}%
    \renewcommand\color[2][]{}%
  }%
  \providecommand\transparent[1]{%
    \errmessage{(Inkscape) Transparency is used (non-zero) for the text in Inkscape, but the package 'transparent.sty' is not loaded}%
    \renewcommand\transparent[1]{}%
  }%
  \providecommand\rotatebox[2]{#2}%
  \newcommand*\fsize{\dimexpr\f@size pt\relax}%
  \newcommand*\lineheight[1]{\fontsize{\fsize}{#1\fsize}\selectfont}%
  \ifx\svgwidth\undefined%
    \setlength{\unitlength}{388.21965026bp}%
    \ifx\svgscale\undefined%
      \relax%
    \else%
      \setlength{\unitlength}{\unitlength * \real{\svgscale}}%
    \fi%
  \else%
    \setlength{\unitlength}{\svgwidth}%
  \fi%
  \global\let\svgwidth\undefined%
  \global\let\svgscale\undefined%
  \makeatother%
  \begin{picture}(1,1.15089499)%
    \lineheight{1}%
    \setlength\tabcolsep{0pt}%
    \put(0,0){\includegraphics[width=\unitlength,page=1]{6k3_1.pdf}}%
  \end{picture}%
\endgroup%

%% file: 6k4_1.pdf_tex
\begingroup%
  \makeatletter%
  \providecommand\color[2][]{%
    \errmessage{(Inkscape) Color is used for the text in Inkscape, but the package 'color.sty' is not loaded}%
    \renewcommand\color[2][]{}%
  }%
  \providecommand\transparent[1]{%
    \errmessage{(Inkscape) Transparency is used (non-zero) for the text in Inkscape, but the package 'transparent.sty' is not loaded}%
    \renewcommand\transparent[1]{}%
  }%
  \providecommand\rotatebox[2]{#2}%
  \newcommand*\fsize{\dimexpr\f@size pt\relax}%
  \newcommand*\lineheight[1]{\fontsize{\fsize}{#1\fsize}\selectfont}%
  \ifx\svgwidth\undefined%
    \setlength{\unitlength}{388.21965026bp}%
    \ifx\svgscale\undefined%
      \relax%
    \else%
      \setlength{\unitlength}{\unitlength * \real{\svgscale}}%
    \fi%
  \else%
    \setlength{\unitlength}{\svgwidth}%
  \fi%
  \global\let\svgwidth\undefined%
  \global\let\svgscale\undefined%
  \makeatother%
  \begin{picture}(1,1.15089499)%
    \lineheight{1}%
    \setlength\tabcolsep{0pt}%
    \put(0,0){\includegraphics[width=\unitlength,page=1]{6k4_1.pdf}}%
  \end{picture}%
\endgroup%

%% file: 6k5_1.pdf_tex
\begingroup%
  \makeatletter%
  \providecommand\color[2][]{%
    \errmessage{(Inkscape) Color is used for the text in Inkscape, but the package 'color.sty' is not loaded}%
    \renewcommand\color[2][]{}%
  }%
  \providecommand\transparent[1]{%
    \errmessage{(Inkscape) Transparency is used (non-zero) for the text in Inkscape, but the package 'transparent.sty' is not loaded}%
    \renewcommand\transparent[1]{}%
  }%
  \providecommand\rotatebox[2]{#2}%
  \newcommand*\fsize{\dimexpr\f@size pt\relax}%
  \newcommand*\lineheight[1]{\fontsize{\fsize}{#1\fsize}\selectfont}%
  \ifx\svgwidth\undefined%
    \setlength{\unitlength}{388.21965026bp}%
    \ifx\svgscale\undefined%
      \relax%
    \else%
      \setlength{\unitlength}{\unitlength * \real{\svgscale}}%
    \fi%
  \else%
    \setlength{\unitlength}{\svgwidth}%
  \fi%
  \global\let\svgwidth\undefined%
  \global\let\svgscale\undefined%
  \makeatother%
  \begin{picture}(1,1.15089499)%
    \lineheight{1}%
    \setlength\tabcolsep{0pt}%
    \put(0,0){\includegraphics[width=\unitlength,page=1]{6k5_1.pdf}}%
  \end{picture}%
\endgroup%

%% file: 7k4_1.pdf_tex
\begingroup%
  \makeatletter%
  \providecommand\color[2][]{%
    \errmessage{(Inkscape) Color is used for the text in Inkscape, but the package 'color.sty' is not loaded}%
    \renewcommand\color[2][]{}%
  }%
  \providecommand\transparent[1]{%
    \errmessage{(Inkscape) Transparency is used (non-zero) for the text in Inkscape, but the package 'transparent.sty' is not loaded}%
    \renewcommand\transparent[1]{}%
  }%
  \providecommand\rotatebox[2]{#2}%
  \newcommand*\fsize{\dimexpr\f@size pt\relax}%
  \newcommand*\lineheight[1]{\fontsize{\fsize}{#1\fsize}\selectfont}%
  \ifx\svgwidth\undefined%
    \setlength{\unitlength}{435.83730026bp}%
    \ifx\svgscale\undefined%
      \relax%
    \else%
      \setlength{\unitlength}{\unitlength * \real{\svgscale}}%
    \fi%
  \else%
    \setlength{\unitlength}{\svgwidth}%
  \fi%
  \global\let\svgwidth\undefined%
  \global\let\svgscale\undefined%
  \makeatother%
  \begin{picture}(1,0.9754772)%
    \lineheight{1}%
    \setlength\tabcolsep{0pt}%
    \put(0,0){\includegraphics[width=\unitlength,page=1]{7k4_1.pdf}}%
  \end{picture}%
\endgroup%

%% file: 7k6_1.pdf_tex
\begingroup%
  \makeatletter%
  \providecommand\color[2][]{%
    \errmessage{(Inkscape) Color is used for the text in Inkscape, but the package 'color.sty' is not loaded}%
    \renewcommand\color[2][]{}%
  }%
  \providecommand\transparent[1]{%
    \errmessage{(Inkscape) Transparency is used (non-zero) for the text in Inkscape, but the package 'transparent.sty' is not loaded}%
    \renewcommand\transparent[1]{}%
  }%
  \providecommand\rotatebox[2]{#2}%
  \newcommand*\fsize{\dimexpr\f@size pt\relax}%
  \newcommand*\lineheight[1]{\fontsize{\fsize}{#1\fsize}\selectfont}%
  \ifx\svgwidth\undefined%
    \setlength{\unitlength}{435.83730026bp}%
    \ifx\svgscale\undefined%
      \relax%
    \else%
      \setlength{\unitlength}{\unitlength * \real{\svgscale}}%
    \fi%
  \else%
    \setlength{\unitlength}{\svgwidth}%
  \fi%
  \global\let\svgwidth\undefined%
  \global\let\svgscale\undefined%
  \makeatother%
  \begin{picture}(1,0.9754772)%
    \lineheight{1}%
    \setlength\tabcolsep{0pt}%
    \put(0,0){\includegraphics[width=\unitlength,page=1]{7k6_1.pdf}}%
  \end{picture}%
\endgroup%

%% file: 8k3_1.pdf_tex
\begingroup%
  \makeatletter%
  \providecommand\color[2][]{%
    \errmessage{(Inkscape) Color is used for the text in Inkscape, but the package 'color.sty' is not loaded}%
    \renewcommand\color[2][]{}%
  }%
  \providecommand\transparent[1]{%
    \errmessage{(Inkscape) Transparency is used (non-zero) for the text in Inkscape, but the package 'transparent.sty' is not loaded}%
    \renewcommand\transparent[1]{}%
  }%
  \providecommand\rotatebox[2]{#2}%
  \newcommand*\fsize{\dimexpr\f@size pt\relax}%
  \newcommand*\lineheight[1]{\fontsize{\fsize}{#1\fsize}\selectfont}%
  \ifx\svgwidth\undefined%
    \setlength{\unitlength}{446.80005026bp}%
    \ifx\svgscale\undefined%
      \relax%
    \else%
      \setlength{\unitlength}{\unitlength * \real{\svgscale}}%
    \fi%
  \else%
    \setlength{\unitlength}{\svgwidth}%
  \fi%
  \global\let\svgwidth\undefined%
  \global\let\svgscale\undefined%
  \makeatother%
  \begin{picture}(1,1)%
    \lineheight{1}%
    \setlength\tabcolsep{0pt}%
    \put(0,0){\includegraphics[width=\unitlength,page=1]{8k3_1.pdf}}%
  \end{picture}%
\endgroup%

%% file: 8k4_6.pdf_tex
\begingroup%
  \makeatletter%
  \providecommand\color[2][]{%
    \errmessage{(Inkscape) Color is used for the text in Inkscape, but the package 'color.sty' is not loaded}%
    \renewcommand\color[2][]{}%
  }%
  \providecommand\transparent[1]{%
    \errmessage{(Inkscape) Transparency is used (non-zero) for the text in Inkscape, but the package 'transparent.sty' is not loaded}%
    \renewcommand\transparent[1]{}%
  }%
  \providecommand\rotatebox[2]{#2}%
  \newcommand*\fsize{\dimexpr\f@size pt\relax}%
  \newcommand*\lineheight[1]{\fontsize{\fsize}{#1\fsize}\selectfont}%
  \ifx\svgwidth\undefined%
    \setlength{\unitlength}{446.80005026bp}%
    \ifx\svgscale\undefined%
      \relax%
    \else%
      \setlength{\unitlength}{\unitlength * \real{\svgscale}}%
    \fi%
  \else%
    \setlength{\unitlength}{\svgwidth}%
  \fi%
  \global\let\svgwidth\undefined%
  \global\let\svgscale\undefined%
  \makeatother%
  \begin{picture}(1,1)%
    \lineheight{1}%
    \setlength\tabcolsep{0pt}%
    \put(0,0){\includegraphics[width=\unitlength,page=1]{8k4_6.pdf}}%
  \end{picture}%
\endgroup%

%% file: 8k5_3.pdf_tex
\begingroup%
  \makeatletter%
  \providecommand\color[2][]{%
    \errmessage{(Inkscape) Color is used for the text in Inkscape, but the package 'color.sty' is not loaded}%
    \renewcommand\color[2][]{}%
  }%
  \providecommand\transparent[1]{%
    \errmessage{(Inkscape) Transparency is used (non-zero) for the text in Inkscape, but the package 'transparent.sty' is not loaded}%
    \renewcommand\transparent[1]{}%
  }%
  \providecommand\rotatebox[2]{#2}%
  \newcommand*\fsize{\dimexpr\f@size pt\relax}%
  \newcommand*\lineheight[1]{\fontsize{\fsize}{#1\fsize}\selectfont}%
  \ifx\svgwidth\undefined%
    \setlength{\unitlength}{446.80005026bp}%
    \ifx\svgscale\undefined%
      \relax%
    \else%
      \setlength{\unitlength}{\unitlength * \real{\svgscale}}%
    \fi%
  \else%
    \setlength{\unitlength}{\svgwidth}%
  \fi%
  \global\let\svgwidth\undefined%
  \global\let\svgscale\undefined%
  \makeatother%
  \begin{picture}(1,1)%
    \lineheight{1}%
    \setlength\tabcolsep{0pt}%
    \put(0,0){\includegraphics[width=\unitlength,page=1]{8k5_3.pdf}}%
  \end{picture}%
\endgroup%

%% file: 8k6_1.pdf_tex
\begingroup%
  \makeatletter%
  \providecommand\color[2][]{%
    \errmessage{(Inkscape) Color is used for the text in Inkscape, but the package 'color.sty' is not loaded}%
    \renewcommand\color[2][]{}%
  }%
  \providecommand\transparent[1]{%
    \errmessage{(Inkscape) Transparency is used (non-zero) for the text in Inkscape, but the package 'transparent.sty' is not loaded}%
    \renewcommand\transparent[1]{}%
  }%
  \providecommand\rotatebox[2]{#2}%
  \newcommand*\fsize{\dimexpr\f@size pt\relax}%
  \newcommand*\lineheight[1]{\fontsize{\fsize}{#1\fsize}\selectfont}%
  \ifx\svgwidth\undefined%
    \setlength{\unitlength}{446.80005026bp}%
    \ifx\svgscale\undefined%
      \relax%
    \else%
      \setlength{\unitlength}{\unitlength * \real{\svgscale}}%
    \fi%
  \else%
    \setlength{\unitlength}{\svgwidth}%
  \fi%
  \global\let\svgwidth\undefined%
  \global\let\svgscale\undefined%
  \makeatother%
  \begin{picture}(1,1)%
    \lineheight{1}%
    \setlength\tabcolsep{0pt}%
    \put(0,0){\includegraphics[width=\unitlength,page=1]{8k6_1.pdf}}%
  \end{picture}%
\endgroup%

%% file: 8k7_1.pdf_tex
\begingroup%
  \makeatletter%
  \providecommand\color[2][]{%
    \errmessage{(Inkscape) Color is used for the text in Inkscape, but the package 'color.sty' is not loaded}%
    \renewcommand\color[2][]{}%
  }%
  \providecommand\transparent[1]{%
    \errmessage{(Inkscape) Transparency is used (non-zero) for the text in Inkscape, but the package 'transparent.sty' is not loaded}%
    \renewcommand\transparent[1]{}%
  }%
  \providecommand\rotatebox[2]{#2}%
  \newcommand*\fsize{\dimexpr\f@size pt\relax}%
  \newcommand*\lineheight[1]{\fontsize{\fsize}{#1\fsize}\selectfont}%
  \ifx\svgwidth\undefined%
    \setlength{\unitlength}{446.80005026bp}%
    \ifx\svgscale\undefined%
      \relax%
    \else%
      \setlength{\unitlength}{\unitlength * \real{\svgscale}}%
    \fi%
  \else%
    \setlength{\unitlength}{\svgwidth}%
  \fi%
  \global\let\svgwidth\undefined%
  \global\let\svgscale\undefined%
  \makeatother%
  \begin{picture}(1,1)%
    \lineheight{1}%
    \setlength\tabcolsep{0pt}%
    \put(0,0){\includegraphics[width=\unitlength,page=1]{8k7_1.pdf}}%
  \end{picture}%
\endgroup%

%% file: 9k4_14.pdf_tex
\begingroup%
  \makeatletter%
  \providecommand\color[2][]{%
    \errmessage{(Inkscape) Color is used for the text in Inkscape, but the package 'color.sty' is not loaded}%
    \renewcommand\color[2][]{}%
  }%
  \providecommand\transparent[1]{%
    \errmessage{(Inkscape) Transparency is used (non-zero) for the text in Inkscape, but the package 'transparent.sty' is not loaded}%
    \renewcommand\transparent[1]{}%
  }%
  \providecommand\rotatebox[2]{#2}%
  \newcommand*\fsize{\dimexpr\f@size pt\relax}%
  \newcommand*\lineheight[1]{\fontsize{\fsize}{#1\fsize}\selectfont}%
  \ifx\svgwidth\undefined%
    \setlength{\unitlength}{440.15730026bp}%
    \ifx\svgscale\undefined%
      \relax%
    \else%
      \setlength{\unitlength}{\unitlength * \real{\svgscale}}%
    \fi%
  \else%
    \setlength{\unitlength}{\svgwidth}%
  \fi%
  \global\let\svgwidth\undefined%
  \global\let\svgscale\undefined%
  \makeatother%
  \begin{picture}(1,0.98513725)%
    \lineheight{1}%
    \setlength\tabcolsep{0pt}%
    \put(0,0){\includegraphics[width=\unitlength,page=1]{9k4_14.pdf}}%
  \end{picture}%
\endgroup%

%% file: 9k6_4.pdf_tex
\begingroup%
  \makeatletter%
  \providecommand\color[2][]{%
    \errmessage{(Inkscape) Color is used for the text in Inkscape, but the package 'color.sty' is not loaded}%
    \renewcommand\color[2][]{}%
  }%
  \providecommand\transparent[1]{%
    \errmessage{(Inkscape) Transparency is used (non-zero) for the text in Inkscape, but the package 'transparent.sty' is not loaded}%
    \renewcommand\transparent[1]{}%
  }%
  \providecommand\rotatebox[2]{#2}%
  \newcommand*\fsize{\dimexpr\f@size pt\relax}%
  \newcommand*\lineheight[1]{\fontsize{\fsize}{#1\fsize}\selectfont}%
  \ifx\svgwidth\undefined%
    \setlength{\unitlength}{440.15730026bp}%
    \ifx\svgscale\undefined%
      \relax%
    \else%
      \setlength{\unitlength}{\unitlength * \real{\svgscale}}%
    \fi%
  \else%
    \setlength{\unitlength}{\svgwidth}%
  \fi%
  \global\let\svgwidth\undefined%
  \global\let\svgscale\undefined%
  \makeatother%
  \begin{picture}(1,0.98513725)%
    \lineheight{1}%
    \setlength\tabcolsep{0pt}%
    \put(0,0){\includegraphics[width=\unitlength,page=1]{9k6_4.pdf}}%
  \end{picture}%
\endgroup%

%% file: 10k3_1.pdf_tex
\begingroup%
  \makeatletter%
  \providecommand\color[2][]{%
    \errmessage{(Inkscape) Color is used for the text in Inkscape, but the package 'color.sty' is not loaded}%
    \renewcommand\color[2][]{}%
  }%
  \providecommand\transparent[1]{%
    \errmessage{(Inkscape) Transparency is used (non-zero) for the text in Inkscape, but the package 'transparent.sty' is not loaded}%
    \renewcommand\transparent[1]{}%
  }%
  \providecommand\rotatebox[2]{#2}%
  \newcommand*\fsize{\dimexpr\f@size pt\relax}%
  \newcommand*\lineheight[1]{\fontsize{\fsize}{#1\fsize}\selectfont}%
  \ifx\svgwidth\undefined%
    \setlength{\unitlength}{492.890625bp}%
    \ifx\svgscale\undefined%
      \relax%
    \else%
      \setlength{\unitlength}{\unitlength * \real{\svgscale}}%
    \fi%
  \else%
    \setlength{\unitlength}{\svgwidth}%
  \fi%
  \global\let\svgwidth\undefined%
  \global\let\svgscale\undefined%
  \makeatother%
  \begin{picture}(1,0.9520447)%
    \lineheight{1}%
    \setlength\tabcolsep{0pt}%
    \put(0,0){\includegraphics[width=\unitlength,page=1]{10k3_1.pdf}}%
  \end{picture}%
\endgroup%

%% file: 10k4_24.pdf_tex
\begingroup%
  \makeatletter%
  \providecommand\color[2][]{%
    \errmessage{(Inkscape) Color is used for the text in Inkscape, but the package 'color.sty' is not loaded}%
    \renewcommand\color[2][]{}%
  }%
  \providecommand\transparent[1]{%
    \errmessage{(Inkscape) Transparency is used (non-zero) for the text in Inkscape, but the package 'transparent.sty' is not loaded}%
    \renewcommand\transparent[1]{}%
  }%
  \providecommand\rotatebox[2]{#2}%
  \newcommand*\fsize{\dimexpr\f@size pt\relax}%
  \newcommand*\lineheight[1]{\fontsize{\fsize}{#1\fsize}\selectfont}%
  \ifx\svgwidth\undefined%
    \setlength{\unitlength}{425.39955026bp}%
    \ifx\svgscale\undefined%
      \relax%
    \else%
      \setlength{\unitlength}{\unitlength * \real{\svgscale}}%
    \fi%
  \else%
    \setlength{\unitlength}{\svgwidth}%
  \fi%
  \global\let\svgwidth\undefined%
  \global\let\svgscale\undefined%
  \makeatother%
  \begin{picture}(1,1.05030682)%
    \lineheight{1}%
    \setlength\tabcolsep{0pt}%
    \put(0,0){\includegraphics[width=\unitlength,page=1]{10k4_24.pdf}}%
  \end{picture}%
\endgroup%

%% file: 10k5_60.pdf_tex
\begingroup%
  \makeatletter%
  \providecommand\color[2][]{%
    \errmessage{(Inkscape) Color is used for the text in Inkscape, but the package 'color.sty' is not loaded}%
    \renewcommand\color[2][]{}%
  }%
  \providecommand\transparent[1]{%
    \errmessage{(Inkscape) Transparency is used (non-zero) for the text in Inkscape, but the package 'transparent.sty' is not loaded}%
    \renewcommand\transparent[1]{}%
  }%
  \providecommand\rotatebox[2]{#2}%
  \newcommand*\fsize{\dimexpr\f@size pt\relax}%
  \newcommand*\lineheight[1]{\fontsize{\fsize}{#1\fsize}\selectfont}%
  \ifx\svgwidth\undefined%
    \setlength{\unitlength}{425.39955026bp}%
    \ifx\svgscale\undefined%
      \relax%
    \else%
      \setlength{\unitlength}{\unitlength * \real{\svgscale}}%
    \fi%
  \else%
    \setlength{\unitlength}{\svgwidth}%
  \fi%
  \global\let\svgwidth\undefined%
  \global\let\svgscale\undefined%
  \makeatother%
  \begin{picture}(1,1.05030682)%
    \lineheight{1}%
    \setlength\tabcolsep{0pt}%
    \put(0,0){\includegraphics[width=\unitlength,page=1]{10k5_60.pdf}}%
  \end{picture}%
\endgroup%

%% file: 10k6_18.pdf_tex
\begingroup%
  \makeatletter%
  \providecommand\color[2][]{%
    \errmessage{(Inkscape) Color is used for the text in Inkscape, but the package 'color.sty' is not loaded}%
    \renewcommand\color[2][]{}%
  }%
  \providecommand\transparent[1]{%
    \errmessage{(Inkscape) Transparency is used (non-zero) for the text in Inkscape, but the package 'transparent.sty' is not loaded}%
    \renewcommand\transparent[1]{}%
  }%
  \providecommand\rotatebox[2]{#2}%
  \newcommand*\fsize{\dimexpr\f@size pt\relax}%
  \newcommand*\lineheight[1]{\fontsize{\fsize}{#1\fsize}\selectfont}%
  \ifx\svgwidth\undefined%
    \setlength{\unitlength}{425.39955026bp}%
    \ifx\svgscale\undefined%
      \relax%
    \else%
      \setlength{\unitlength}{\unitlength * \real{\svgscale}}%
    \fi%
  \else%
    \setlength{\unitlength}{\svgwidth}%
  \fi%
  \global\let\svgwidth\undefined%
  \global\let\svgscale\undefined%
  \makeatother%
  \begin{picture}(1,1.05030682)%
    \lineheight{1}%
    \setlength\tabcolsep{0pt}%
    \put(0,0){\includegraphics[width=\unitlength,page=1]{10k6_18.pdf}}%
  \end{picture}%
\endgroup%

%% file: 10k7_2.pdf_tex
\begingroup%
  \makeatletter%
  \providecommand\color[2][]{%
    \errmessage{(Inkscape) Color is used for the text in Inkscape, but the package 'color.sty' is not loaded}%
    \renewcommand\color[2][]{}%
  }%
  \providecommand\transparent[1]{%
    \errmessage{(Inkscape) Transparency is used (non-zero) for the text in Inkscape, but the package 'transparent.sty' is not loaded}%
    \renewcommand\transparent[1]{}%
  }%
  \providecommand\rotatebox[2]{#2}%
  \newcommand*\fsize{\dimexpr\f@size pt\relax}%
  \newcommand*\lineheight[1]{\fontsize{\fsize}{#1\fsize}\selectfont}%
  \ifx\svgwidth\undefined%
    \setlength{\unitlength}{425.39955026bp}%
    \ifx\svgscale\undefined%
      \relax%
    \else%
      \setlength{\unitlength}{\unitlength * \real{\svgscale}}%
    \fi%
  \else%
    \setlength{\unitlength}{\svgwidth}%
  \fi%
  \global\let\svgwidth\undefined%
  \global\let\svgscale\undefined%
  \makeatother%
  \begin{picture}(1,1.05030682)%
    \lineheight{1}%
    \setlength\tabcolsep{0pt}%
    \put(0,0){\includegraphics[width=\unitlength,page=1]{10k7_2.pdf}}%
  \end{picture}%
\endgroup%

%% file: 11k4_37.pdf_tex
\begingroup%
  \makeatletter%
  \providecommand\color[2][]{%
    \errmessage{(Inkscape) Color is used for the text in Inkscape, but the package 'color.sty' is not loaded}%
    \renewcommand\color[2][]{}%
  }%
  \providecommand\transparent[1]{%
    \errmessage{(Inkscape) Transparency is used (non-zero) for the text in Inkscape, but the package 'transparent.sty' is not loaded}%
    \renewcommand\transparent[1]{}%
  }%
  \providecommand\rotatebox[2]{#2}%
  \newcommand*\fsize{\dimexpr\f@size pt\relax}%
  \newcommand*\lineheight[1]{\fontsize{\fsize}{#1\fsize}\selectfont}%
  \ifx\svgwidth\undefined%
    \setlength{\unitlength}{442.34947526bp}%
    \ifx\svgscale\undefined%
      \relax%
    \else%
      \setlength{\unitlength}{\unitlength * \real{\svgscale}}%
    \fi%
  \else%
    \setlength{\unitlength}{\svgwidth}%
  \fi%
  \global\let\svgwidth\undefined%
  \global\let\svgscale\undefined%
  \makeatother%
  \begin{picture}(1,0.99004119)%
    \lineheight{1}%
    \setlength\tabcolsep{0pt}%
    \put(0,0){\includegraphics[width=\unitlength,page=1]{11k4_37.pdf}}%
  \end{picture}%
\endgroup%

%% file: 11k6_49.pdf_tex
\begingroup%
  \makeatletter%
  \providecommand\color[2][]{%
    \errmessage{(Inkscape) Color is used for the text in Inkscape, but the package 'color.sty' is not loaded}%
    \renewcommand\color[2][]{}%
  }%
  \providecommand\transparent[1]{%
    \errmessage{(Inkscape) Transparency is used (non-zero) for the text in Inkscape, but the package 'transparent.sty' is not loaded}%
    \renewcommand\transparent[1]{}%
  }%
  \providecommand\rotatebox[2]{#2}%
  \newcommand*\fsize{\dimexpr\f@size pt\relax}%
  \newcommand*\lineheight[1]{\fontsize{\fsize}{#1\fsize}\selectfont}%
  \ifx\svgwidth\undefined%
    \setlength{\unitlength}{442.34947526bp}%
    \ifx\svgscale\undefined%
      \relax%
    \else%
      \setlength{\unitlength}{\unitlength * \real{\svgscale}}%
    \fi%
  \else%
    \setlength{\unitlength}{\svgwidth}%
  \fi%
  \global\let\svgwidth\undefined%
  \global\let\svgscale\undefined%
  \makeatother%
  \begin{picture}(1,0.99004119)%
    \lineheight{1}%
    \setlength\tabcolsep{0pt}%
    \put(0,0){\includegraphics[width=\unitlength,page=1]{11k6_49.pdf}}%
  \end{picture}%
\endgroup%

%% file: 12k3_2.pdf_tex
\begingroup%
  \makeatletter%
  \providecommand\color[2][]{%
    \errmessage{(Inkscape) Color is used for the text in Inkscape, but the package 'color.sty' is not loaded}%
    \renewcommand\color[2][]{}%
  }%
  \providecommand\transparent[1]{%
    \errmessage{(Inkscape) Transparency is used (non-zero) for the text in Inkscape, but the package 'transparent.sty' is not loaded}%
    \renewcommand\transparent[1]{}%
  }%
  \providecommand\rotatebox[2]{#2}%
  \newcommand*\fsize{\dimexpr\f@size pt\relax}%
  \newcommand*\lineheight[1]{\fontsize{\fsize}{#1\fsize}\selectfont}%
  \ifx\svgwidth\undefined%
    \setlength{\unitlength}{446.80005026bp}%
    \ifx\svgscale\undefined%
      \relax%
    \else%
      \setlength{\unitlength}{\unitlength * \real{\svgscale}}%
    \fi%
  \else%
    \setlength{\unitlength}{\svgwidth}%
  \fi%
  \global\let\svgwidth\undefined%
  \global\let\svgscale\undefined%
  \makeatother%
  \begin{picture}(1,1)%
    \lineheight{1}%
    \setlength\tabcolsep{0pt}%
    \put(0,0){\includegraphics[width=\unitlength,page=1]{12k3_2.pdf}}%
  \end{picture}%
\endgroup%

%% file: 12k4_26.pdf_tex
\begingroup%
  \makeatletter%
  \providecommand\color[2][]{%
    \errmessage{(Inkscape) Color is used for the text in Inkscape, but the package 'color.sty' is not loaded}%
    \renewcommand\color[2][]{}%
  }%
  \providecommand\transparent[1]{%
    \errmessage{(Inkscape) Transparency is used (non-zero) for the text in Inkscape, but the package 'transparent.sty' is not loaded}%
    \renewcommand\transparent[1]{}%
  }%
  \providecommand\rotatebox[2]{#2}%
  \newcommand*\fsize{\dimexpr\f@size pt\relax}%
  \newcommand*\lineheight[1]{\fontsize{\fsize}{#1\fsize}\selectfont}%
  \ifx\svgwidth\undefined%
    \setlength{\unitlength}{431.90115026bp}%
    \ifx\svgscale\undefined%
      \relax%
    \else%
      \setlength{\unitlength}{\unitlength * \real{\svgscale}}%
    \fi%
  \else%
    \setlength{\unitlength}{\svgwidth}%
  \fi%
  \global\let\svgwidth\undefined%
  \global\let\svgscale\undefined%
  \makeatother%
  \begin{picture}(1,1)%
    \lineheight{1}%
    \setlength\tabcolsep{0pt}%
    \put(0,0){\includegraphics[width=\unitlength,page=1]{12k4_26.pdf}}%
  \end{picture}%
\endgroup%

%% file: 12k5_81.pdf_tex
\begingroup%
  \makeatletter%
  \providecommand\color[2][]{%
    \errmessage{(Inkscape) Color is used for the text in Inkscape, but the package 'color.sty' is not loaded}%
    \renewcommand\color[2][]{}%
  }%
  \providecommand\transparent[1]{%
    \errmessage{(Inkscape) Transparency is used (non-zero) for the text in Inkscape, but the package 'transparent.sty' is not loaded}%
    \renewcommand\transparent[1]{}%
  }%
  \providecommand\rotatebox[2]{#2}%
  \newcommand*\fsize{\dimexpr\f@size pt\relax}%
  \newcommand*\lineheight[1]{\fontsize{\fsize}{#1\fsize}\selectfont}%
  \ifx\svgwidth\undefined%
    \setlength{\unitlength}{446.80005026bp}%
    \ifx\svgscale\undefined%
      \relax%
    \else%
      \setlength{\unitlength}{\unitlength * \real{\svgscale}}%
    \fi%
  \else%
    \setlength{\unitlength}{\svgwidth}%
  \fi%
  \global\let\svgwidth\undefined%
  \global\let\svgscale\undefined%
  \makeatother%
  \begin{picture}(1,1)%
    \lineheight{1}%
    \setlength\tabcolsep{0pt}%
    \put(0,0){\includegraphics[width=\unitlength,page=1]{12k5_81.pdf}}%
  \end{picture}%
\endgroup%

%% file: 12k6_7849.pdf_tex
\begingroup%
  \makeatletter%
  \providecommand\color[2][]{%
    \errmessage{(Inkscape) Color is used for the text in Inkscape, but the package 'color.sty' is not loaded}%
    \renewcommand\color[2][]{}%
  }%
  \providecommand\transparent[1]{%
    \errmessage{(Inkscape) Transparency is used (non-zero) for the text in Inkscape, but the package 'transparent.sty' is not loaded}%
    \renewcommand\transparent[1]{}%
  }%
  \providecommand\rotatebox[2]{#2}%
  \newcommand*\fsize{\dimexpr\f@size pt\relax}%
  \newcommand*\lineheight[1]{\fontsize{\fsize}{#1\fsize}\selectfont}%
  \ifx\svgwidth\undefined%
    \setlength{\unitlength}{446.80005026bp}%
    \ifx\svgscale\undefined%
      \relax%
    \else%
      \setlength{\unitlength}{\unitlength * \real{\svgscale}}%
    \fi%
  \else%
    \setlength{\unitlength}{\svgwidth}%
  \fi%
  \global\let\svgwidth\undefined%
  \global\let\svgscale\undefined%
  \makeatother%
  \begin{picture}(1,1)%
    \lineheight{1}%
    \setlength\tabcolsep{0pt}%
    \put(0,0){\includegraphics[width=\unitlength,page=1]{12k6_7849.pdf}}%
  \end{picture}%
\endgroup%

%% file: 12k7_220.pdf_tex
\begingroup%
  \makeatletter%
  \providecommand\color[2][]{%
    \errmessage{(Inkscape) Color is used for the text in Inkscape, but the package 'color.sty' is not loaded}%
    \renewcommand\color[2][]{}%
  }%
  \providecommand\transparent[1]{%
    \errmessage{(Inkscape) Transparency is used (non-zero) for the text in Inkscape, but the package 'transparent.sty' is not loaded}%
    \renewcommand\transparent[1]{}%
  }%
  \providecommand\rotatebox[2]{#2}%
  \newcommand*\fsize{\dimexpr\f@size pt\relax}%
  \newcommand*\lineheight[1]{\fontsize{\fsize}{#1\fsize}\selectfont}%
  \ifx\svgwidth\undefined%
    \setlength{\unitlength}{446.80005026bp}%
    \ifx\svgscale\undefined%
      \relax%
    \else%
      \setlength{\unitlength}{\unitlength * \real{\svgscale}}%
    \fi%
  \else%
    \setlength{\unitlength}{\svgwidth}%
  \fi%
  \global\let\svgwidth\undefined%
  \global\let\svgscale\undefined%
  \makeatother%
  \begin{picture}(1,1)%
    \lineheight{1}%
    \setlength\tabcolsep{0pt}%
    \put(0,0){\includegraphics[width=\unitlength,page=1]{12k7_220.pdf}}%
  \end{picture}%
\endgroup%

%% file: table2-color.tex
\begin{longtable}[c]{|c|cr||c|cr|}
\caption{The optimal graphs for $N=14, 16, ...,32$ and $k=3$ (left) and for $N=13, 14, ..., 21, 32$ and $k=4$ (right).}
\label{tab:table2}\\
\hline
   \diagbox{$N$}{$k$}                  & \multicolumn{2}{c||}{3}                                                                                &               \diagbox{$N$}{$k$}       & \multicolumn{2}{c|}{4}                                                                   \\ \hline
\endfirsthead
\multicolumn{5}{l}%
{\small{\bfseries Table \thetable.} The optimal graphs for $N=14, 16, ...,32$ } \\
\multicolumn{5}{l}%
{\small and $k=3$ (left) and for $N=13, 14, ..., 21, 32$ and $k=4$}\\
\multicolumn{5}{l}%
{\small (right) (continued). }\\
\hline
         \diagbox{$N$}{$k$}            & \multicolumn{2}{c||}{3}                                                                                &              \diagbox{$N$}{$k$}        & \multicolumn{2}{c|}{4}                                                                   \\ \hline
\endhead
\hline
\endfoot
\endlastfoot
                     & \multicolumn{1}{l}{\cellcolor[HTML]{ffffc7}2.0769} & \cellcolor[HTML]{ffffc7}1                        &                      & \multicolumn{1}{l}{\cellcolor[HTML]{d0ffd5}1.6667}      & \cellcolor[HTML]{d0ffd5}1      \\*
                     & \multicolumn{2}{c||}{\cellcolor[HTML]{ffffc7}\def\svgwidth{1.5cm}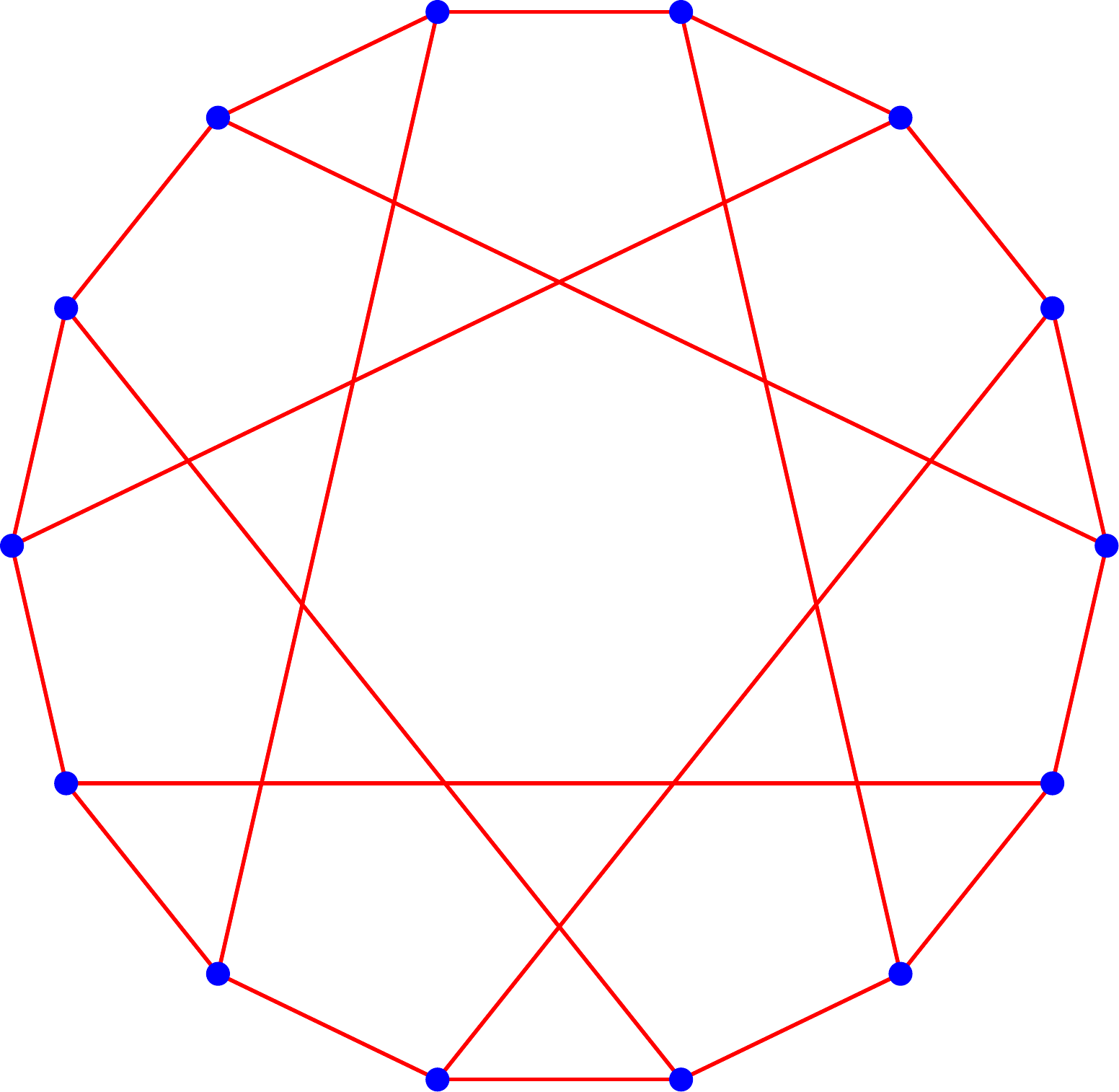}                 &                      & \multicolumn{2}{c|}{\cellcolor[HTML]{d0ffd5}\def\svgwidth{1.5cm}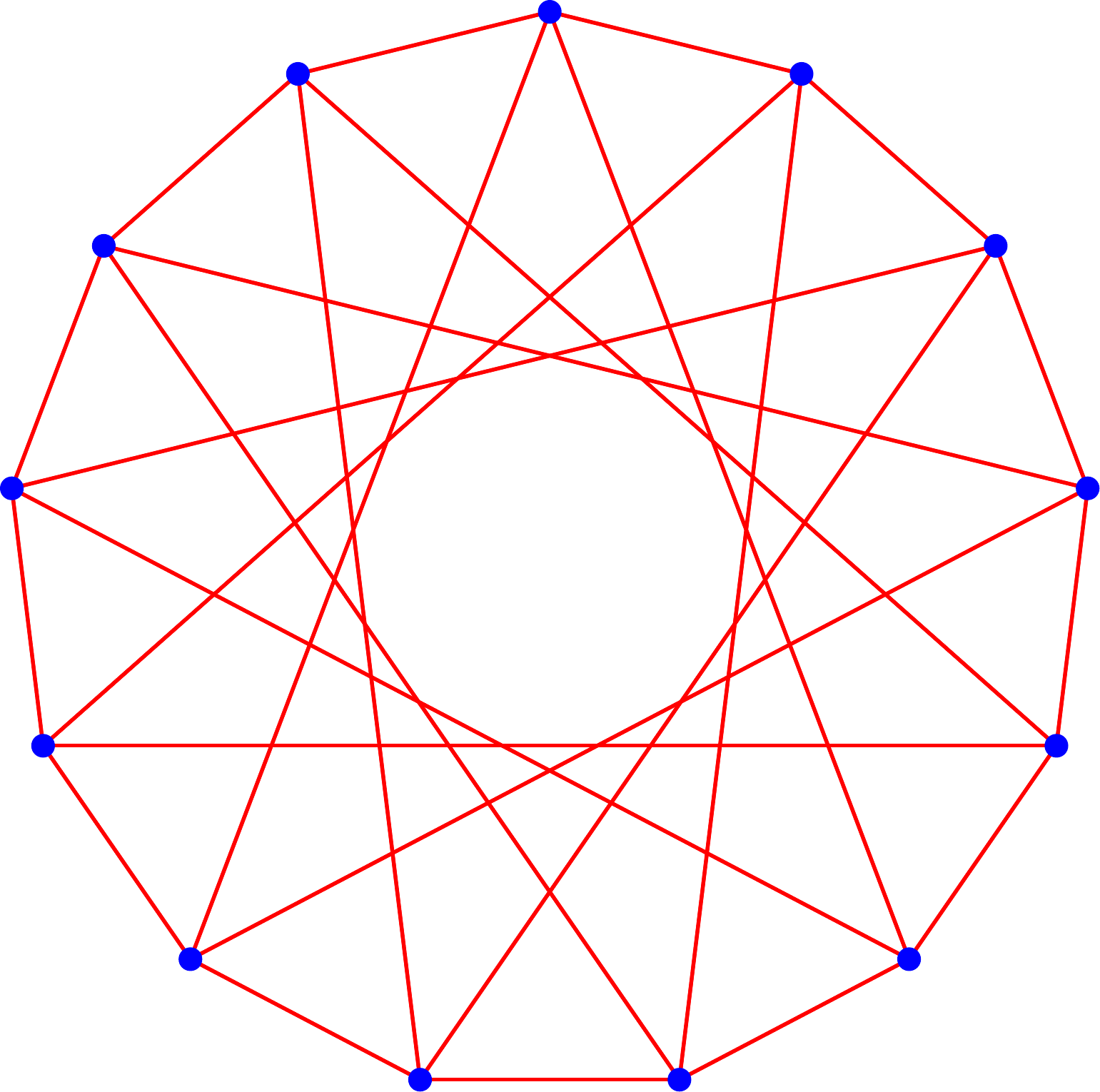}   \\*
                     & \multicolumn{1}{l}{\cellcolor[HTML]{ffffc7}7}      & \cellcolor[HTML]{ffffc7}336                      &                      & \multicolumn{1}{l}{\cellcolor[HTML]{d0ffd5}10}          & \cellcolor[HTML]{d0ffd5}52     \\*
\multirow{-4}{*}[0.5cm]{14} & \multicolumn{2}{c||}{\cellcolor[HTML]{ffffc7}$\mathrm{PGL}(2,7)$}                                      & \multirow{-4}{*}[0.5cm]{13} & \multicolumn{2}{c|}{\cellcolor[HTML]{d0ffd5}$C_{13} \rtimes C_{4}$}                      \\ \hline
                     & \multicolumn{1}{l}{\cellcolor[HTML]{ffffc7}2.2000} & \cellcolor[HTML]{ffffc7}2                        &                      & \multicolumn{1}{l}{\cellcolor[HTML]{d0ffd5}1.6923}      & \cellcolor[HTML]{d0ffd5}1      \\*
                     & \multicolumn{2}{c||}{\cellcolor[HTML]{ffffc7}\def\svgwidth{1.5cm}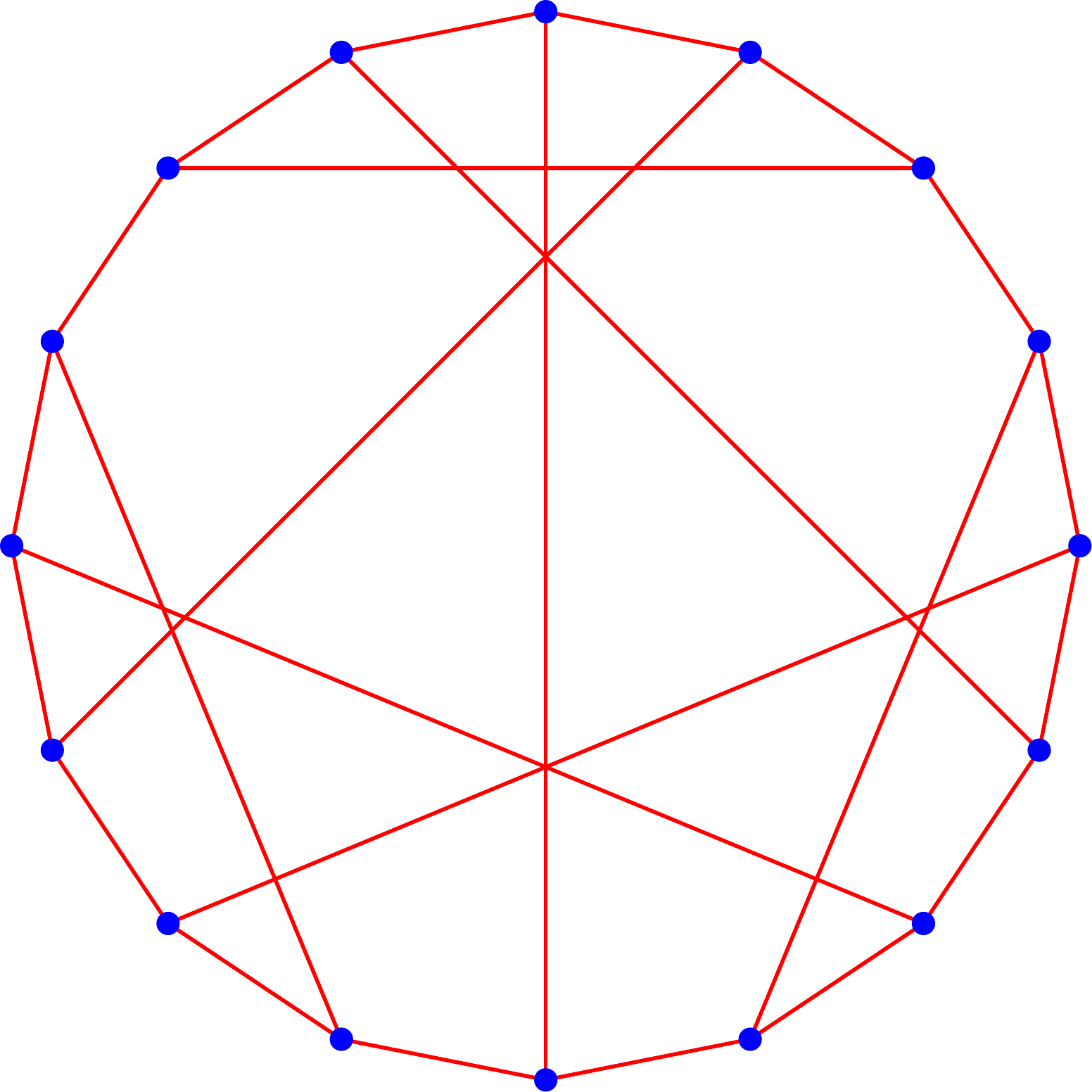}                 &                      & \multicolumn{2}{c|}{\cellcolor[HTML]{d0ffd5}\def\svgwidth{1.5cm}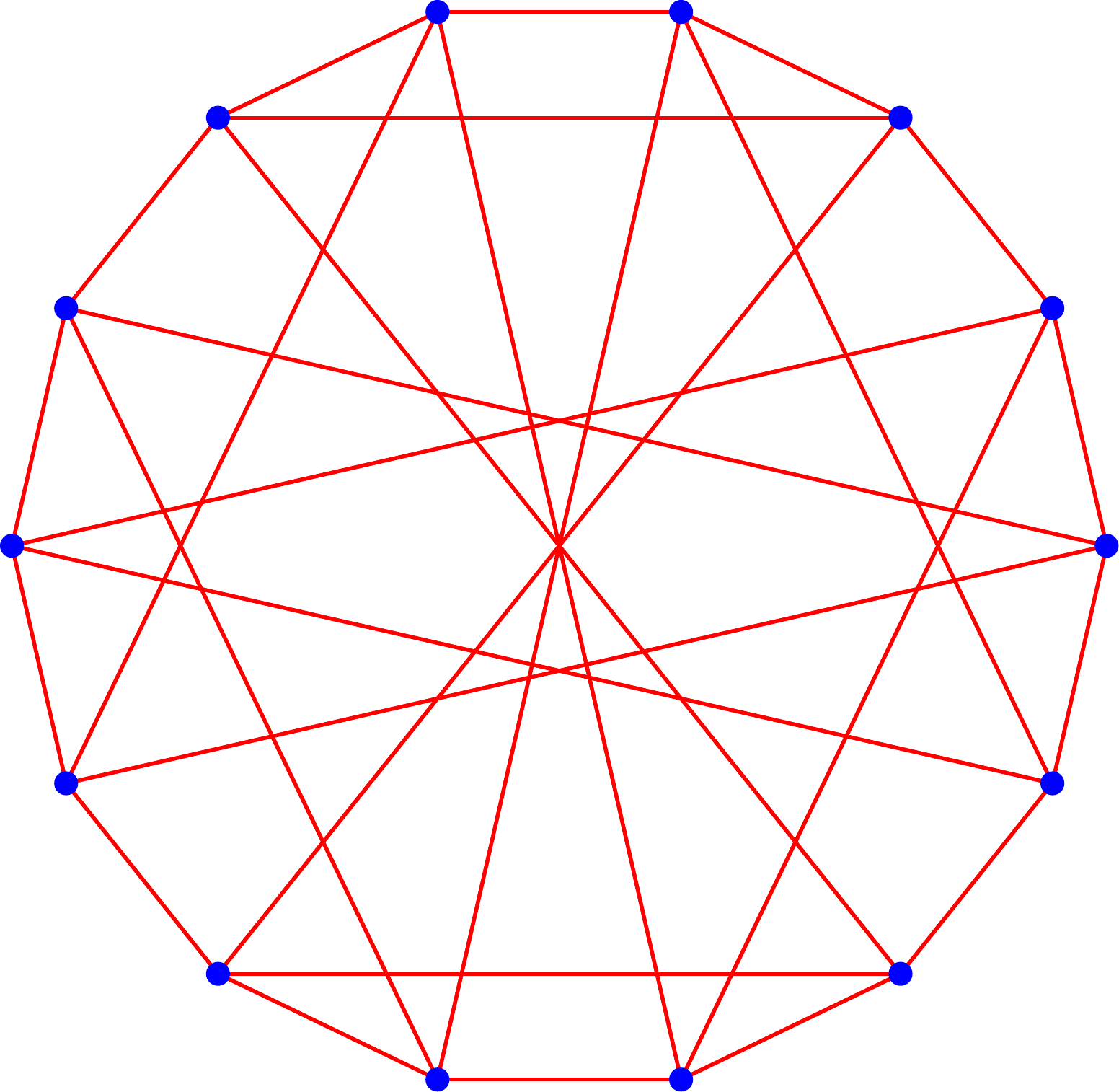}    \\*
                     & \multicolumn{1}{l}{\cellcolor[HTML]{ffffc7}6}      & \cellcolor[HTML]{ffffc7}6                        &                      & \multicolumn{1}{l}{\cellcolor[HTML]{d0ffd5}10}          & \cellcolor[HTML]{d0ffd5}96     \\*
\multirow{-4}{*}[0.5cm]{16} & \multicolumn{2}{c||}{\cellcolor[HTML]{ffffc7}$S_{3}$}                                                  & \multirow{-4}{*}[0.5cm]{14} & \multicolumn{2}{c|}{\cellcolor[HTML]{d0ffd5}${C_{2}}^{2}\times S_4$}           \\ \hline
                     & \multicolumn{1}{l}{\cellcolor[HTML]{ffffc7}2.2941} & \cellcolor[HTML]{ffffc7}1                        &                      & \multicolumn{1}{l}{\cellcolor[HTML]{d0ffd5}1.7143}      & \cellcolor[HTML]{d0ffd5}1      \\*
                     & \multicolumn{2}{c||}{\cellcolor[HTML]{ffffc7}\def\svgwidth{1.5cm}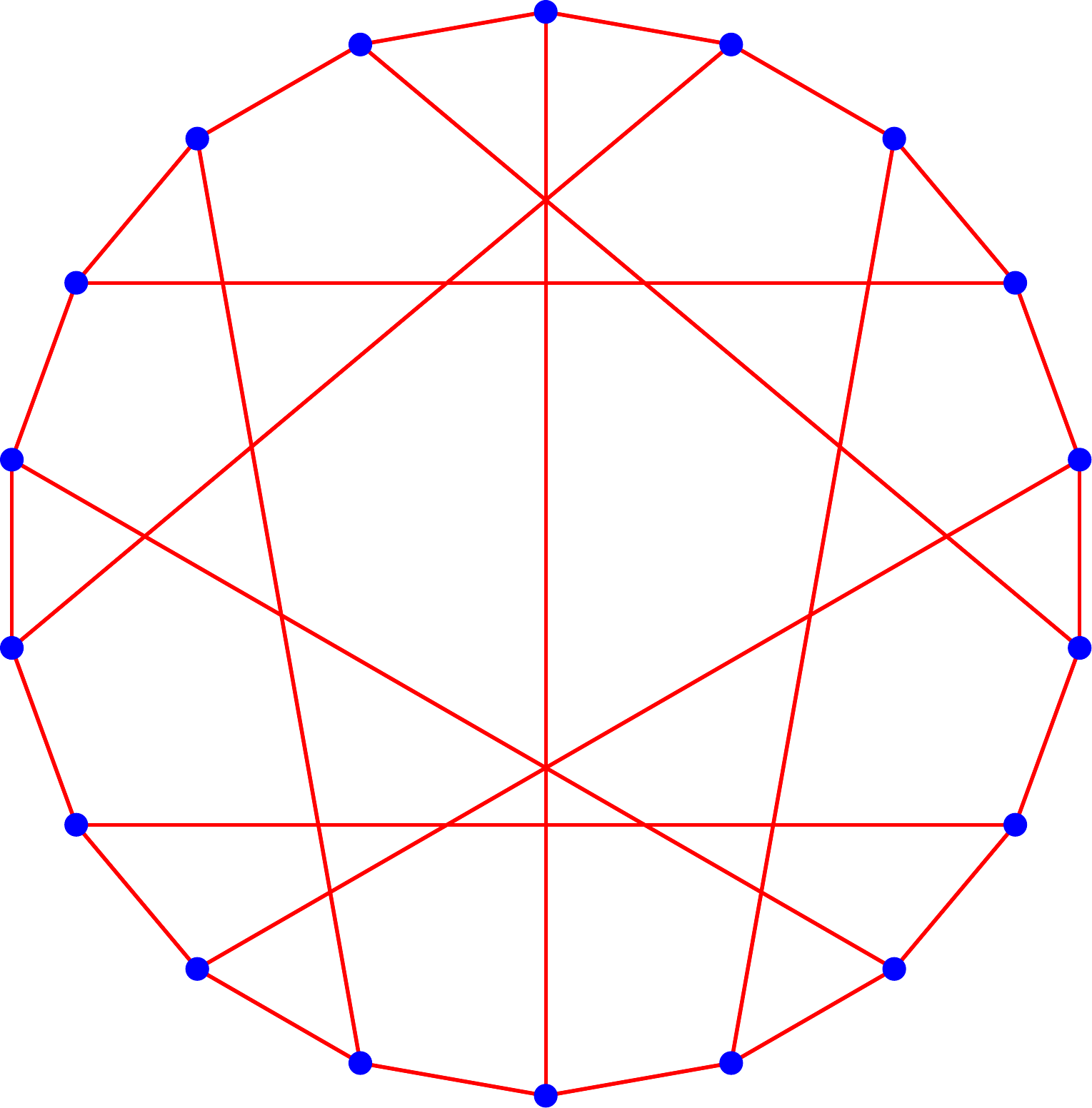}                 &                      & \multicolumn{2}{c|}{\cellcolor[HTML]{d0ffd5}\def\svgwidth{1.5cm}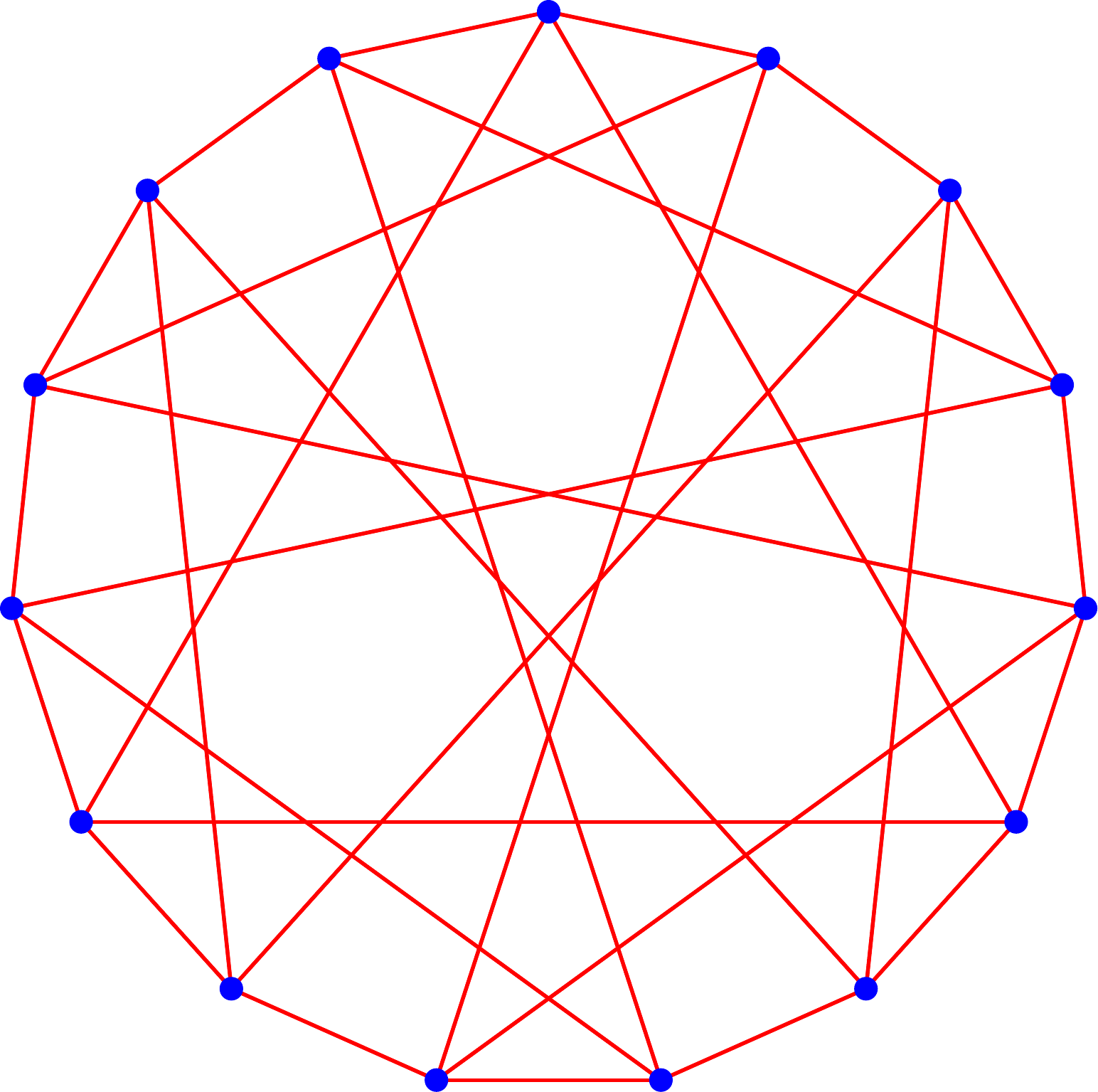}    \\*
                     & \multicolumn{1}{l}{\cellcolor[HTML]{ffffc7}7}      & \cellcolor[HTML]{ffffc7}8                        &                      & \multicolumn{1}{l}{\cellcolor[HTML]{d0ffd5}10}          & \cellcolor[HTML]{d0ffd5}12     \\*
\multirow{-4}{*}[0.5cm]{18} & \multicolumn{2}{c||}{\cellcolor[HTML]{ffffc7}$D_{4}$}                                                  & \multirow{-4}{*}[0.5cm]{15} & \multicolumn{2}{c|}{\cellcolor[HTML]{d0ffd5}$D_{6}$}                                     \\ \hline
                     & \multicolumn{1}{l}{\cellcolor[HTML]{ffffc7}2.3684} & \cellcolor[HTML]{ffffc7}1                        &                      & \multicolumn{1}{l}{\cellcolor[HTML]{ffffc7}1.7500*}      & \cellcolor[HTML]{ffffc7}1      \\*
                     & \multicolumn{2}{c||}{\cellcolor[HTML]{ffffc7}\def\svgwidth{1.5cm}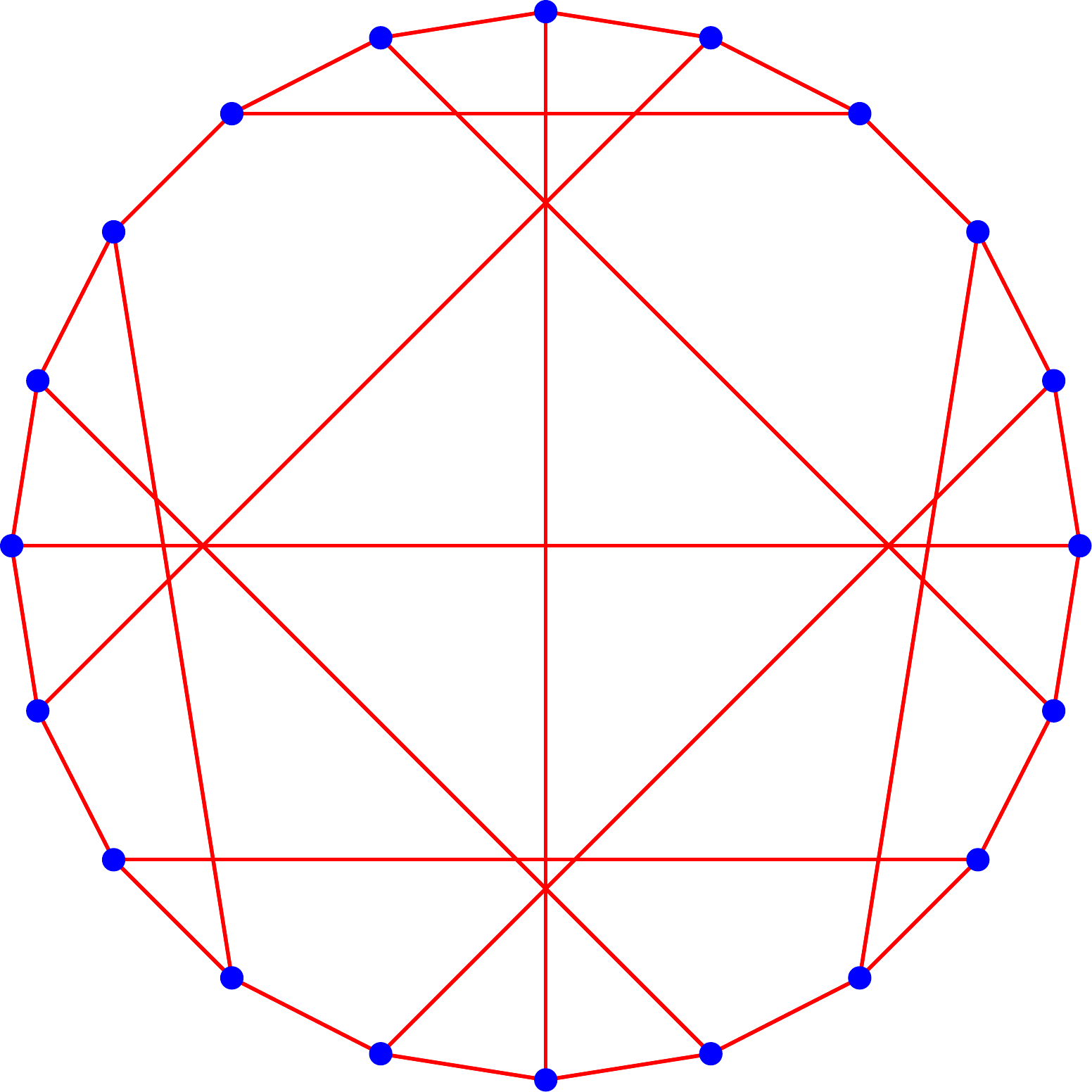}                 &                      & \multicolumn{2}{c|}{\cellcolor[HTML]{ffffc7}\def\svgwidth{1.5cm}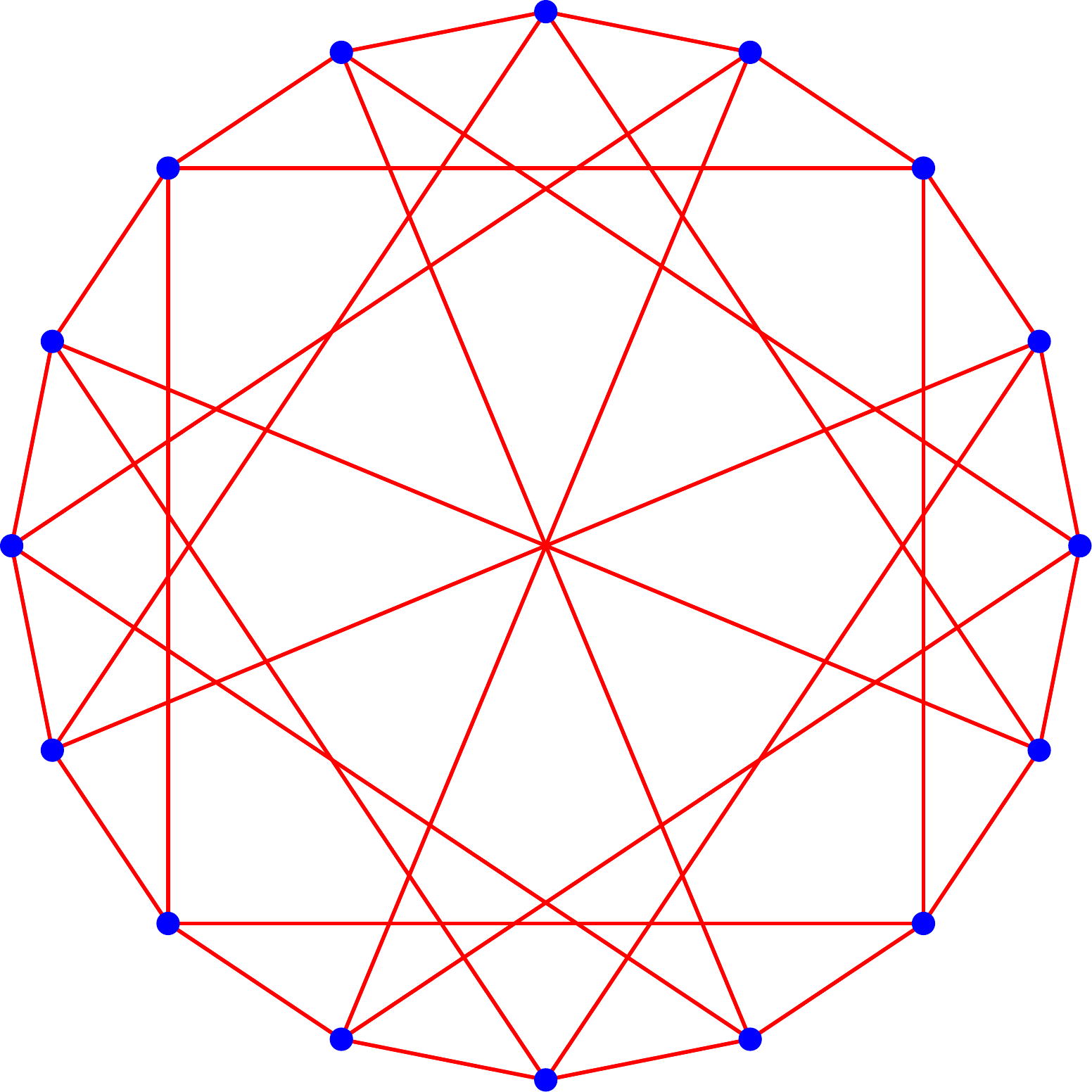}    \\*
                     & \multicolumn{1}{l}{\cellcolor[HTML]{ffffc7}8}      & \cellcolor[HTML]{ffffc7}20                       &                      & \multicolumn{1}{l}{\cellcolor[HTML]{ffffc7}12}          & \cellcolor[HTML]{ffffc7}32     \\*
\multirow{-4}{*}[0.5cm]{20} & \multicolumn{2}{c||}{\cellcolor[HTML]{ffffc7}$D_{10}$}                                                 & \multirow{-4}{*}[0.5cm]{16} & \multicolumn{2}{c|}{\cellcolor[HTML]{ffffc7}${C_{2}}^{2} \wr C_{2}$}                 \\ \hline
                     & \multicolumn{1}{l}{\cellcolor[HTML]{ffdd9b}2.4805*} & \cellcolor[HTML]{ffdd9b}1                        &                      & \multicolumn{1}{l}{\cellcolor[HTML]{ffffc7}1.8162*}      & \cellcolor[HTML]{ffffc7}1      \\*
                     & \multicolumn{2}{c||}{\cellcolor[HTML]{ffdd9b}\def\svgwidth{1.5cm}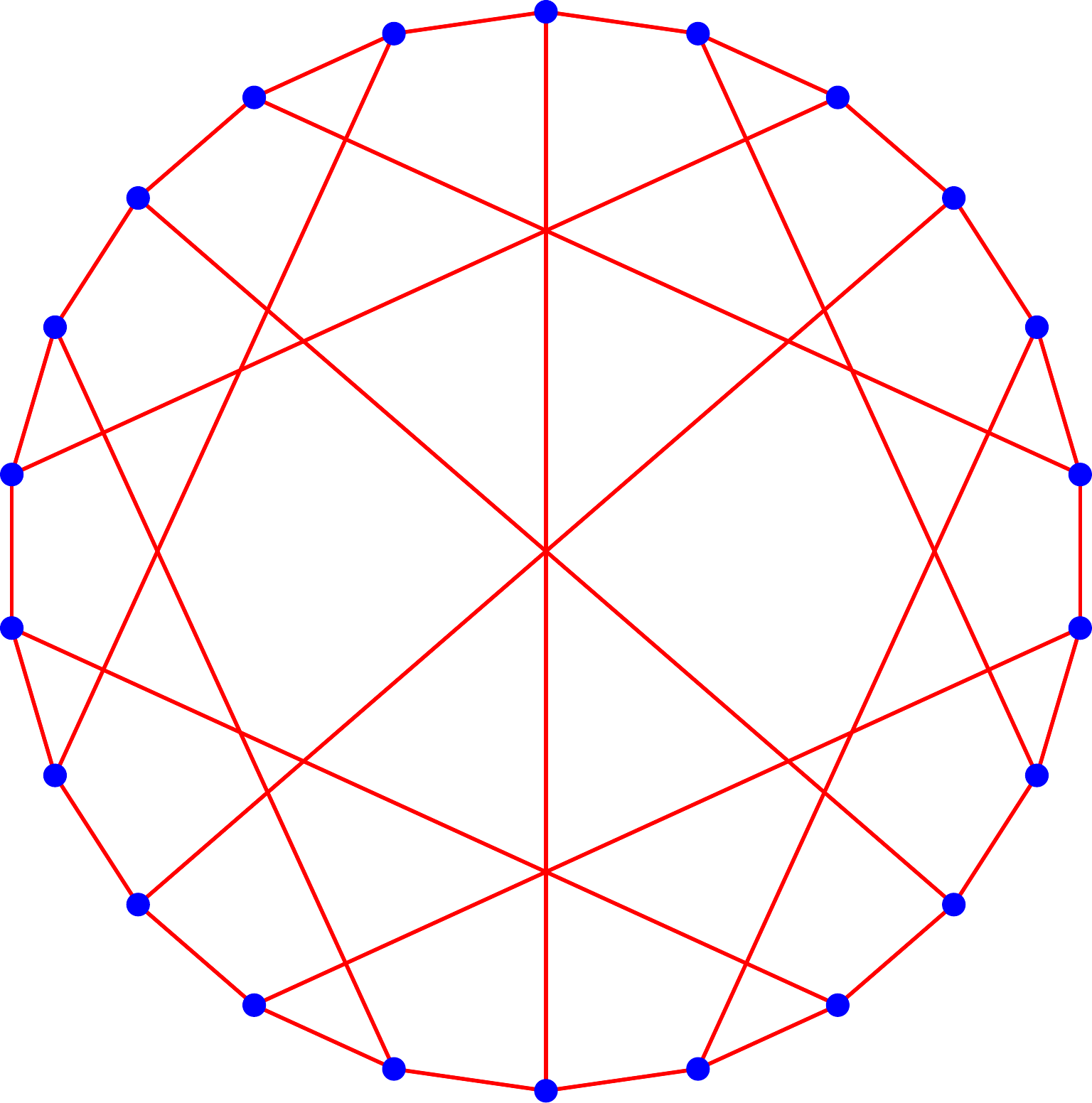}                 &                      & \multicolumn{2}{c|}{\cellcolor[HTML]{ffffc7}\def\svgwidth{1.5cm}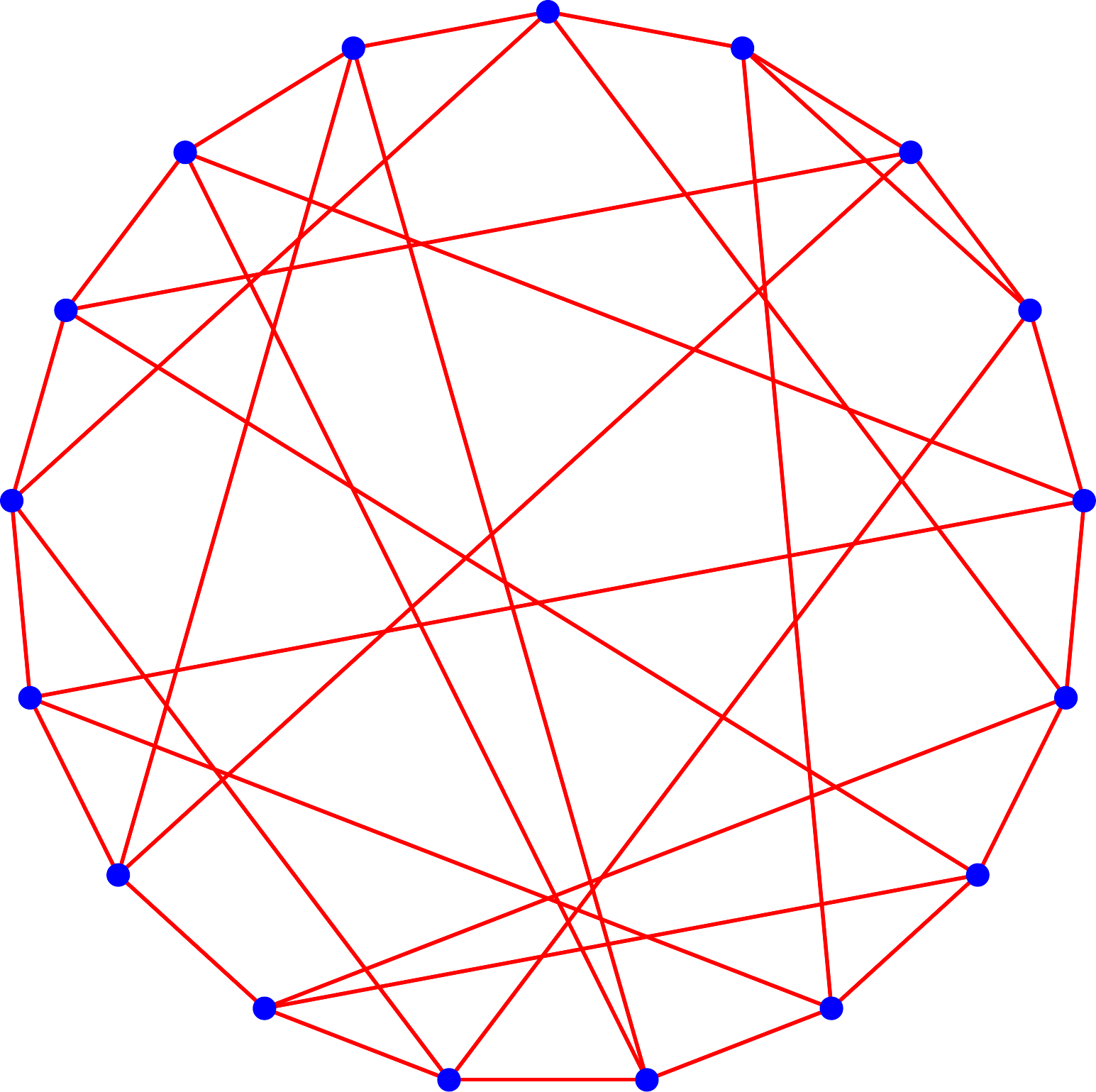}    \\*
                     & \multicolumn{1}{l}{\cellcolor[HTML]{ffdd9b}9}      & \cellcolor[HTML]{ffdd9b}16                       &                      & \multicolumn{1}{l}{\cellcolor[HTML]{ffffc7}12}          & \cellcolor[HTML]{ffffc7}36     \\*
\multirow{-4}{*}[0.5cm]{22} & \multicolumn{2}{c||}{\cellcolor[HTML]{ffdd9b}$C_{2} \times D_{4}$}                                     & \multirow{-4}{*}[0.5cm]{17} & \multicolumn{2}{c|}{\cellcolor[HTML]{ffffc7}${S_{3}}^{2}$}                               \\ \hline
                     & \multicolumn{1}{l}{\cellcolor[HTML]{ffdd9b}2.5652} & \cellcolor[HTML]{ffdd9b}1                        &                      & \multicolumn{1}{l}{\cellcolor[HTML]{ffffc7}1.8627*}      & \cellcolor[HTML]{ffffc7}1      \\*
                     & \multicolumn{2}{c||}{\cellcolor[HTML]{ffdd9b}\def\svgwidth{1.5cm}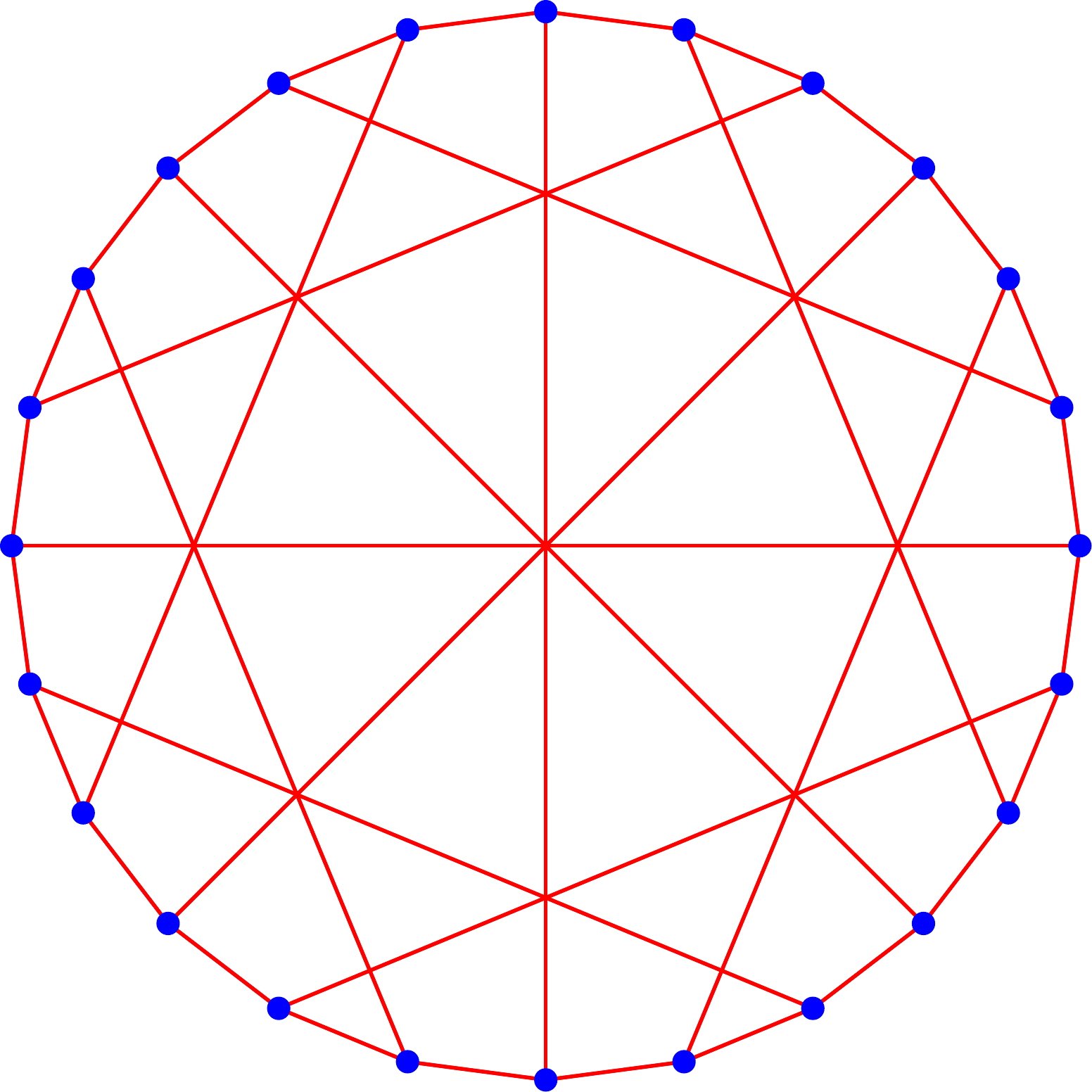}                 &                      & \multicolumn{2}{c|}{\cellcolor[HTML]{ffffc7}\def\svgwidth{1.5cm}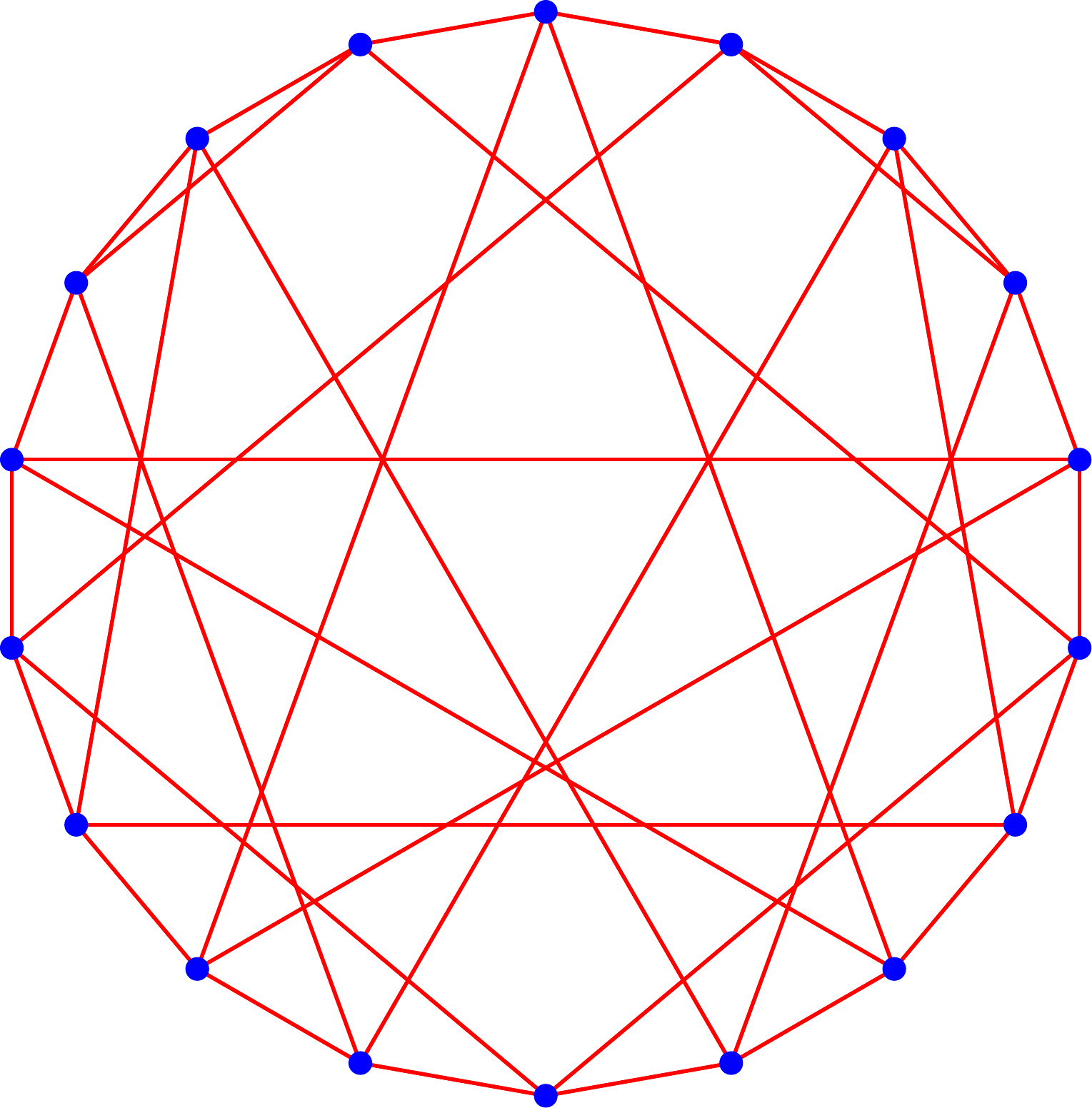}    \\*
                     & \multicolumn{1}{l}{\cellcolor[HTML]{ffdd9b}8}      & \cellcolor[HTML]{ffdd9b}32                       &                      & \multicolumn{1}{l}{\cellcolor[HTML]{ffffc7}12}          & \cellcolor[HTML]{ffffc7}12     \\*
\multirow{-5}{*}[0.5cm]{24} & \multicolumn{2}{c||}{\cellcolor[HTML]{ffdd9b}$\mathrm{Hol}(C_8)$}                                & \multirow{-4}{*}[0.5cm]{18} & \multicolumn{2}{c|}{\cellcolor[HTML]{ffffc7}$D_{6}$}                                 \\ \hline
                     & \multicolumn{1}{l}{\cellcolor[HTML]{ffdd9b}2.6800} & \cellcolor[HTML]{ffdd9b}1                        &                      & \multicolumn{1}{l}{\cellcolor[HTML]{ffffc7}1.8889}      & \cellcolor[HTML]{ffffc7}1      \\*
                     & \multicolumn{2}{c||}{\cellcolor[HTML]{ffdd9b}\def\svgwidth{1.5cm}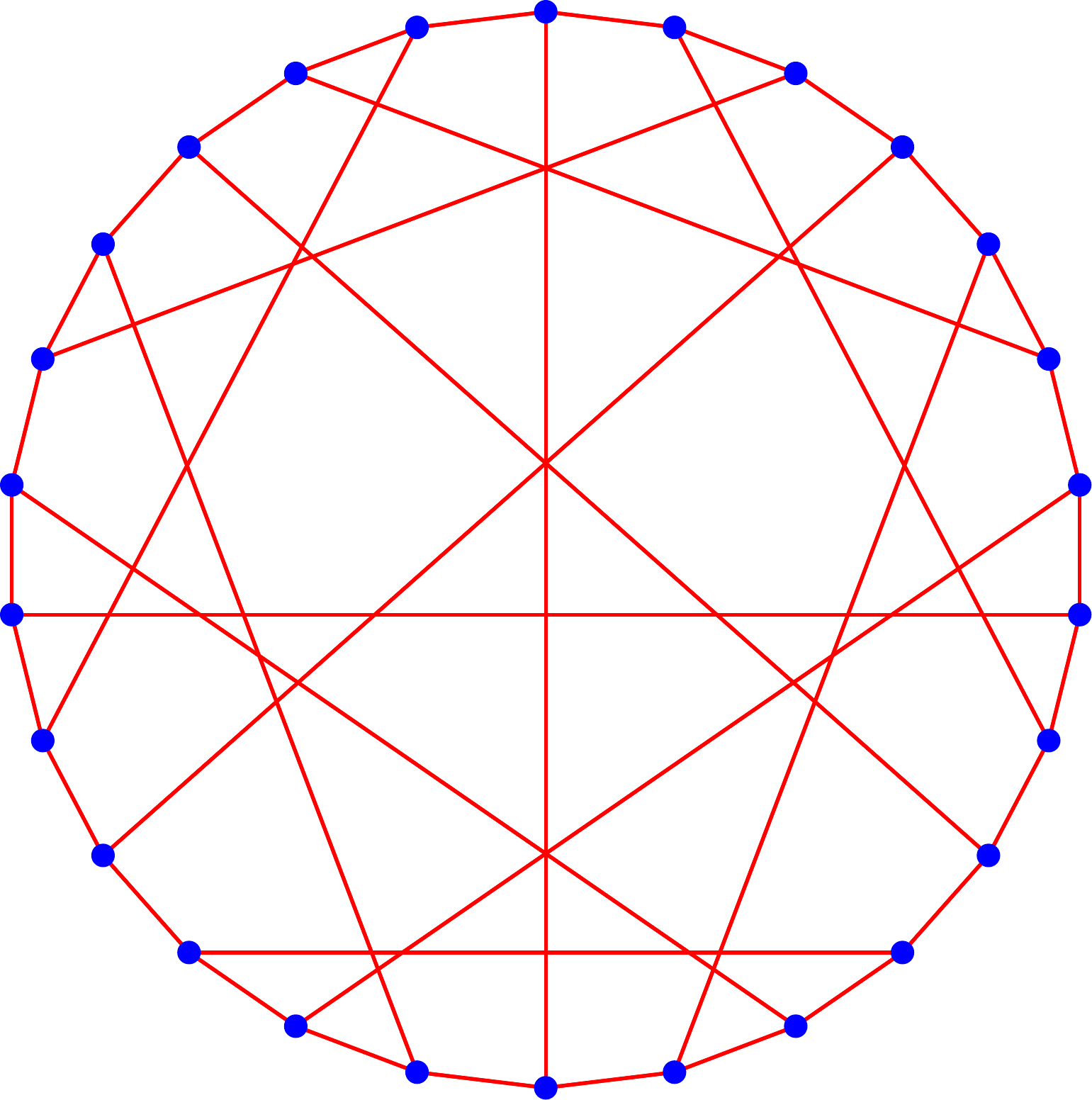}                 &                      & \multicolumn{2}{c|}{\cellcolor[HTML]{ffffc7}\def\svgwidth{1.5cm}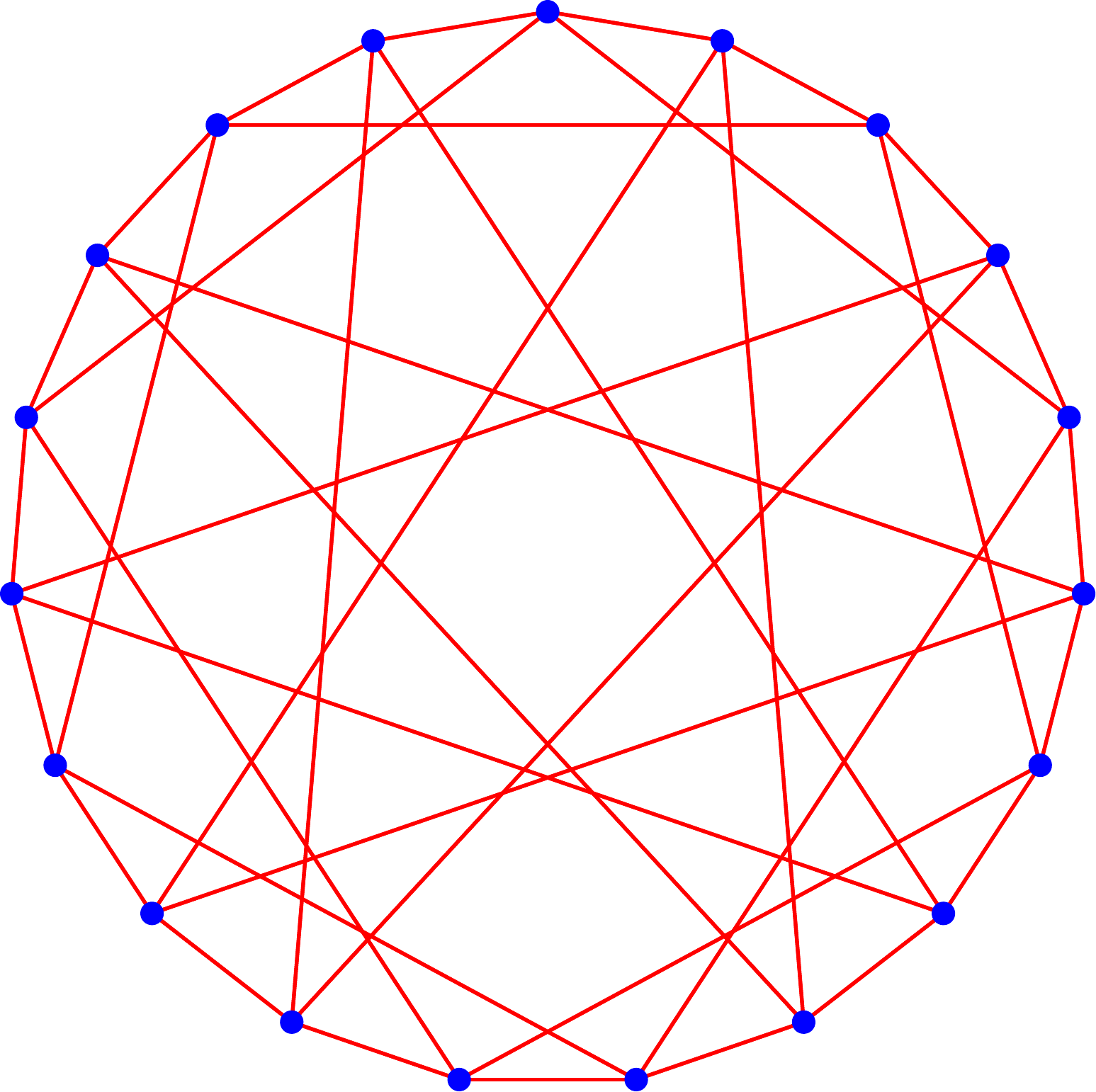}    \\*
                     & \multicolumn{1}{l}{\cellcolor[HTML]{ffdd9b}9}      & \cellcolor[HTML]{ffdd9b}52                       &                      & \multicolumn{1}{l}{\cellcolor[HTML]{ffffc7}12}          & \cellcolor[HTML]{ffffc7}24     \\*
\multirow{-4}{*}[0.5cm]{26} & \multicolumn{2}{c||}{\cellcolor[HTML]{ffdd9b}$C_{13} \rtimes C_{4}$}                                   & \multirow{-4}{*}[0.5cm]{19} & \multicolumn{2}{c|}{\cellcolor[HTML]{ffffc7}$D_{12}$}                                    \\ \hline
                     & \multicolumn{1}{l}{\cellcolor[HTML]{ffdd9b}2.7778} & \cellcolor[HTML]{ffdd9b}1                        &                      & \multicolumn{1}{l}{\cellcolor[HTML]{ffffc7}1.9474}      & \cellcolor[HTML]{ffffc7}1      \\*
                     & \multicolumn{2}{c||}{\cellcolor[HTML]{ffdd9b}\def\svgwidth{1.5cm}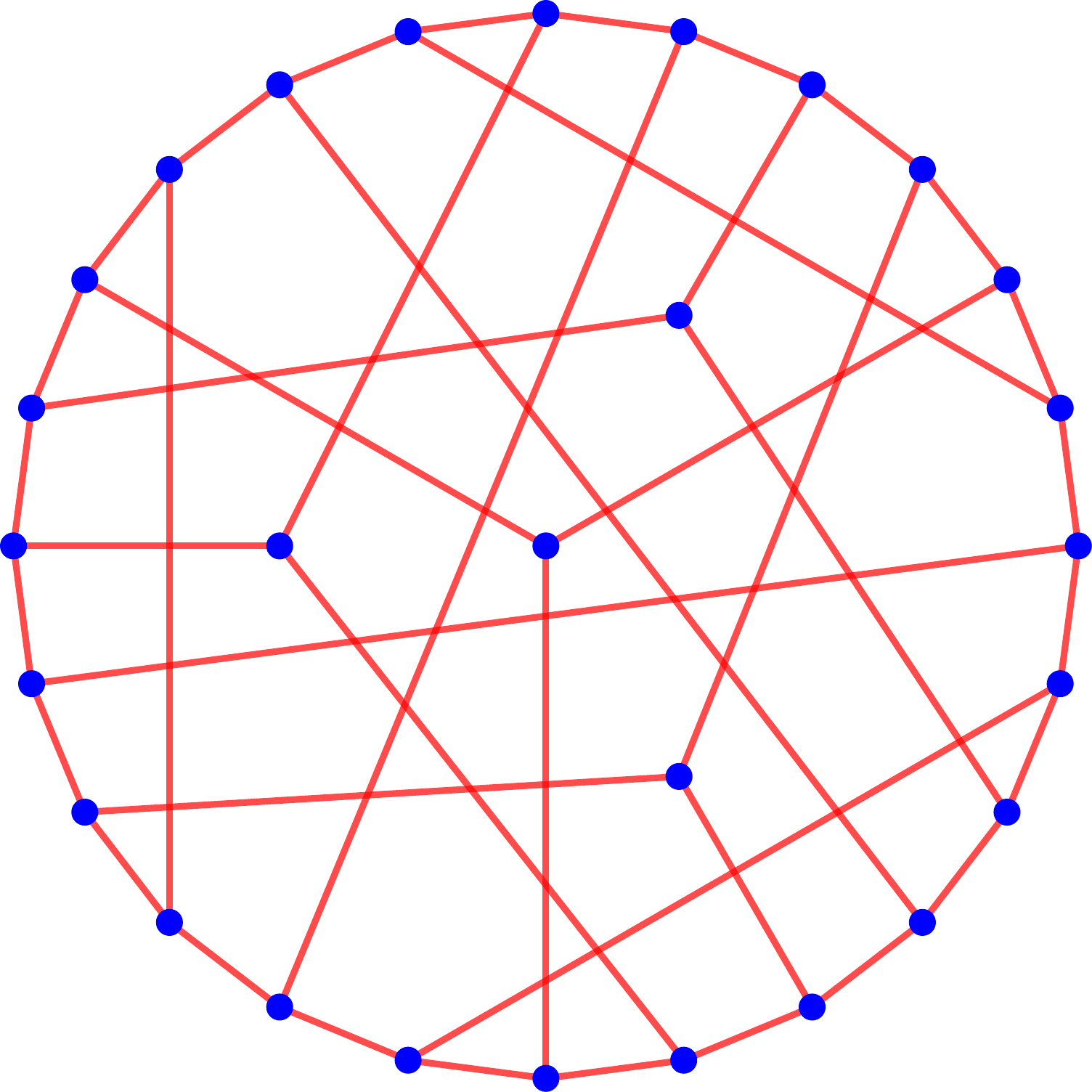}                 &                      & \multicolumn{2}{c|}{\cellcolor[HTML]{ffffc7}\def\svgwidth{1.5cm}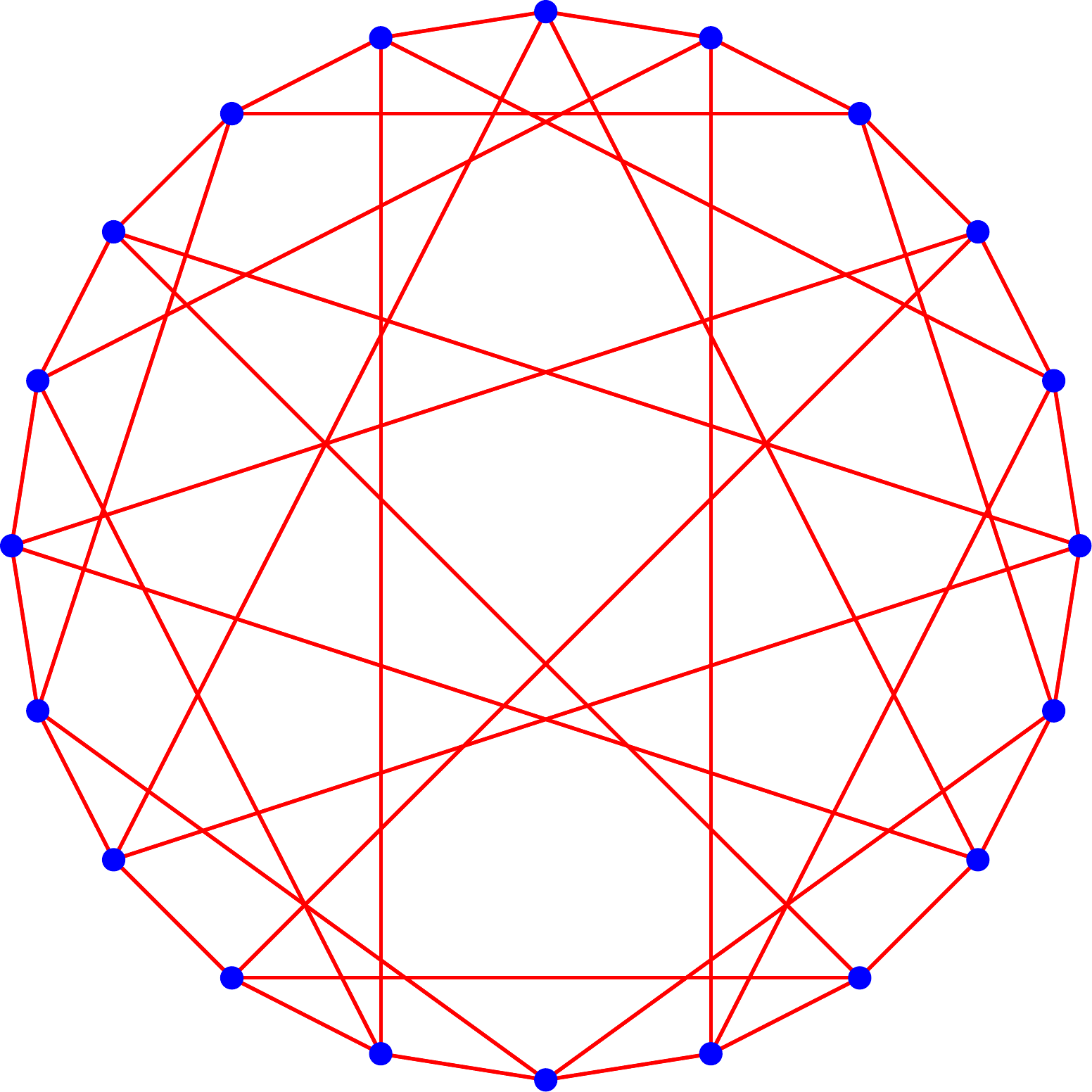}    \\*
                     & \multicolumn{1}{l}{\cellcolor[HTML]{ffdd9b}10}     & \multicolumn{1}{r||}{\cellcolor[HTML]{ffdd9b}336} &                      & \multicolumn{1}{l}{\cellcolor[HTML]{ffffc7}14}          & \cellcolor[HTML]{ffffc7}96     \\*
\multirow{-4}{*}[0.5cm]{28} & \multicolumn{2}{c||}{\cellcolor[HTML]{ffdd9b}$\mathrm{PGL}(2,7)$}                                      & \multirow{-4}{*}[0.5cm]{20} & \multicolumn{2}{c|}{\cellcolor[HTML]{ffffc7}$\mathrm{GL}(2,3) \rtimes C_{2}$}            \\ \hline
                     & \multicolumn{1}{l}{\cellcolor[HTML]{ffdd9b}2.8621} & \cellcolor[HTML]{ffdd9b}1                        &                      & \multicolumn{1}{l}{\cellcolor[HTML]{ffffc7}2.0000}      & \cellcolor[HTML]{ffffc7}2      \\*
                     & \multicolumn{2}{c||}{\cellcolor[HTML]{ffdd9b}\def\svgwidth{1.5cm}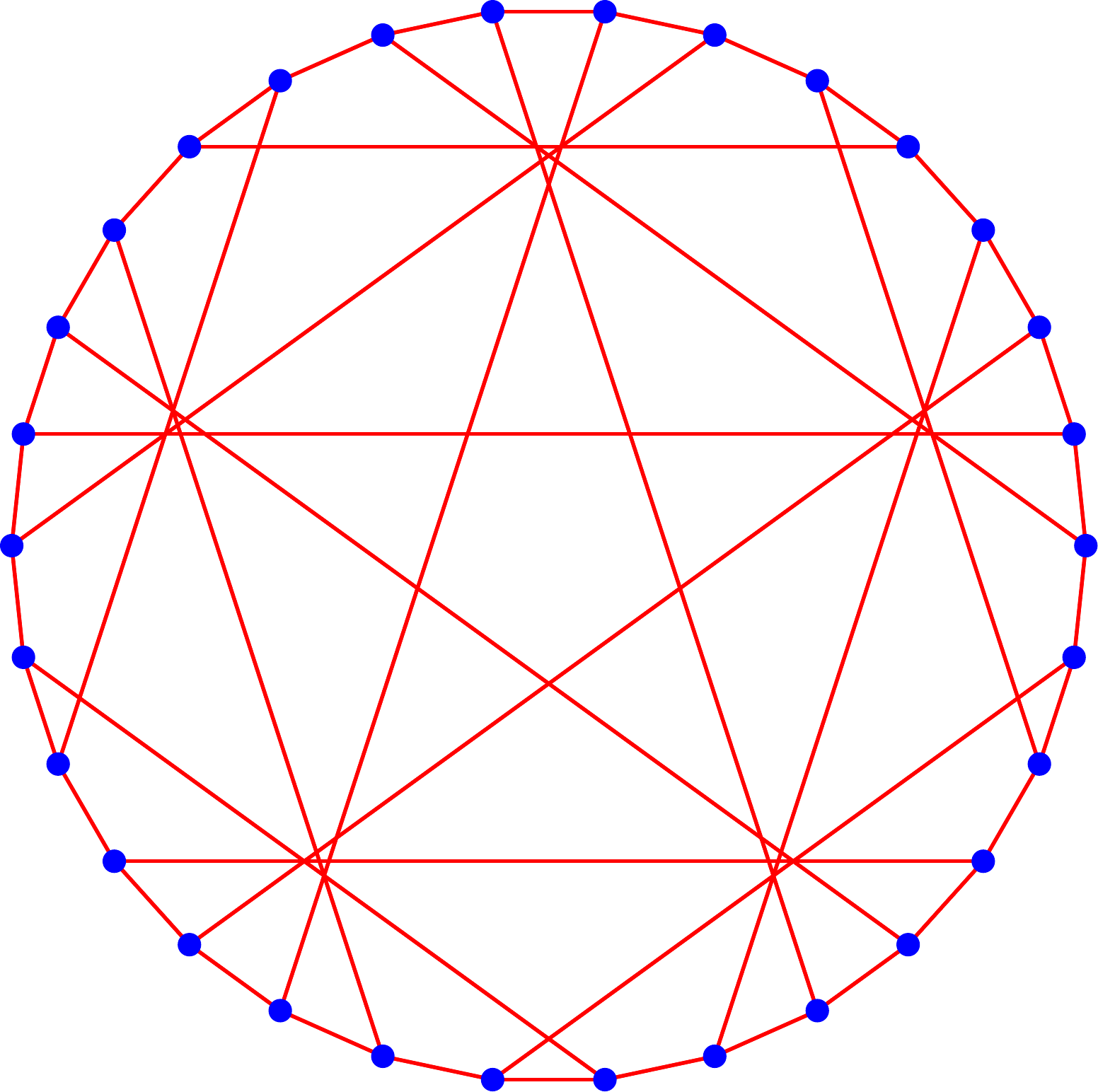}                &                      & \multicolumn{2}{c|}{\cellcolor[HTML]{ffffc7}\def\svgwidth{1.5cm}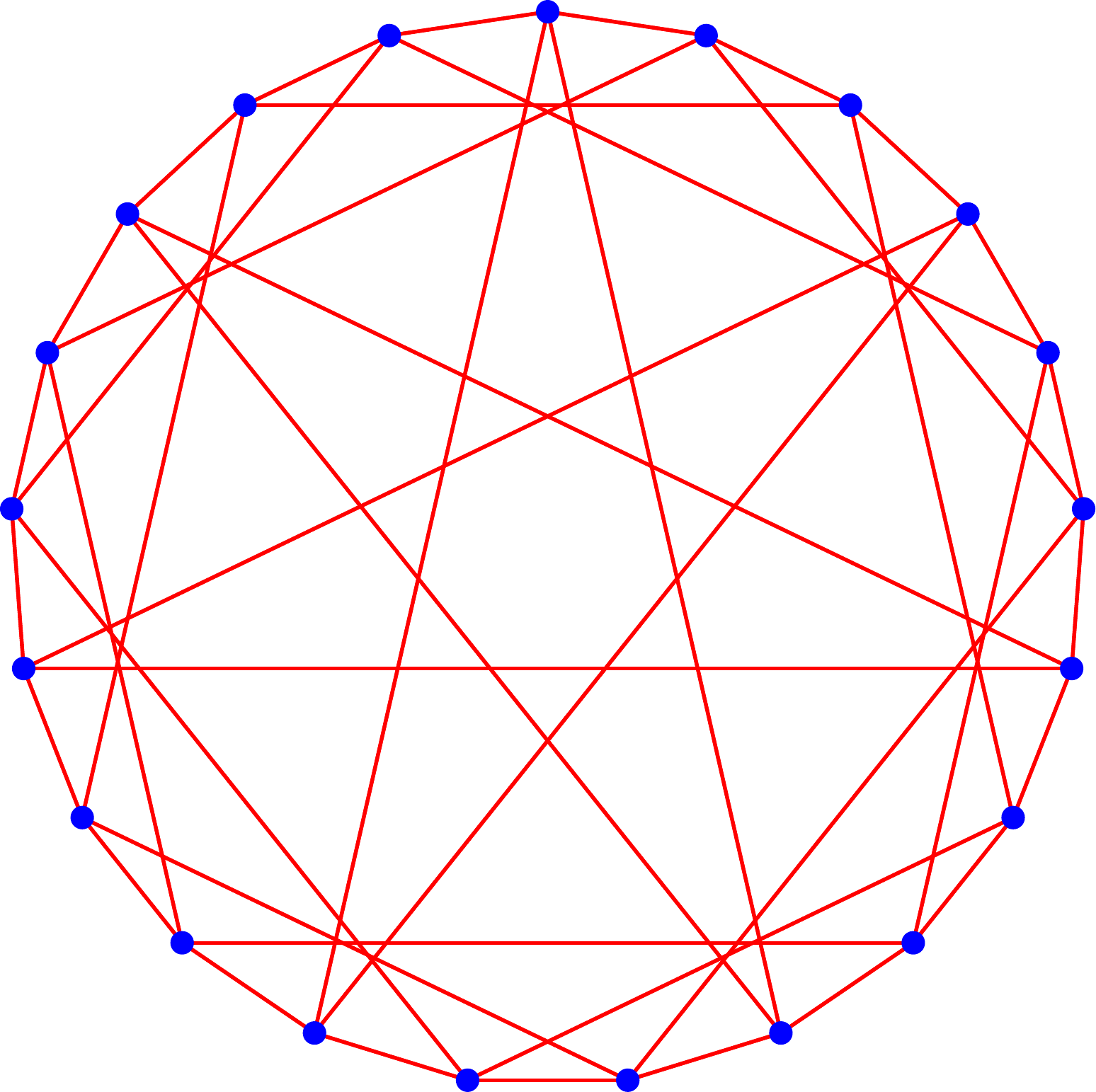}    \\*
                     & \multicolumn{1}{l}{\cellcolor[HTML]{ffdd9b}9}      & \cellcolor[HTML]{ffdd9b}1440                     &                      & \multicolumn{1}{l}{\cellcolor[HTML]{ffffc7}14}          & \cellcolor[HTML]{ffffc7}14     \\*
\multirow{-4}{*}[0.5cm]{30} & \multicolumn{2}{c||}{\cellcolor[HTML]{ffdd9b}$\mathrm{Aut}(S_6)$}                                    & \multirow{-4}{*}[0.5cm]{21} & \multicolumn{2}{c|}{\cellcolor[HTML]{ffffc7}$D_{7}$}                                     \\ \hline
                     & \multicolumn{1}{l}{\cellcolor[HTML]{ffdd9b}2.9355} & \cellcolor[HTML]{ffdd9b}1                        &                      & \multicolumn{1}{l}{\cellcolor[HTML]{ffffc7}2.3548}      & \cellcolor[HTML]{ffffc7}3      \\*
                     & \multicolumn{2}{c||}{\cellcolor[HTML]{ffdd9b}\def\svgwidth{1.5cm}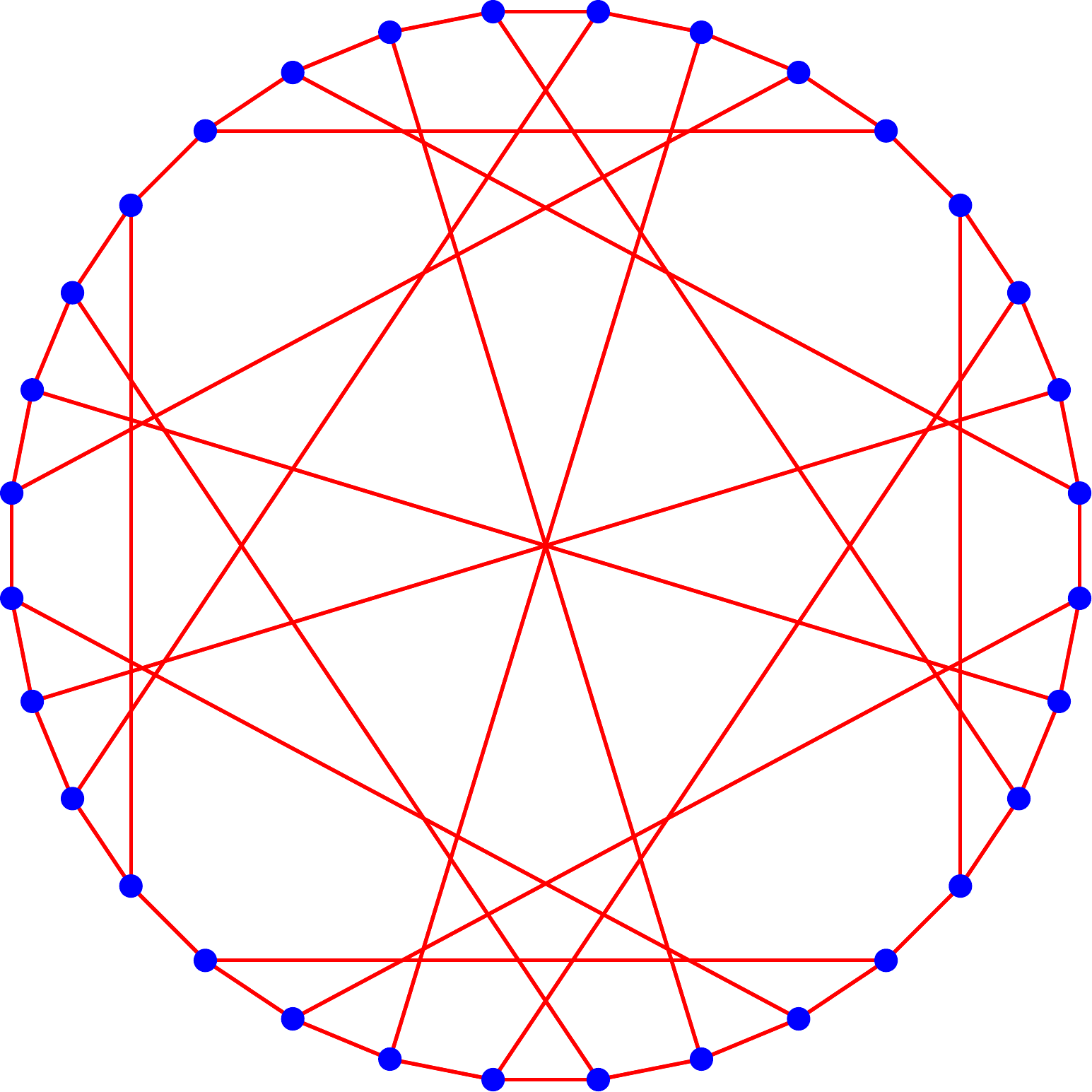}                &                      & \multicolumn{2}{c|}{\cellcolor[HTML]{ffffc7}\def\svgwidth{1.5cm}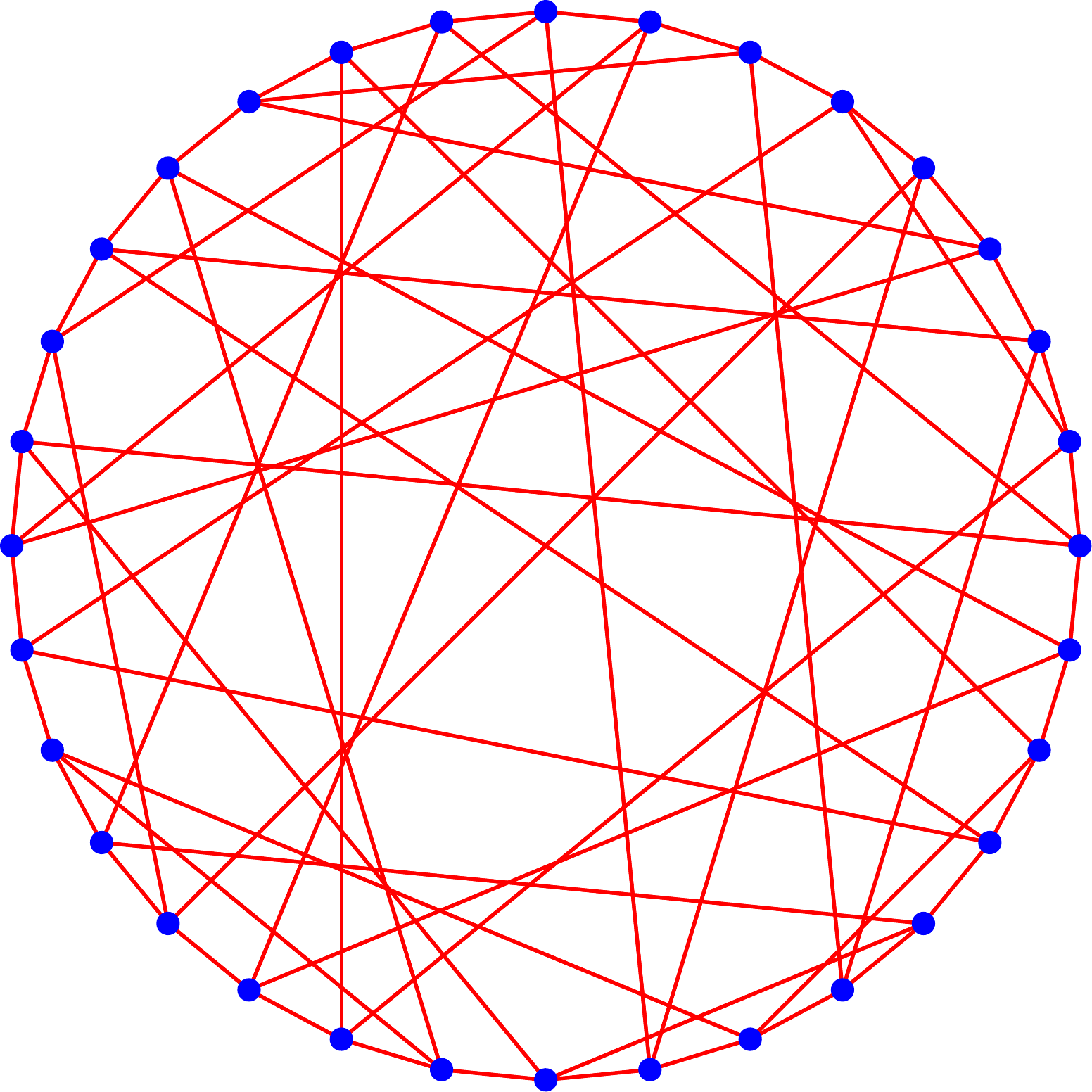} \\*
                     & \multicolumn{1}{l}{\cellcolor[HTML]{ffdd9b}10}     & \cellcolor[HTML]{ffdd9b}12                       &                      & \multicolumn{1}{l}{\cellcolor[HTML]{ffffc7}18}          & \cellcolor[HTML]{ffffc7}3      \\*
\multirow{-4}{*}[0.5cm]{32} & \multicolumn{2}{c||}{\cellcolor[HTML]{ffdd9b}$A_{4}$}                                                  & \multirow{-4}{*}[0.5cm]{32} & \multicolumn{2}{c|}{\cellcolor[HTML]{ffffc7}$C_{3}$}                                     \\ \hline
\end{longtable}

%% file: 14k3_1.pdf_tex
\begingroup%
  \makeatletter%
  \providecommand\color[2][]{%
    \errmessage{(Inkscape) Color is used for the text in Inkscape, but the package 'color.sty' is not loaded}%
    \renewcommand\color[2][]{}%
  }%
  \providecommand\transparent[1]{%
    \errmessage{(Inkscape) Transparency is used (non-zero) for the text in Inkscape, but the package 'transparent.sty' is not loaded}%
    \renewcommand\transparent[1]{}%
  }%
  \providecommand\rotatebox[2]{#2}%
  \newcommand*\fsize{\dimexpr\f@size pt\relax}%
  \newcommand*\lineheight[1]{\fontsize{\fsize}{#1\fsize}\selectfont}%
  \ifx\svgwidth\undefined%
    \setlength{\unitlength}{446.80005026bp}%
    \ifx\svgscale\undefined%
      \relax%
    \else%
      \setlength{\unitlength}{\unitlength * \real{\svgscale}}%
    \fi%
  \else%
    \setlength{\unitlength}{\svgwidth}%
  \fi%
  \global\let\svgwidth\undefined%
  \global\let\svgscale\undefined%
  \makeatother%
  \begin{picture}(1,0.97546386)%
    \lineheight{1}%
    \setlength\tabcolsep{0pt}%
    \put(0,0){\includegraphics[width=\unitlength,page=1]{14k3_1.pdf}}%
  \end{picture}%
\endgroup%

%% file: 13k4_10.pdf_tex
\begingroup%
  \makeatletter%
  \providecommand\color[2][]{%
    \errmessage{(Inkscape) Color is used for the text in Inkscape, but the package 'color.sty' is not loaded}%
    \renewcommand\color[2][]{}%
  }%
  \providecommand\transparent[1]{%
    \errmessage{(Inkscape) Transparency is used (non-zero) for the text in Inkscape, but the package 'transparent.sty' is not loaded}%
    \renewcommand\transparent[1]{}%
  }%
  \providecommand\rotatebox[2]{#2}%
  \newcommand*\fsize{\dimexpr\f@size pt\relax}%
  \newcommand*\lineheight[1]{\fontsize{\fsize}{#1\fsize}\selectfont}%
  \ifx\svgwidth\undefined%
    \setlength{\unitlength}{443.61195026bp}%
    \ifx\svgscale\undefined%
      \relax%
    \else%
      \setlength{\unitlength}{\unitlength * \real{\svgscale}}%
    \fi%
  \else%
    \setlength{\unitlength}{\svgwidth}%
  \fi%
  \global\let\svgwidth\undefined%
  \global\let\svgscale\undefined%
  \makeatother%
  \begin{picture}(1,0.99286589)%
    \lineheight{1}%
    \setlength\tabcolsep{0pt}%
    \put(0,0){\includegraphics[width=\unitlength,page=1]{13k4_10.pdf}}%
  \end{picture}%
\endgroup%

%% file: 16k3_2.pdf_tex
\begingroup%
  \makeatletter%
  \providecommand\color[2][]{%
    \errmessage{(Inkscape) Color is used for the text in Inkscape, but the package 'color.sty' is not loaded}%
    \renewcommand\color[2][]{}%
  }%
  \providecommand\transparent[1]{%
    \errmessage{(Inkscape) Transparency is used (non-zero) for the text in Inkscape, but the package 'transparent.sty' is not loaded}%
    \renewcommand\transparent[1]{}%
  }%
  \providecommand\rotatebox[2]{#2}%
  \newcommand*\fsize{\dimexpr\f@size pt\relax}%
  \newcommand*\lineheight[1]{\fontsize{\fsize}{#1\fsize}\selectfont}%
  \ifx\svgwidth\undefined%
    \setlength{\unitlength}{446.80005026bp}%
    \ifx\svgscale\undefined%
      \relax%
    \else%
      \setlength{\unitlength}{\unitlength * \real{\svgscale}}%
    \fi%
  \else%
    \setlength{\unitlength}{\svgwidth}%
  \fi%
  \global\let\svgwidth\undefined%
  \global\let\svgscale\undefined%
  \makeatother%
  \begin{picture}(1,1)%
    \lineheight{1}%
    \setlength\tabcolsep{0pt}%
    \put(0,0){\includegraphics[width=\unitlength,page=1]{16k3_2.pdf}}%
  \end{picture}%
\endgroup%

%% file: 14k4_1.pdf_tex
\begingroup%
  \makeatletter%
  \providecommand\color[2][]{%
    \errmessage{(Inkscape) Color is used for the text in Inkscape, but the package 'color.sty' is not loaded}%
    \renewcommand\color[2][]{}%
  }%
  \providecommand\transparent[1]{%
    \errmessage{(Inkscape) Transparency is used (non-zero) for the text in Inkscape, but the package 'transparent.sty' is not loaded}%
    \renewcommand\transparent[1]{}%
  }%
  \providecommand\rotatebox[2]{#2}%
  \newcommand*\fsize{\dimexpr\f@size pt\relax}%
  \newcommand*\lineheight[1]{\fontsize{\fsize}{#1\fsize}\selectfont}%
  \ifx\svgwidth\undefined%
    \setlength{\unitlength}{446.80005026bp}%
    \ifx\svgscale\undefined%
      \relax%
    \else%
      \setlength{\unitlength}{\unitlength * \real{\svgscale}}%
    \fi%
  \else%
    \setlength{\unitlength}{\svgwidth}%
  \fi%
  \global\let\svgwidth\undefined%
  \global\let\svgscale\undefined%
  \makeatother%
  \begin{picture}(1,0.97546386)%
    \lineheight{1}%
    \setlength\tabcolsep{0pt}%
    \put(0,0){\includegraphics[width=\unitlength,page=1]{14k4_1.pdf}}%
  \end{picture}%
\endgroup%

%% file: 18k3_1.pdf_tex
\begingroup%
  \makeatletter%
  \providecommand\color[2][]{%
    \errmessage{(Inkscape) Color is used for the text in Inkscape, but the package 'color.sty' is not loaded}%
    \renewcommand\color[2][]{}%
  }%
  \providecommand\transparent[1]{%
    \errmessage{(Inkscape) Transparency is used (non-zero) for the text in Inkscape, but the package 'transparent.sty' is not loaded}%
    \renewcommand\transparent[1]{}%
  }%
  \providecommand\rotatebox[2]{#2}%
  \newcommand*\fsize{\dimexpr\f@size pt\relax}%
  \newcommand*\lineheight[1]{\fontsize{\fsize}{#1\fsize}\selectfont}%
  \ifx\svgwidth\undefined%
    \setlength{\unitlength}{440.15730026bp}%
    \ifx\svgscale\undefined%
      \relax%
    \else%
      \setlength{\unitlength}{\unitlength * \real{\svgscale}}%
    \fi%
  \else%
    \setlength{\unitlength}{\svgwidth}%
  \fi%
  \global\let\svgwidth\undefined%
  \global\let\svgscale\undefined%
  \makeatother%
  \begin{picture}(1,1.01509176)%
    \lineheight{1}%
    \setlength\tabcolsep{0pt}%
    \put(0,0){\includegraphics[width=\unitlength,page=1]{18k3_1.pdf}}%
  \end{picture}%
\endgroup%

%% file: 15k4_1.pdf_tex
\begingroup%
  \makeatletter%
  \providecommand\color[2][]{%
    \errmessage{(Inkscape) Color is used for the text in Inkscape, but the package 'color.sty' is not loaded}%
    \renewcommand\color[2][]{}%
  }%
  \providecommand\transparent[1]{%
    \errmessage{(Inkscape) Transparency is used (non-zero) for the text in Inkscape, but the package 'transparent.sty' is not loaded}%
    \renewcommand\transparent[1]{}%
  }%
  \providecommand\rotatebox[2]{#2}%
  \newcommand*\fsize{\dimexpr\f@size pt\relax}%
  \newcommand*\lineheight[1]{\fontsize{\fsize}{#1\fsize}\selectfont}%
  \ifx\svgwidth\undefined%
    \setlength{\unitlength}{444.40470026bp}%
    \ifx\svgscale\undefined%
      \relax%
    \else%
      \setlength{\unitlength}{\unitlength * \real{\svgscale}}%
    \fi%
  \else%
    \setlength{\unitlength}{\svgwidth}%
  \fi%
  \global\let\svgwidth\undefined%
  \global\let\svgscale\undefined%
  \makeatother%
  \begin{picture}(1,0.99463968)%
    \lineheight{1}%
    \setlength\tabcolsep{0pt}%
    \put(0,0){\includegraphics[width=\unitlength,page=1]{15k4_1.pdf}}%
  \end{picture}%
\endgroup%

%% file: 20k3_1.pdf_tex
\begingroup%
  \makeatletter%
  \providecommand\color[2][]{%
    \errmessage{(Inkscape) Color is used for the text in Inkscape, but the package 'color.sty' is not loaded}%
    \renewcommand\color[2][]{}%
  }%
  \providecommand\transparent[1]{%
    \errmessage{(Inkscape) Transparency is used (non-zero) for the text in Inkscape, but the package 'transparent.sty' is not loaded}%
    \renewcommand\transparent[1]{}%
  }%
  \providecommand\rotatebox[2]{#2}%
  \newcommand*\fsize{\dimexpr\f@size pt\relax}%
  \newcommand*\lineheight[1]{\fontsize{\fsize}{#1\fsize}\selectfont}%
  \ifx\svgwidth\undefined%
    \setlength{\unitlength}{446.80005026bp}%
    \ifx\svgscale\undefined%
      \relax%
    \else%
      \setlength{\unitlength}{\unitlength * \real{\svgscale}}%
    \fi%
  \else%
    \setlength{\unitlength}{\svgwidth}%
  \fi%
  \global\let\svgwidth\undefined%
  \global\let\svgscale\undefined%
  \makeatother%
  \begin{picture}(1,1)%
    \lineheight{1}%
    \setlength\tabcolsep{0pt}%
    \put(0,0){\includegraphics[width=\unitlength,page=1]{20k3_1.pdf}}%
  \end{picture}%
\endgroup%

%% file: 16k4_1.pdf_tex
\begingroup%
  \makeatletter%
  \providecommand\color[2][]{%
    \errmessage{(Inkscape) Color is used for the text in Inkscape, but the package 'color.sty' is not loaded}%
    \renewcommand\color[2][]{}%
  }%
  \providecommand\transparent[1]{%
    \errmessage{(Inkscape) Transparency is used (non-zero) for the text in Inkscape, but the package 'transparent.sty' is not loaded}%
    \renewcommand\transparent[1]{}%
  }%
  \providecommand\rotatebox[2]{#2}%
  \newcommand*\fsize{\dimexpr\f@size pt\relax}%
  \newcommand*\lineheight[1]{\fontsize{\fsize}{#1\fsize}\selectfont}%
  \ifx\svgwidth\undefined%
    \setlength{\unitlength}{446.80005026bp}%
    \ifx\svgscale\undefined%
      \relax%
    \else%
      \setlength{\unitlength}{\unitlength * \real{\svgscale}}%
    \fi%
  \else%
    \setlength{\unitlength}{\svgwidth}%
  \fi%
  \global\let\svgwidth\undefined%
  \global\let\svgscale\undefined%
  \makeatother%
  \begin{picture}(1,1)%
    \lineheight{1}%
    \setlength\tabcolsep{0pt}%
    \put(0,0){\includegraphics[width=\unitlength,page=1]{16k4_1.pdf}}%
  \end{picture}%
\endgroup%

%% file: 22k3_1.pdf_tex
\begingroup%
  \makeatletter%
  \providecommand\color[2][]{%
    \errmessage{(Inkscape) Color is used for the text in Inkscape, but the package 'color.sty' is not loaded}%
    \renewcommand\color[2][]{}%
  }%
  \providecommand\transparent[1]{%
    \errmessage{(Inkscape) Transparency is used (non-zero) for the text in Inkscape, but the package 'transparent.sty' is not loaded}%
    \renewcommand\transparent[1]{}%
  }%
  \providecommand\rotatebox[2]{#2}%
  \newcommand*\fsize{\dimexpr\f@size pt\relax}%
  \newcommand*\lineheight[1]{\fontsize{\fsize}{#1\fsize}\selectfont}%
  \ifx\svgwidth\undefined%
    \setlength{\unitlength}{442.34947526bp}%
    \ifx\svgscale\undefined%
      \relax%
    \else%
      \setlength{\unitlength}{\unitlength * \real{\svgscale}}%
    \fi%
  \else%
    \setlength{\unitlength}{\svgwidth}%
  \fi%
  \global\let\svgwidth\undefined%
  \global\let\svgscale\undefined%
  \makeatother%
  \begin{picture}(1,1.01006122)%
    \lineheight{1}%
    \setlength\tabcolsep{0pt}%
    \put(0,0){\includegraphics[width=\unitlength,page=1]{22k3_1.pdf}}%
  \end{picture}%
\endgroup%

%% file: 17k4_2.pdf_tex
\begingroup%
  \makeatletter%
  \providecommand\color[2][]{%
    \errmessage{(Inkscape) Color is used for the text in Inkscape, but the package 'color.sty' is not loaded}%
    \renewcommand\color[2][]{}%
  }%
  \providecommand\transparent[1]{%
    \errmessage{(Inkscape) Transparency is used (non-zero) for the text in Inkscape, but the package 'transparent.sty' is not loaded}%
    \renewcommand\transparent[1]{}%
  }%
  \providecommand\rotatebox[2]{#2}%
  \newcommand*\fsize{\dimexpr\f@size pt\relax}%
  \newcommand*\lineheight[1]{\fontsize{\fsize}{#1\fsize}\selectfont}%
  \ifx\svgwidth\undefined%
    \setlength{\unitlength}{444.93480026bp}%
    \ifx\svgscale\undefined%
      \relax%
    \else%
      \setlength{\unitlength}{\unitlength * \real{\svgscale}}%
    \fi%
  \else%
    \setlength{\unitlength}{\svgwidth}%
  \fi%
  \global\let\svgwidth\undefined%
  \global\let\svgscale\undefined%
  \makeatother%
  \begin{picture}(1,0.99582585)%
    \lineheight{1}%
    \setlength\tabcolsep{0pt}%
    \put(0,0){\includegraphics[width=\unitlength,page=1]{17k4_2.pdf}}%
  \end{picture}%
\endgroup%

%% file: 24k3_1.pdf_tex
\begingroup%
  \makeatletter%
  \providecommand\color[2][]{%
    \errmessage{(Inkscape) Color is used for the text in Inkscape, but the package 'color.sty' is not loaded}%
    \renewcommand\color[2][]{}%
  }%
  \providecommand\transparent[1]{%
    \errmessage{(Inkscape) Transparency is used (non-zero) for the text in Inkscape, but the package 'transparent.sty' is not loaded}%
    \renewcommand\transparent[1]{}%
  }%
  \providecommand\rotatebox[2]{#2}%
  \newcommand*\fsize{\dimexpr\f@size pt\relax}%
  \newcommand*\lineheight[1]{\fontsize{\fsize}{#1\fsize}\selectfont}%
  \ifx\svgwidth\undefined%
    \setlength{\unitlength}{446.80005026bp}%
    \ifx\svgscale\undefined%
      \relax%
    \else%
      \setlength{\unitlength}{\unitlength * \real{\svgscale}}%
    \fi%
  \else%
    \setlength{\unitlength}{\svgwidth}%
  \fi%
  \global\let\svgwidth\undefined%
  \global\let\svgscale\undefined%
  \makeatother%
  \begin{picture}(1,1)%
    \lineheight{1}%
    \setlength\tabcolsep{0pt}%
    \put(0,0){\includegraphics[width=\unitlength,page=1]{24k3_1.pdf}}%
  \end{picture}%
\endgroup%

%% file: 18k4_1.pdf_tex
\begingroup%
  \makeatletter%
  \providecommand\color[2][]{%
    \errmessage{(Inkscape) Color is used for the text in Inkscape, but the package 'color.sty' is not loaded}%
    \renewcommand\color[2][]{}%
  }%
  \providecommand\transparent[1]{%
    \errmessage{(Inkscape) Transparency is used (non-zero) for the text in Inkscape, but the package 'transparent.sty' is not loaded}%
    \renewcommand\transparent[1]{}%
  }%
  \providecommand\rotatebox[2]{#2}%
  \newcommand*\fsize{\dimexpr\f@size pt\relax}%
  \newcommand*\lineheight[1]{\fontsize{\fsize}{#1\fsize}\selectfont}%
  \ifx\svgwidth\undefined%
    \setlength{\unitlength}{440.15730026bp}%
    \ifx\svgscale\undefined%
      \relax%
    \else%
      \setlength{\unitlength}{\unitlength * \real{\svgscale}}%
    \fi%
  \else%
    \setlength{\unitlength}{\svgwidth}%
  \fi%
  \global\let\svgwidth\undefined%
  \global\let\svgscale\undefined%
  \makeatother%
  \begin{picture}(1,1.01509176)%
    \lineheight{1}%
    \setlength\tabcolsep{0pt}%
    \put(0,0){\includegraphics[width=\unitlength,page=1]{18k4_1.pdf}}%
  \end{picture}%
\endgroup%

%% file: 26k3_2.pdf_tex
\begingroup%
  \makeatletter%
  \providecommand\color[2][]{%
    \errmessage{(Inkscape) Color is used for the text in Inkscape, but the package 'color.sty' is not loaded}%
    \renewcommand\color[2][]{}%
  }%
  \providecommand\transparent[1]{%
    \errmessage{(Inkscape) Transparency is used (non-zero) for the text in Inkscape, but the package 'transparent.sty' is not loaded}%
    \renewcommand\transparent[1]{}%
  }%
  \providecommand\rotatebox[2]{#2}%
  \newcommand*\fsize{\dimexpr\f@size pt\relax}%
  \newcommand*\lineheight[1]{\fontsize{\fsize}{#1\fsize}\selectfont}%
  \ifx\svgwidth\undefined%
    \setlength{\unitlength}{443.61195026bp}%
    \ifx\svgscale\undefined%
      \relax%
    \else%
      \setlength{\unitlength}{\unitlength * \real{\svgscale}}%
    \fi%
  \else%
    \setlength{\unitlength}{\svgwidth}%
  \fi%
  \global\let\svgwidth\undefined%
  \global\let\svgscale\undefined%
  \makeatother%
  \begin{picture}(1,1.00718669)%
    \lineheight{1}%
    \setlength\tabcolsep{0pt}%
    \put(0,0){\includegraphics[width=\unitlength,page=1]{26k3_2.pdf}}%
  \end{picture}%
\endgroup%

%% file: 19k4_1.pdf_tex
\begingroup%
  \makeatletter%
  \providecommand\color[2][]{%
    \errmessage{(Inkscape) Color is used for the text in Inkscape, but the package 'color.sty' is not loaded}%
    \renewcommand\color[2][]{}%
  }%
  \providecommand\transparent[1]{%
    \errmessage{(Inkscape) Transparency is used (non-zero) for the text in Inkscape, but the package 'transparent.sty' is not loaded}%
    \renewcommand\transparent[1]{}%
  }%
  \providecommand\rotatebox[2]{#2}%
  \newcommand*\fsize{\dimexpr\f@size pt\relax}%
  \newcommand*\lineheight[1]{\fontsize{\fsize}{#1\fsize}\selectfont}%
  \ifx\svgwidth\undefined%
    \setlength{\unitlength}{445.30665026bp}%
    \ifx\svgscale\undefined%
      \relax%
    \else%
      \setlength{\unitlength}{\unitlength * \real{\svgscale}}%
    \fi%
  \else%
    \setlength{\unitlength}{\svgwidth}%
  \fi%
  \global\let\svgwidth\undefined%
  \global\let\svgscale\undefined%
  \makeatother%
  \begin{picture}(1,0.99665781)%
    \lineheight{1}%
    \setlength\tabcolsep{0pt}%
    \put(0,0){\includegraphics[width=\unitlength,page=1]{19k4_1.pdf}}%
  \end{picture}%
\endgroup%

%% file: 28k3_9.pdf_tex
\begingroup%
  \makeatletter%
  \providecommand\color[2][]{%
    \errmessage{(Inkscape) Color is used for the text in Inkscape, but the package 'color.sty' is not loaded}%
    \renewcommand\color[2][]{}%
  }%
  \providecommand\transparent[1]{%
    \errmessage{(Inkscape) Transparency is used (non-zero) for the text in Inkscape, but the package 'transparent.sty' is not loaded}%
    \renewcommand\transparent[1]{}%
  }%
  \providecommand\rotatebox[2]{#2}%
  \newcommand*\fsize{\dimexpr\f@size pt\relax}%
  \newcommand*\lineheight[1]{\fontsize{\fsize}{#1\fsize}\selectfont}%
  \ifx\svgwidth\undefined%
    \setlength{\unitlength}{433.08594551bp}%
    \ifx\svgscale\undefined%
      \relax%
    \else%
      \setlength{\unitlength}{\unitlength * \real{\svgscale}}%
    \fi%
  \else%
    \setlength{\unitlength}{\svgwidth}%
  \fi%
  \global\let\svgwidth\undefined%
  \global\let\svgscale\undefined%
  \makeatother%
  \begin{picture}(1,1)%
    \lineheight{1}%
    \setlength\tabcolsep{0pt}%
    \put(0,0){\includegraphics[width=\unitlength,page=1]{28k3_9.pdf}}%
  \end{picture}%
\endgroup%

%% file: 20k4_2.pdf_tex
\begingroup%
  \makeatletter%
  \providecommand\color[2][]{%
    \errmessage{(Inkscape) Color is used for the text in Inkscape, but the package 'color.sty' is not loaded}%
    \renewcommand\color[2][]{}%
  }%
  \providecommand\transparent[1]{%
    \errmessage{(Inkscape) Transparency is used (non-zero) for the text in Inkscape, but the package 'transparent.sty' is not loaded}%
    \renewcommand\transparent[1]{}%
  }%
  \providecommand\rotatebox[2]{#2}%
  \newcommand*\fsize{\dimexpr\f@size pt\relax}%
  \newcommand*\lineheight[1]{\fontsize{\fsize}{#1\fsize}\selectfont}%
  \ifx\svgwidth\undefined%
    \setlength{\unitlength}{446.80005026bp}%
    \ifx\svgscale\undefined%
      \relax%
    \else%
      \setlength{\unitlength}{\unitlength * \real{\svgscale}}%
    \fi%
  \else%
    \setlength{\unitlength}{\svgwidth}%
  \fi%
  \global\let\svgwidth\undefined%
  \global\let\svgscale\undefined%
  \makeatother%
  \begin{picture}(1,1)%
    \lineheight{1}%
    \setlength\tabcolsep{0pt}%
    \put(0,0){\includegraphics[width=\unitlength,page=1]{20k4_2.pdf}}%
  \end{picture}%
\endgroup%

%% file: 30k3_39.pdf_tex
\begingroup%
  \makeatletter%
  \providecommand\color[2][]{%
    \errmessage{(Inkscape) Color is used for the text in Inkscape, but the package 'color.sty' is not loaded}%
    \renewcommand\color[2][]{}%
  }%
  \providecommand\transparent[1]{%
    \errmessage{(Inkscape) Transparency is used (non-zero) for the text in Inkscape, but the package 'transparent.sty' is not loaded}%
    \renewcommand\transparent[1]{}%
  }%
  \providecommand\rotatebox[2]{#2}%
  \newcommand*\fsize{\dimexpr\f@size pt\relax}%
  \newcommand*\lineheight[1]{\fontsize{\fsize}{#1\fsize}\selectfont}%
  \ifx\svgwidth\undefined%
    \setlength{\unitlength}{446.80005026bp}%
    \ifx\svgscale\undefined%
      \relax%
    \else%
      \setlength{\unitlength}{\unitlength * \real{\svgscale}}%
    \fi%
  \else%
    \setlength{\unitlength}{\svgwidth}%
  \fi%
  \global\let\svgwidth\undefined%
  \global\let\svgscale\undefined%
  \makeatother%
  \begin{picture}(1,0.99463888)%
    \lineheight{1}%
    \setlength\tabcolsep{0pt}%
    \put(0,0){\includegraphics[width=\unitlength,page=1]{30k3_39.pdf}}%
  \end{picture}%
\endgroup%

%% file: 21k4_3.pdf_tex
\begingroup%
  \makeatletter%
  \providecommand\color[2][]{%
    \errmessage{(Inkscape) Color is used for the text in Inkscape, but the package 'color.sty' is not loaded}%
    \renewcommand\color[2][]{}%
  }%
  \providecommand\transparent[1]{%
    \errmessage{(Inkscape) Transparency is used (non-zero) for the text in Inkscape, but the package 'transparent.sty' is not loaded}%
    \renewcommand\transparent[1]{}%
  }%
  \providecommand\rotatebox[2]{#2}%
  \newcommand*\fsize{\dimexpr\f@size pt\relax}%
  \newcommand*\lineheight[1]{\fontsize{\fsize}{#1\fsize}\selectfont}%
  \ifx\svgwidth\undefined%
    \setlength{\unitlength}{445.57740026bp}%
    \ifx\svgscale\undefined%
      \relax%
    \else%
      \setlength{\unitlength}{\unitlength * \real{\svgscale}}%
    \fi%
  \else%
    \setlength{\unitlength}{\svgwidth}%
  \fi%
  \global\let\svgwidth\undefined%
  \global\let\svgscale\undefined%
  \makeatother%
  \begin{picture}(1,0.99726378)%
    \lineheight{1}%
    \setlength\tabcolsep{0pt}%
    \put(0,0){\includegraphics[width=\unitlength,page=1]{21k4_3.pdf}}%
  \end{picture}%
\endgroup%

%% file: 32k3_52.pdf_tex
\begingroup%
  \makeatletter%
  \providecommand\color[2][]{%
    \errmessage{(Inkscape) Color is used for the text in Inkscape, but the package 'color.sty' is not loaded}%
    \renewcommand\color[2][]{}%
  }%
  \providecommand\transparent[1]{%
    \errmessage{(Inkscape) Transparency is used (non-zero) for the text in Inkscape, but the package 'transparent.sty' is not loaded}%
    \renewcommand\transparent[1]{}%
  }%
  \providecommand\rotatebox[2]{#2}%
  \newcommand*\fsize{\dimexpr\f@size pt\relax}%
  \newcommand*\lineheight[1]{\fontsize{\fsize}{#1\fsize}\selectfont}%
  \ifx\svgwidth\undefined%
    \setlength{\unitlength}{444.69450026bp}%
    \ifx\svgscale\undefined%
      \relax%
    \else%
      \setlength{\unitlength}{\unitlength * \real{\svgscale}}%
    \fi%
  \else%
    \setlength{\unitlength}{\svgwidth}%
  \fi%
  \global\let\svgwidth\undefined%
  \global\let\svgscale\undefined%
  \makeatother%
  \begin{picture}(1,1)%
    \lineheight{1}%
    \setlength\tabcolsep{0pt}%
    \put(0,0){\includegraphics[width=\unitlength,page=1]{32k3_52.pdf}}%
  \end{picture}%
\endgroup%

%% file: 32k4_8544.pdf_tex
\begingroup%
  \makeatletter%
  \providecommand\color[2][]{%
    \errmessage{(Inkscape) Color is used for the text in Inkscape, but the package 'color.sty' is not loaded}%
    \renewcommand\color[2][]{}%
  }%
  \providecommand\transparent[1]{%
    \errmessage{(Inkscape) Transparency is used (non-zero) for the text in Inkscape, but the package 'transparent.sty' is not loaded}%
    \renewcommand\transparent[1]{}%
  }%
  \providecommand\rotatebox[2]{#2}%
  \newcommand*\fsize{\dimexpr\f@size pt\relax}%
  \newcommand*\lineheight[1]{\fontsize{\fsize}{#1\fsize}\selectfont}%
  \ifx\svgwidth\undefined%
    \setlength{\unitlength}{446.80005026bp}%
    \ifx\svgscale\undefined%
      \relax%
    \else%
      \setlength{\unitlength}{\unitlength * \real{\svgscale}}%
    \fi%
  \else%
    \setlength{\unitlength}{\svgwidth}%
  \fi%
  \global\let\svgwidth\undefined%
  \global\let\svgscale\undefined%
  \makeatother%
  \begin{picture}(1,1)%
    \lineheight{1}%
    \setlength\tabcolsep{0pt}%
    \put(0,0){\includegraphics[width=\unitlength,page=1]{32k4_8544.pdf}}%
  \end{picture}%
\endgroup%

%% file: table3-color.tex
\begin{table}[htbp]
\centering
\caption{Other optimal graphs.}
\label{rest}
\begin{tabular}{|cr|cr|cr|cr|}
\hline
\multicolumn{2}{|c|}{$(16,3)$}                                                           & \multicolumn{2}{c|}{$(21,4)$}                                                           & \multicolumn{2}{c|}{$(32,4)$}                                                               & \multicolumn{2}{c|}{$(32,4)$}                                                               \\ \hline
\rowcolor[HTML]{ffffc7} 
\multicolumn{1}{|l}{\cellcolor[HTML]{ffffc7}2.2000}                  & 2                 & \multicolumn{1}{l}{\cellcolor[HTML]{ffffc7}2.0000}                 & 2                  & \multicolumn{1}{l}{\cellcolor[HTML]{ffffc7}2.3548}                    & 3                   & \multicolumn{1}{l}{\cellcolor[HTML]{ffffc7}2.3548}                    & 3                   \\
\rowcolor[HTML]{ffffc7} 
\multicolumn{2}{|c|}{\cellcolor[HTML]{ffffc7}\def\svgwidth{1.5cm}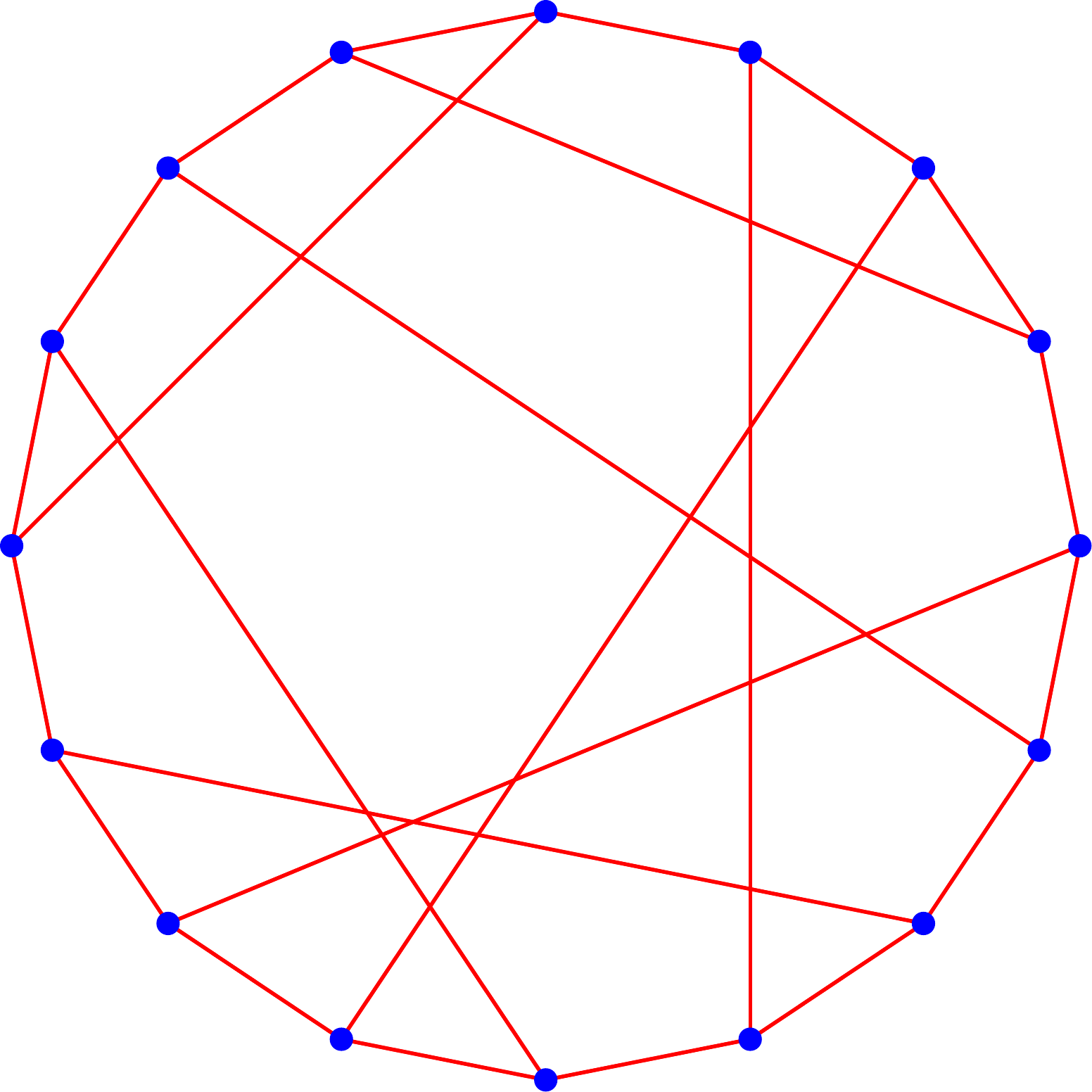} & \multicolumn{2}{c|}{\cellcolor[HTML]{ffffc7}\def\svgwidth{1.5cm}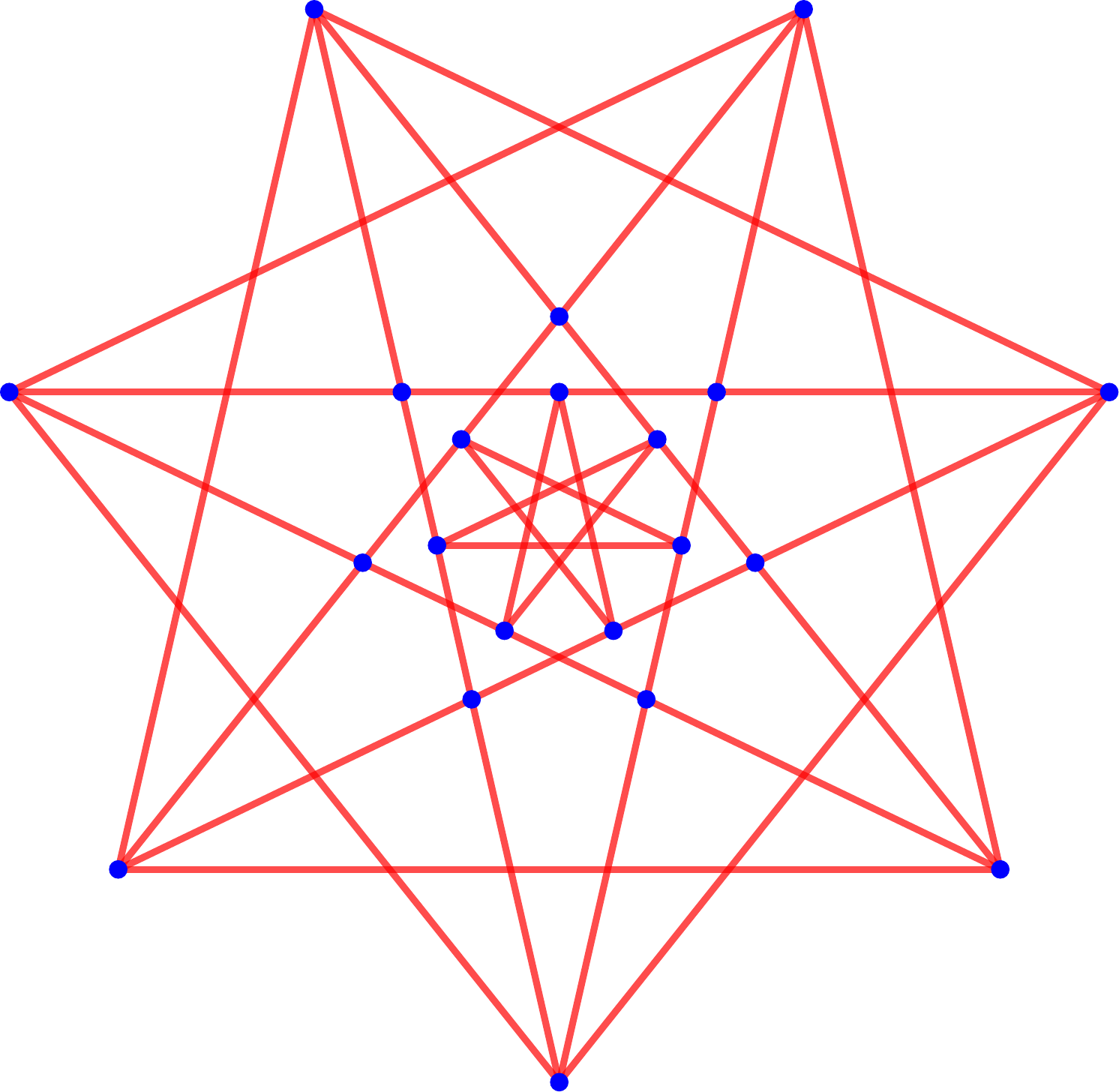} & \multicolumn{2}{c|}{\cellcolor[HTML]{ffffc7}\def\svgwidth{1.5cm}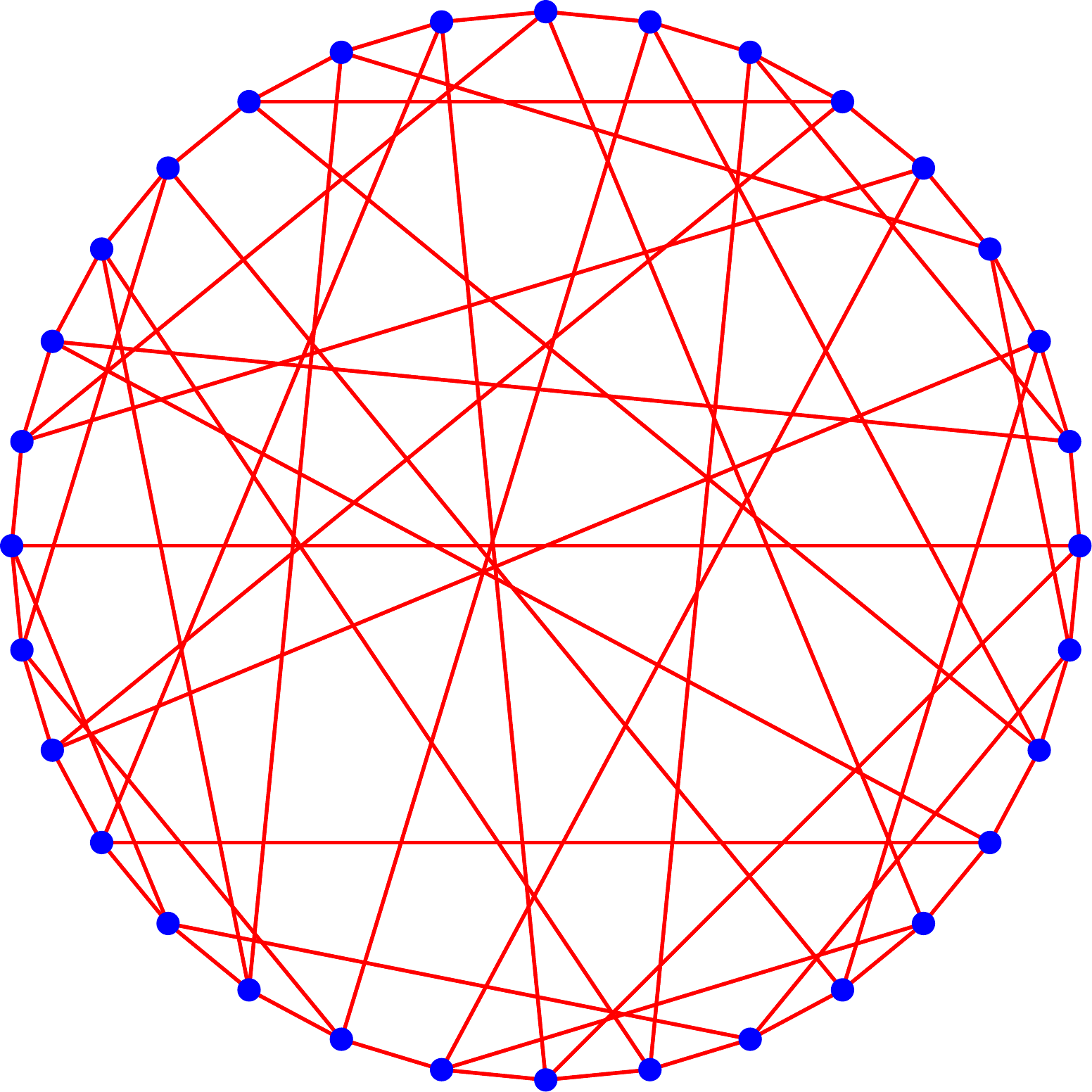} & \multicolumn{2}{c|}{\cellcolor[HTML]{ffffc7}\def\svgwidth{1.5cm}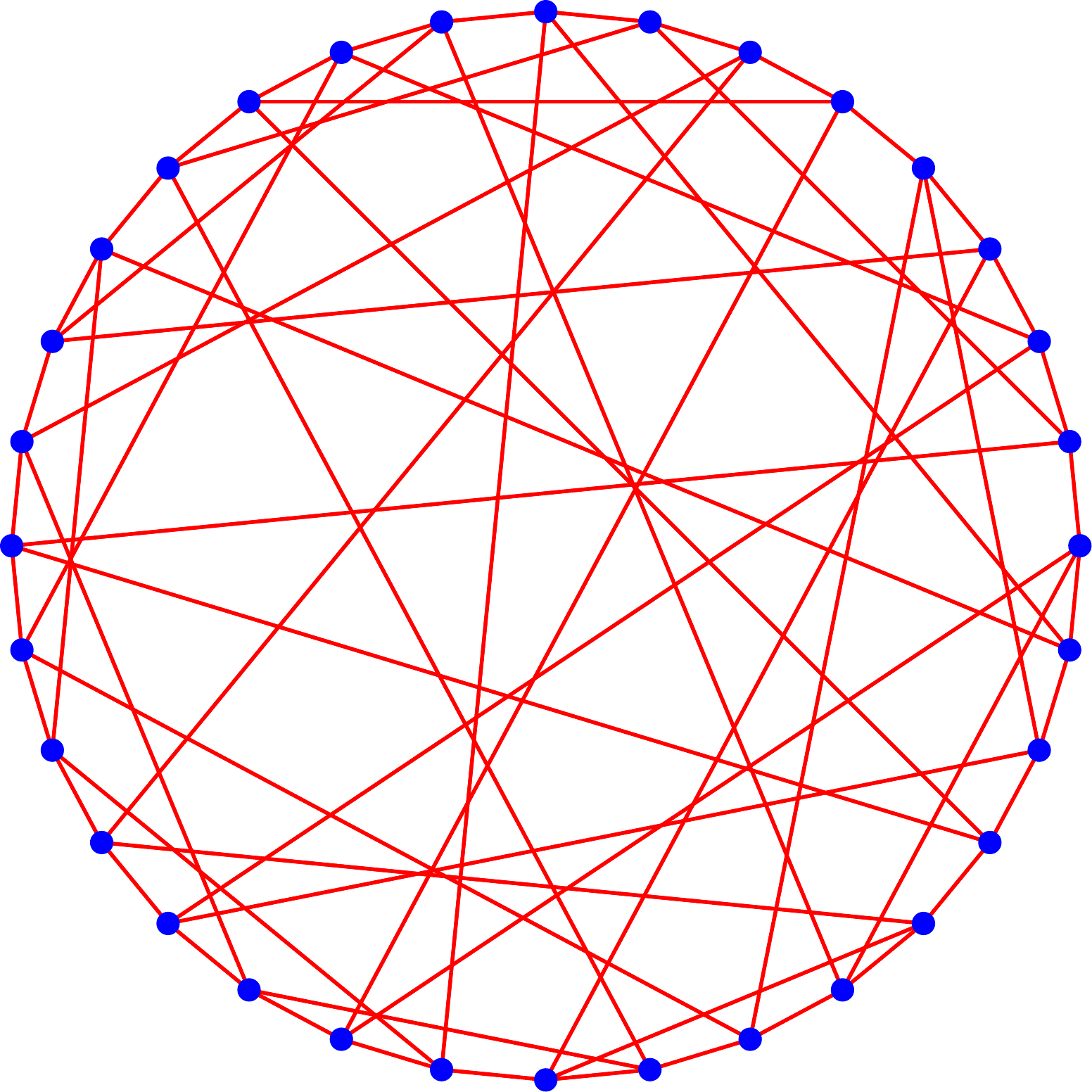} \\
\rowcolor[HTML]{ffffc7} 
\multicolumn{1}{|l}{\cellcolor[HTML]{ffffc7}6}                       & 6                 & \multicolumn{1}{l}{\cellcolor[HTML]{ffffc7}14}                     & 14                 & \multicolumn{1}{l}{\cellcolor[HTML]{ffffc7}18}                        & 3                   & \multicolumn{1}{l}{\cellcolor[HTML]{ffffc7}18}                        & 3                   \\
\rowcolor[HTML]{ffffc7} 
\multicolumn{2}{|c|}{\cellcolor[HTML]{ffffc7} $S_{3}$}                              & \multicolumn{2}{c|}{\cellcolor[HTML]{ffffc7} $D_{7}$}                              & \multicolumn{2}{c|}{\cellcolor[HTML]{ffffc7} $C_{3}$}                                  & \multicolumn{2}{c|}{\cellcolor[HTML]{ffffc7} $C_{3}$}                                  \\ \hline
\end{tabular}
\end{table}

%% file: 16k3_4.pdf_tex
\begingroup%
  \makeatletter%
  \providecommand\color[2][]{%
    \errmessage{(Inkscape) Color is used for the text in Inkscape, but the package 'color.sty' is not loaded}%
    \renewcommand\color[2][]{}%
  }%
  \providecommand\transparent[1]{%
    \errmessage{(Inkscape) Transparency is used (non-zero) for the text in Inkscape, but the package 'transparent.sty' is not loaded}%
    \renewcommand\transparent[1]{}%
  }%
  \providecommand\rotatebox[2]{#2}%
  \newcommand*\fsize{\dimexpr\f@size pt\relax}%
  \newcommand*\lineheight[1]{\fontsize{\fsize}{#1\fsize}\selectfont}%
  \ifx\svgwidth\undefined%
    \setlength{\unitlength}{446.80005026bp}%
    \ifx\svgscale\undefined%
      \relax%
    \else%
      \setlength{\unitlength}{\unitlength * \real{\svgscale}}%
    \fi%
  \else%
    \setlength{\unitlength}{\svgwidth}%
  \fi%
  \global\let\svgwidth\undefined%
  \global\let\svgscale\undefined%
  \makeatother%
  \begin{picture}(1,1)%
    \lineheight{1}%
    \setlength\tabcolsep{0pt}%
    \put(0,0){\includegraphics[width=\unitlength,page=1]{16k3_4.pdf}}%
  \end{picture}%
\endgroup%

%% file: 21k4_5.pdf_tex
\begingroup%
  \makeatletter%
  \providecommand\color[2][]{%
    \errmessage{(Inkscape) Color is used for the text in Inkscape, but the package 'color.sty' is not loaded}%
    \renewcommand\color[2][]{}%
  }%
  \providecommand\transparent[1]{%
    \errmessage{(Inkscape) Transparency is used (non-zero) for the text in Inkscape, but the package 'transparent.sty' is not loaded}%
    \renewcommand\transparent[1]{}%
  }%
  \providecommand\rotatebox[2]{#2}%
  \newcommand*\fsize{\dimexpr\f@size pt\relax}%
  \newcommand*\lineheight[1]{\fontsize{\fsize}{#1\fsize}\selectfont}%
  \ifx\svgwidth\undefined%
    \setlength{\unitlength}{429.51561738bp}%
    \ifx\svgscale\undefined%
      \relax%
    \else%
      \setlength{\unitlength}{\unitlength * \real{\svgscale}}%
    \fi%
  \else%
    \setlength{\unitlength}{\svgwidth}%
  \fi%
  \global\let\svgwidth\undefined%
  \global\let\svgscale\undefined%
  \makeatother%
  \begin{picture}(1,0.97577213)%
    \lineheight{1}%
    \setlength\tabcolsep{0pt}%
    \put(0,0){\includegraphics[width=\unitlength,page=1]{21k4_5.pdf}}%
  \end{picture}%
\endgroup%

%% file: 32k4_10457.pdf_tex
\begingroup%
  \makeatletter%
  \providecommand\color[2][]{%
    \errmessage{(Inkscape) Color is used for the text in Inkscape, but the package 'color.sty' is not loaded}%
    \renewcommand\color[2][]{}%
  }%
  \providecommand\transparent[1]{%
    \errmessage{(Inkscape) Transparency is used (non-zero) for the text in Inkscape, but the package 'transparent.sty' is not loaded}%
    \renewcommand\transparent[1]{}%
  }%
  \providecommand\rotatebox[2]{#2}%
  \newcommand*\fsize{\dimexpr\f@size pt\relax}%
  \newcommand*\lineheight[1]{\fontsize{\fsize}{#1\fsize}\selectfont}%
  \ifx\svgwidth\undefined%
    \setlength{\unitlength}{446.80005026bp}%
    \ifx\svgscale\undefined%
      \relax%
    \else%
      \setlength{\unitlength}{\unitlength * \real{\svgscale}}%
    \fi%
  \else%
    \setlength{\unitlength}{\svgwidth}%
  \fi%
  \global\let\svgwidth\undefined%
  \global\let\svgscale\undefined%
  \makeatother%
  \begin{picture}(1,1)%
    \lineheight{1}%
    \setlength\tabcolsep{0pt}%
    \put(0,0){\includegraphics[width=\unitlength,page=1]{32k4_10457.pdf}}%
  \end{picture}%
\endgroup%

%% file: 32k4_10460.pdf_tex
\begingroup%
  \makeatletter%
  \providecommand\color[2][]{%
    \errmessage{(Inkscape) Color is used for the text in Inkscape, but the package 'color.sty' is not loaded}%
    \renewcommand\color[2][]{}%
  }%
  \providecommand\transparent[1]{%
    \errmessage{(Inkscape) Transparency is used (non-zero) for the text in Inkscape, but the package 'transparent.sty' is not loaded}%
    \renewcommand\transparent[1]{}%
  }%
  \providecommand\rotatebox[2]{#2}%
  \newcommand*\fsize{\dimexpr\f@size pt\relax}%
  \newcommand*\lineheight[1]{\fontsize{\fsize}{#1\fsize}\selectfont}%
  \ifx\svgwidth\undefined%
    \setlength{\unitlength}{446.80005026bp}%
    \ifx\svgscale\undefined%
      \relax%
    \else%
      \setlength{\unitlength}{\unitlength * \real{\svgscale}}%
    \fi%
  \else%
    \setlength{\unitlength}{\svgwidth}%
  \fi%
  \global\let\svgwidth\undefined%
  \global\let\svgscale\undefined%
  \makeatother%
  \begin{picture}(1,1)%
    \lineheight{1}%
    \setlength\tabcolsep{0pt}%
    \put(0,0){\includegraphics[width=\unitlength,page=1]{32k4_10460.pdf}}%
  \end{picture}%
\endgroup%